\pdfoutput=1 

\documentclass{revtex4-2}
\usepackage[utf8]{inputenc}
\usepackage[english]{babel}

\usepackage[hidelinks]{hyperref}

\usepackage{graphicx}

\usepackage{floatrow}
\usepackage[caption=false]{subfig}

\usepackage{parskip}

\usepackage{float}
\usepackage{booktabs}
\usepackage{makecell}
\usepackage{multirow}

\usepackage{amsmath}
\usepackage{amssymb}
\usepackage{amsfonts}
\usepackage{mathtools}

\usepackage{upgreek}
\usepackage{xcolor}
\usepackage[normalem]{ulem}

\AtBeginDocument{}
\AtBeginDocument{}

\begin{document}

\title{Magnetized Thick Disks around Boson Stars}
\date{\today}

\author{Kristian Gjorgjieski}
 \email{kristian.gjorgjieski@uol.de}
\author{Jutta Kunz}
 \email{jutta.kunz@uni-oldenburg.de}
\author{Matheus C. Teodoro}
 \email{matheus.do.carmo.teodoro@uni-oldenburg.de}
\affiliation{Department of Physics, Carl von Ossietzky University of Oldenburg, 26111 Oldenburg, Germany}

\author{Lucas G. Collodel}
 \email{lucas.gardai-collodel@uni-tuebingen.de}
\affiliation{Theoretical Astrophysics, University of T\"ubingen, 72076 T\"ubingen, Germany,}

\author{Petya Nedkova}
 \email{pnedkova@phys.uni-sofia.bg}
\affiliation{Department of Theoretical Physics, Sofia University, Sofia 1164, Bulgaria}

\begin{abstract}
    {\small The effects of magnetic fields on accretion disks around compact objects are of high importance in the study of their general properties and dynamics. Here we analyze the influence of magnetic fields on thick accretion disks around rotating boson stars. We assume a uniform constant specific angular momentum distribution and a polytropic equation of state. The purely hydrodynamical thick disk solutions are extended to magnetized solutions by adding a toroidal magnetic field and then analyzed in terms of a magnetization parameter $\beta_{mc}$. We consider one-centered solutions as well as two-centered solutions and focus on retrograde tori, since they are more distinctive due to their unique properties. Our computed solutions indicate that strong magnetic fields influence the characteristics of thick disks around rotating boson stars and possibly affect their unique features.}
\end{abstract}

\maketitle


\section{Introduction}

Accretion disks are formed around various astrophysical objects ranging from supermassive active galactic nuclei to coalescent stellar mass black holes and neutron star binaries. Converting gravitational energy into radiation they harbor many high-energy astrophysical phenomena and thus can serve as a natural arena for probing the gravitational physics around compact objects. With the development of the imaging techniques we are now able to observe directly the accretion disks in the nearby galactic targets M87 and Sgr A*, which provides a unique opportunity for gaining knowledge about the nature of the compact objects in their centers and the physical processes in the accreting plasma \cite{EHT_1, EHT_10, EHT_2}.

The interpretation of the astrophysical observations relies considerably on the accretion disk modeling. Currently general relativistic
magneto-hydrodynamic (GRMHD) simulations are developed which are able to integrate numerically the coupled Einstein-Euler equations thus providing a precise description of the self-gravitating accreting system. Although the computations can be simplified by assuming axial symmetry or using a test-fluid approximation, these simulations produce a complicated picture depending on many parameters which can be hard to disentangle and interpret. In this respect simplified semi-analytical models can be extremely useful since they may capture the main effects from the non-linear treatment with less computational cost and offer more physical intuition and predictability.

Some of the basic constructions which describe geometrically thick accretion disks are the so called \textit{Polish doughnuts}. They represent the simplest solutions to the relativistic Euler equations assuming a non-selfgravitating perfect fluid and neglecting the electromagnetic, viscosity and radiation terms. Thus, they provide the equilibrium states of the relativistic matter orbiting on non-geodesic trajectories around the compact object, which result from the balance of the pressure gradients and the gravitational and centrifugal forces.

The theory of the equilibrium tori has a long history dating back to the classical works \cite{Fishbone1,
Abramowicz, Kozlowski} where the equilibrium tori for the Kerr black hole were obtained. Assuming an isentropic or barotropic equation of state for the perfect fluid it was demonstrated that in this case the constant angular velocity and the constant angular momentum surfaces coincide. This statement, known as the von Zeipel theorem, represents an integrability condition for the relativistic Euler equation and therefore it allows the construction of analytical solutions.

Various equilibrium tori can be constructed possessing a different profile of the specific angular momentum \cite{Qian, Daigne}. Solutions with a constant  angular momentum  play a fundamental role since despite their simplicity they already capture the main qualitative features of the more general configurations. The physical conditions can be further extended by adding a toroidal magnetic field which allows for the construction of analytical solutions under the same integrability conditions \cite{Komissarov:2006nz} (see also \cite{Font} and \cite{Wielgus} for non-constant angular momentum configurations). This generalization is particularly important for the astrophysical applications since magnetic fields play a dynamical role in astrophysical scenarios and the magnetized equilibrium configurations provide suitable initial conditions for the GRMHD simulations of the accretion process or jet formation (see e. g. \cite{Igumenshchev, McKinney}). Magnetized tori can be further generalized by including shear viscosity effects \cite{Lahiri_2020}.

Although the accretion disk theory was developed considering mainly the Kerr black hole, fundamental physics suggests that more diverse compact objects may exist in nature. This motivates the construction of accretion disk models in more exotic spacetimes with the view of confronting the simulations with observational data and searching for astrophysical signatures of new fundamental objects \cite{Vincent_2016, Olivares,Vincent_2021, EHT_Sgr}. In this line of research thick accretion disks around black holes in the modified theories of gravity were studied \cite{ Font_2019, Font_2021, Teodoro_2021, Gimeno-Soler}, equilibrium configurations in de Sitter background \cite{Stuchlik_2000, Slany} or interacting with an external matter distribution \cite{Faraji_2021a, Faraji_2021b},  as well as thick disks in  naked singularity spacetimes \cite{Adamek, Stuchlik_2015}.

In our work we concentrate on boson stars. Boson stars represent compact scalar field condensates which  form as equilibrium states resulting from the gravitational collapse of a self-gravitating massive scalar field \cite{Jetzer, Schunck, Liebling}. They produce a strong gravitational field and negligible electromagnetic emission, in this way they are mimicking the properties of black holes. Various boson star configurations have been constructed \cite{Kaup, Feinblum, Ruffini, Kobayashi, Yoshida, Schunck_1998, Kleihaus, Kleihaus_2008, Collodel_2017, Collodel_2019}, and their properties were investigated such as stability \cite{Lee, Kusmartsev, Sanchis, Siemonsen:2020hcg},  geodesic motion \cite{Eilers, Brihaye, Grandclement, Meliani, Grould, Collodel_2017a} and tidal effects in the motion of gas clouds around them \cite{Meliani_2017, Teodoro_2021a}. Since they are horizonless compact objects, which are not characterized by a solid boundary but  rather  by a decaying profile of the scalar field density towards infinity, the particles and light propagation in their vicinity is specific leading to phenomenological signatures. For example, boson stars may possess only bound circular orbits which extend into the inner regions with high density of the scalar field distribution \cite{Meliani}. In addition, photon rings may be absent which prevents the formation of a shadow in the strict sense. Yet, if we assume the presence of an accretion disk, a central dark region will be observed similar to black holes which corresponds to the lensed image of an accretion disk's inner edge \cite{Vincent_2016, Olivares, Vincent_2021, EHT_Sgr}.

The perfect fluid equilibrium tori around boson stars also possess distinctive morphology compared to the Kerr black hole \cite{Meliani, Teodoro_2021b}. The equilibrium tori in the Kerr spacetime are characterized by two different types of solutions, which are distinguished by the presence of a cusp located at the inner edge. Solutions with a cusp emerge if the specific angular momentum of the disk lies between the specific angular momenta corresponding to the marginally stable and the marginally bound orbit \cite{Abramowicz, Font_2002}. In contrast, the equilibrium configurations for boson stars constructed in the literature do not form an inner cusp. This distinction has implications on the possible scenarios for dynamical evolution of the accretion disk such as the formation of run-away instabilities \cite{Abramowicz_1983, Daigne, Font_2002}. Boson stars can further support topologically non-trivial configurations like two-centered tori which can be disjoint or connected by a cusp. These topologies are absent for the Kerr black hole, although they occur for other exotic compact objects like naked singularities \cite{Stuchlik_2000, Slany}. Another interesting feature is the formation of static surfaces for counter-rotating equilibrium tori around boson stars \cite{Collodel_2017, Teodoro_2021b}. These surfaces are defined as the cross-sections where the fluid is at rest with respect to a zero angular momentum observer (ZAMO) at infinity. Thus, they serve as a boundary separating regions where the fluid moves in prograde and retrograde direction, respectively.

The aim of our work is to study the equilibrium tori in boson star spacetimes in the presence of a toroidal magnetic field. We construct constant angular momentum configurations and analyse how their properties are modified with respect to the purely hydrodynamical case. In particular we describe the qualitative effects which are induced by the magnetic field and may  have observational implications. These include modifications in the location of the predominant disk density and its compactness, as well as variation of  the characteristic geometry of the equidensity surfaces. Another interesting phenomenon is that the magnetic field can trigger topological transitions between different thick disk configurations. Thus, after a certain magnitude of the magnetic field, two-centered disks may lose their outer center and become one-centered.

The paper is organized as follows. In the next section we describe the boson star solutions which we consider and some of their relevant properties. In section 3 we briefly review the equilibrium tori configurations, which are possible in boson star spacetimes in the purely hydrodynamical case. In section 4 we describe the construction procedure of magnetized tori for a perfect fluid with polytropic equation of state. In section 5 we present our results and the analysis of the properties of the magnetized thick disk, which are illustrated on a range of representative examples. Section 6 contains our conclusions.

\section{Boson Stars}
We consider a complex scalar field without self-interaction. The boson stars (BSs) are obtained from the Einstein-Klein-Gordon equations, derived from the action $\mathcal{S}$
\begin{align}
    \mathcal{S} = \int \sqrt{-g} \left(\frac{R}{16 \pi G} - \mathcal{L}_m \right) \mathrm{d}^4x, \label{eq:Action}
\end{align}
where $g$ is the metric determinant, $R$ is the Ricci scalar, $G$ is Newton's constant and $\mathcal{L}_m$ is the Lagrangian of the complex scalar field $\phi$ with mass $m$
\begin{align}
    \mathcal{L}_m = |\partial_\mu \phi|^2 + m^2|\phi|^2.
\end{align}

The action (\ref{eq:Action}) is invariant under transformations of the global U(1)-symmetry group of the complex scalar field, $\phi \rightarrow \phi e^{i\tau}$, with constant $\tau$. According to Noether's theorem this invariance implies the existence of a conserved current, $\nabla_\mu j^\mu =0$, with
\begin{align}
    j_\mu = i \left(\phi \partial_\mu \phi^* - \phi^* \partial_\mu \phi \right), \label{eq:current}
\end{align}
and a conserved Noether charge $Q$, which is the bosonic particle number given by $Q = \int \sqrt{-g} j^t \mathrm{d}^3 x$. 

The variation of the action (\ref{eq:Action}) leads to the coupled set of Einstein-Klein-Gordon equations,
\begin{align}
    R_{\mu\nu} - \frac{1}{2} R g_{\mu\nu} &= 8 \pi G T_{\mu\nu}, \\
    \left(\Box - m^2 \right)\phi &= 0,
\end{align}
where $T_{\mu\nu} \equiv \left(\partial_\mu \phi\partial_\nu \phi^* + \partial_\nu \phi \partial_\mu \phi^* \right) - \mathcal{L}_m g_{\mu\nu}$ is the stress-energy tensor and $\Box$ denotes the covariant d'Alembert operator. In order to obtain rotating BSs, the scalar field should depend on all four coordinates as follows \cite{Schunck_1998},
\begin{align}
    \phi \equiv \phi_0(r,\theta) e^{i\left(\omega t - k \varphi \right)}.
    \label{has}
\end{align}
A harmonic time-dependence is already needed for the classical non-rotating BSs in order to obtain stable localized solutions. The presence of rotation implies an additional harmonic dependence of the scalar field on the azimuthal angle $\varphi$. 
In the Ansatz (\ref{has}) $\omega$ is the angular frequency of the scalar field and $k$ is the azimuthal winding number.
Due to the condition $\phi(\varphi=0) = \phi(\varphi=2\pi)$ the winding number $k$ must be an integer ($k$ signals the strength of the angular excitations since it counts the nodes, $2k$, of the real and imaginary parts of the scalar field along the azimuthal direction).

The angular momentum $J$ of the BSs is given by a quantisation law \cite{Schunck_1998},
\begin{align}
    J = k Q,
\end{align}
thus it is an integer multiple of the charge $Q$. This law follows directly from the relation $T^t_\varphi= -k j^t$ with $\partial_\varphi \phi = -i k \phi$.
The harmonic Ansatz for the scalar field yields a stress-energy tensor that does not depend on the coordinates $t$ and $\varphi$.
Thus solutions with a stationary and axially symmetric metric result, implying the presence of two Killing vectors of the metric associated with these two coordinates. The line element of the BS spacetime metric can then be written as,
\begin{align}
    \mathrm{d}s^2 = &- \alpha^2 \mathrm{d}t^2  + A^2 \left(\mathrm{d}r^2 + r^2 \mathrm{d}\theta^2  \right) 
    \\ &+ B^2 r^2 \sin^2 \theta  \left(\mathrm{d}\varphi  + \beta^\varphi \mathrm{d}t \right)^2,
\end{align}
with $\alpha$ the lapse function, $\beta^\varphi$ the shift function, and the functions $A, B, \alpha, \beta$ which depend only on $r$ and $\theta$. 

Substituting the harmonic Ansatz (\ref{has}) into the field equations leads to a coupled set of partial differential equations for the functions, which were solved numerically. The BSs constructed are asymptotically flat.Thus the metric approaches asymptotically Minkowski spacetime and the scalar field vanishes exponentially proportional to $e^{-\sqrt{m^2 - \omega^2}r}/r$. The solutions were computed with the FIDISOL/CADSOL package, which is a PDE solver which employs a finite difference method of discretization together with a Newton-Raphson scheme to linearize the resulting system of algebraic equations \cite{Schauder:1992}. Solutions exist only for a set of angular frequencies $\omega$.

Here we consider only rotating BSs with winding number $k = 1$. Furthermore, we focus only on solutions without ergoregions. Therefore the minimal angular frequency taken into account is $\omega =0.655$. Solutions with smaller angular frequencies would possess ergoregions \cite{Grandclement}. It should be noted, that the equations feature a scaling symmetry, which we exploit to go to dimensionless quantities by scaling with the boson mass $m$, i.e., $\tilde r= rm$, $\tilde\omega =\frac{\omega}{m}$ and $\tilde M = Mm$. In the following we omit the tilde again for brevity.

\begin{figure}
\centering
   \subfloat[\label{BS_mass}]{\includegraphics[width=0.485\columnwidth]{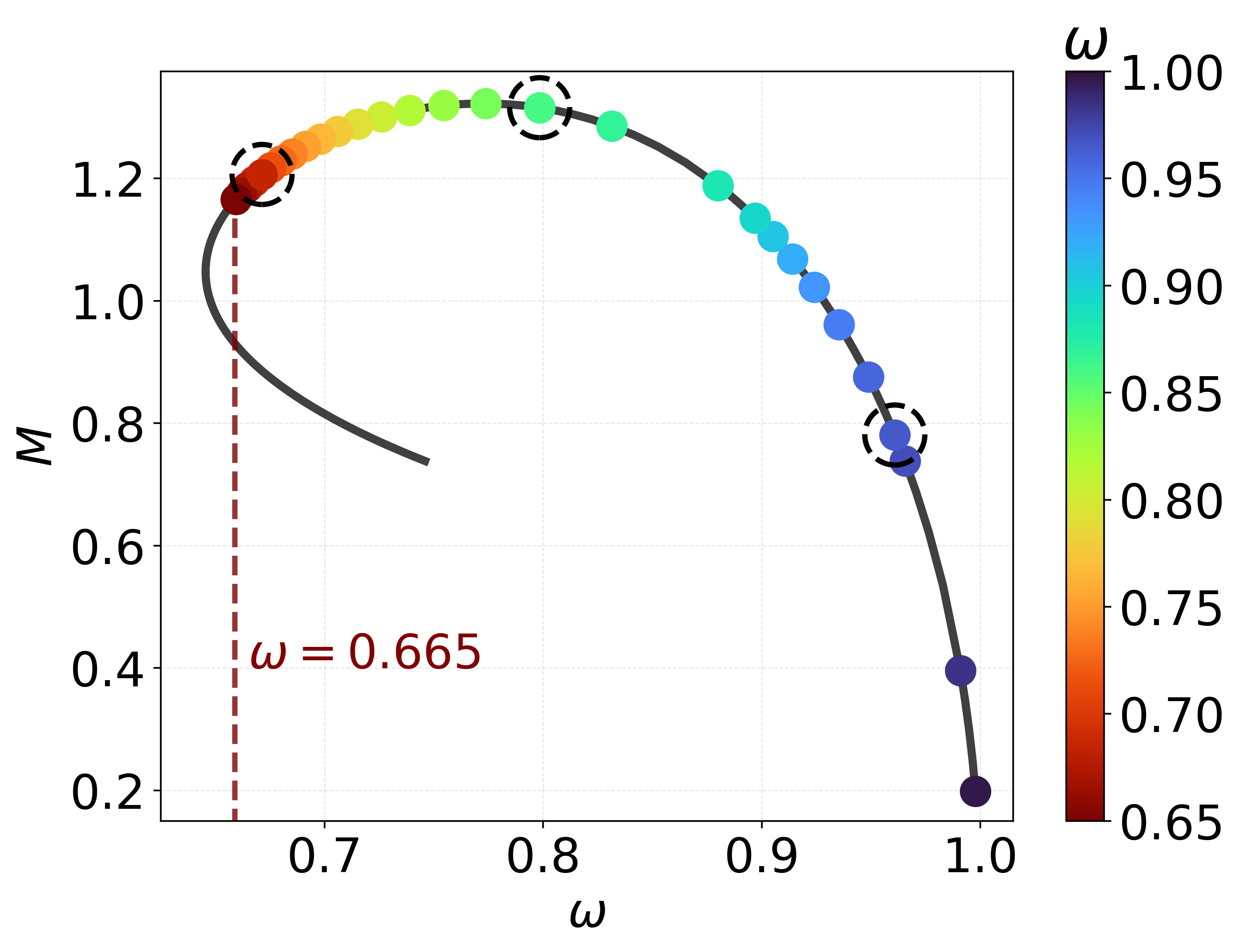}}\hfill
   \subfloat[\label{BS_modulus}]{\includegraphics[width=0.485\columnwidth]{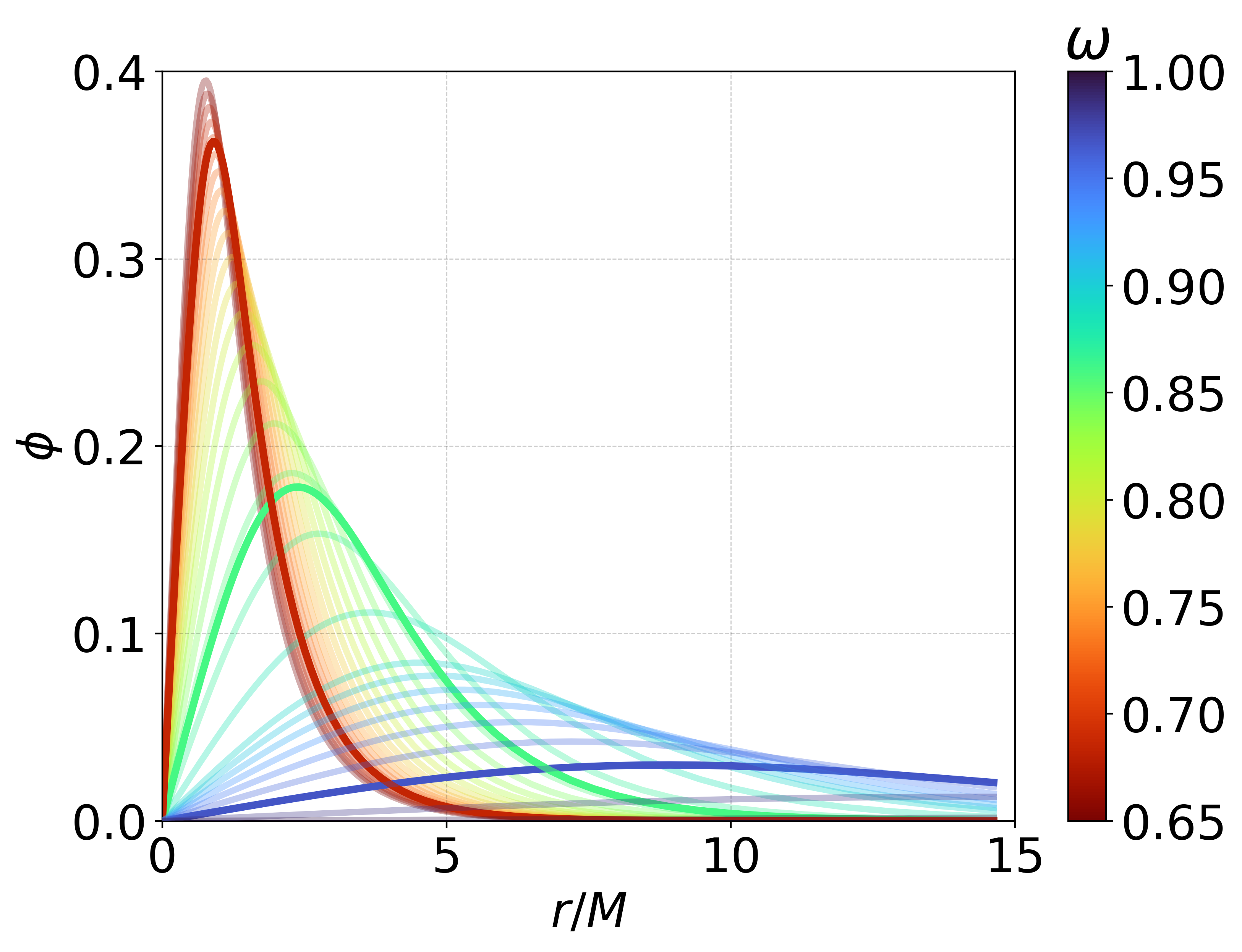}}
\caption{(a) Mass $M$ of BSs versus boson field frequency $\omega$. The dashed circles mark the solutions used for the computation of the magnetized disks. The dashed red line marks the threshold value of $\omega$ below which the solutions contain ergoregions. (b) The amplitude of the scalar field $\phi$ versus the normalized radial coordinate $r/M$ for the BS solutions. The highlighted curves show the solutions marked in (a), namely $\omega = \{0.671,0.798,0.960\}$.}
\label{fig:BS_Mass_Modulus}
\end{figure}

The solutions with lower angular frequencies $\omega$ are more relativistic solutions, whereas the angular frequencies near 1 are close to vacuum solutions. 
Since the amplitude of the scalar field $\phi$ becomes higher for lower angular frequencies and for small values of the radial coordinate $r$, the corresponding accretion disk solutions can be more compact and located closer to the center compared to those of BSs with higher angular frequencies.

\section{Effective potential and Keplerian angular momentum}
Unmagnetized thick disks can be computed from an effective potential $\mathcal{W}$, which acts as a combination of the gravitational and centrifugal potential of a fluid particle rotating around a central gravitating object. $\mathcal{W}$ can be derived by integrating the relativistic Euler equations and assuming the von Zeipel theorem as a necessary integrability condition,
\begin{align}
    \mathcal{W} - \mathcal{W}_{in} &\coloneqq \ln|u_t| - \ln|(u_t)_{in}| - \int_{\ell_{in}}^\ell \frac{\Omega}{1 - \Omega \ell'} \mathrm{d} \ell' \\
    &= - \int_0^p \frac{1}{\rho h} \mathrm{d}p',
\end{align}
with $u_t$ as the covariant four-velocity (and $-u_t$ the mass-normalized energy), $\ell$ the specific angular momentum, $\rho$ the rest-mass density, $h$ the specific enthalpy and $p$ the thermodynamic pressure. The effective potential at the inner edge of the disk, $\mathcal{W}_{in}$, is taken as a free parameter, as its value determines the boundary of the disk. Choosing $\mathcal{W}_{in} < 0$ leads to tori for which the outermost equipotential surface is closed, while $\mathcal{W}_{in} > 0$ represents open equipotential surfaces. In this work we choose $\mathcal{W}_{in} = 0$ which corresponds to an equipotential surface closed at infinity.
Assuming a constant specific angular momentum distribution the integral term containing $\ell$ vanishes, and by assuming furthermore a polytropic equation of state, $p = K \rho^\Gamma$, with $K$ and $\Gamma$ being the polytropic constant and exponent, the rest-mass density can be rewritten to
\begin{align}
    \rho = \left( \frac{(\mathrm{e}^{{W}_{in} - \mathcal{W}} - 1) (\Gamma - 1)}{K \Gamma} \right)^\frac{1}{1-\Gamma},
\end{align}
where $\rho$ depends only on $\mathcal{W}$. 

\begin{figure}[h!]
\centering
  \subfloat[\label{Kerr_Kep_Momentum}]{\includegraphics[width=0.466\columnwidth]{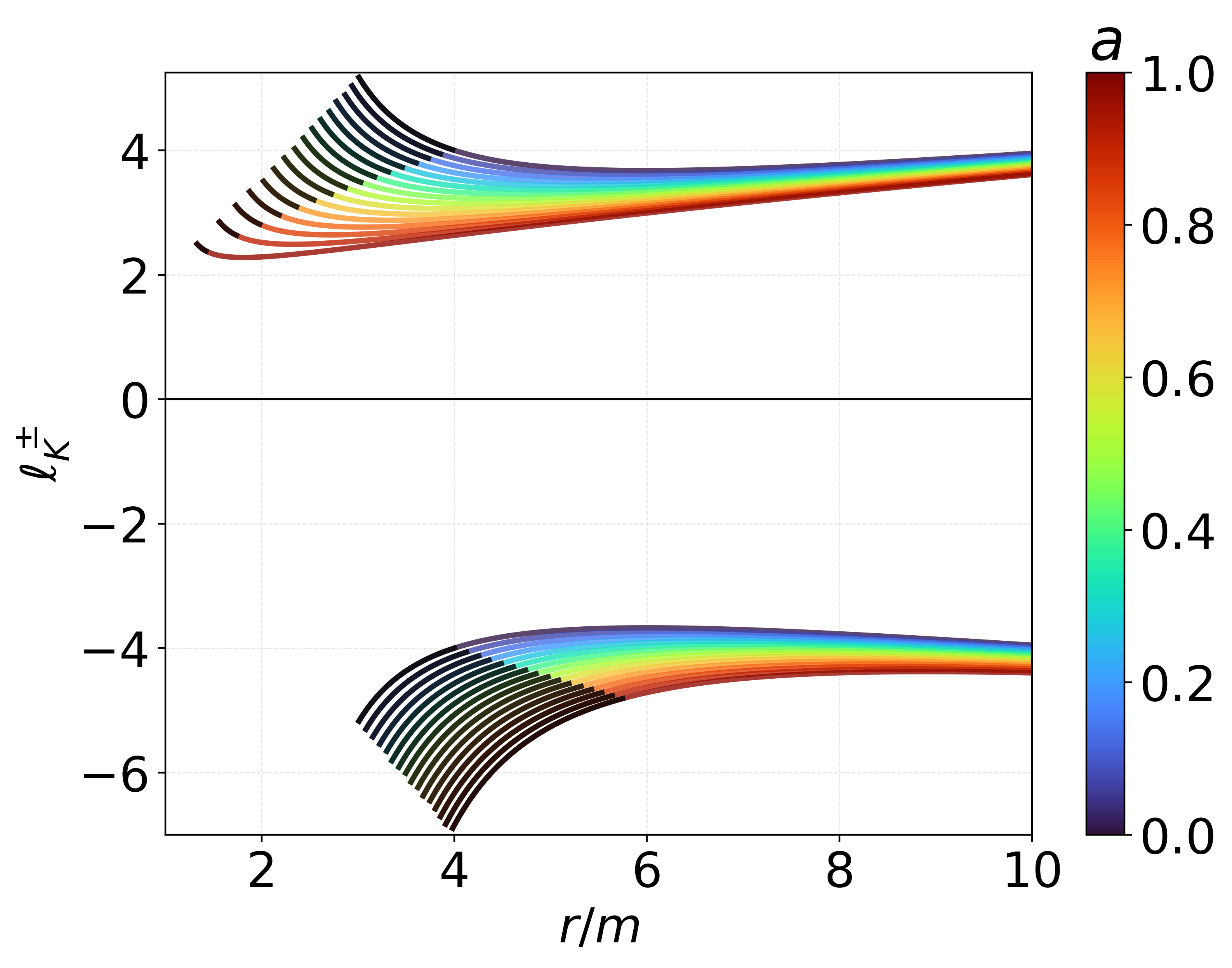}}\hfill
  \subfloat[\label{BS_Kep_Momentum}]{\includegraphics[width=0.504\columnwidth]{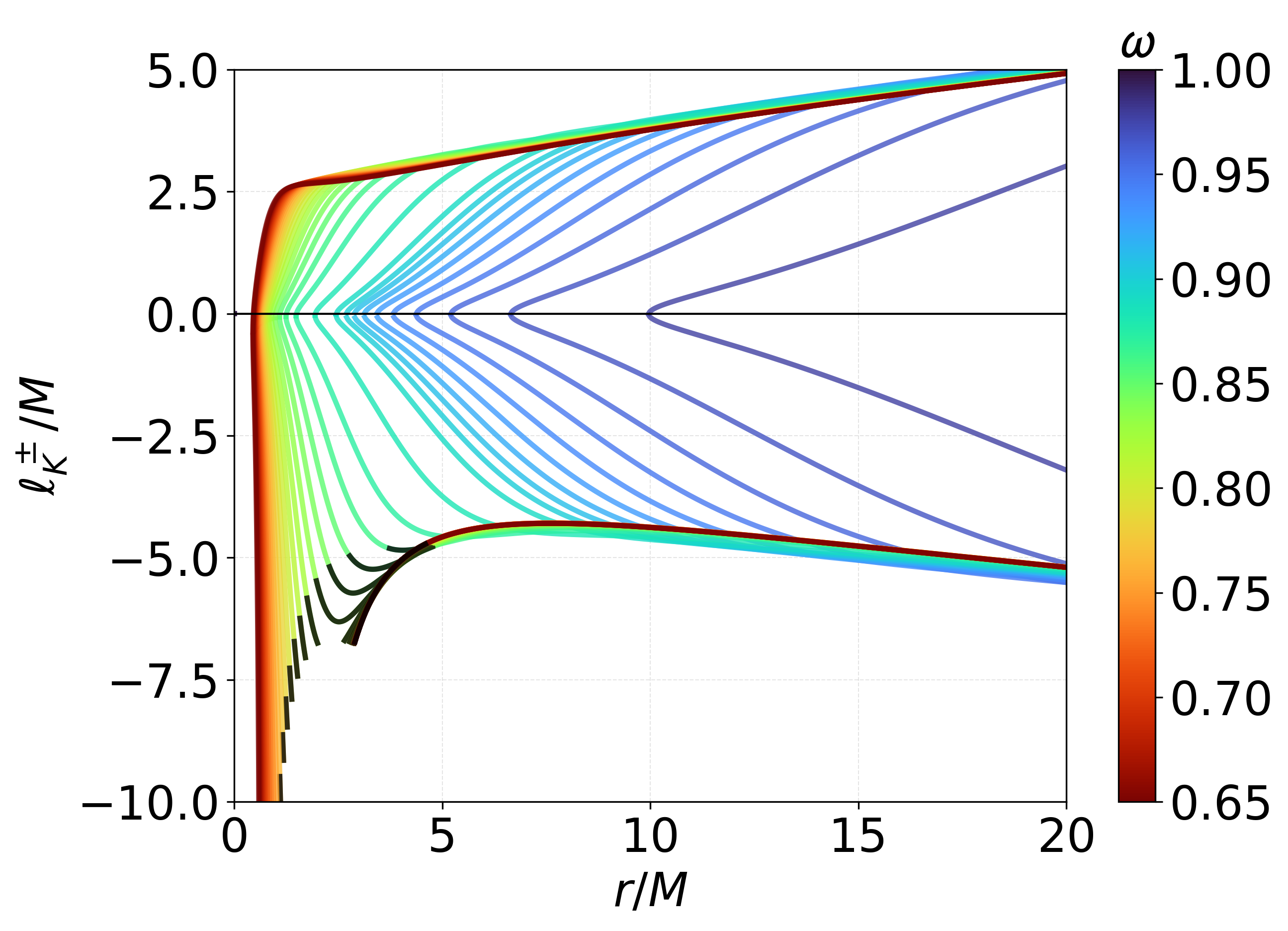}}
\caption{(a) Distributions of $\ell^\pm_K$ for a set of Kerr black holes with different spin parameter $a$. 
The special case $a=0$ is representing a Schwarzschild black hole. 
(b) Normalized $\ell^\pm_K$ distributions for the analyzed set of BS solutions. 
The black curve sections display the radial coordinates for which no bound orbits are possible.} 
\label{fig:KeplerEll_Kerr_BS}
\end{figure}

The equidensity surfaces coincide with the equipotential surfaces and therefore the disk geometry can be studied by analyzing the effective potential. 
The location of the accretion disk center $r_c$ is given by the maximum of the rest-mass density and therefore the minimum of $\mathcal{W}$. 
A local maximum of $\mathcal{W}$ corresponds to a self-intersection of an equipotential surface and is called a cusp. 
Since $\frac{\partial \mathcal{W}}{\partial r} |_{(r=\{r_c,r_{cusp}\},\theta = \frac{\pi}{2})}=0$ the motion at the accretion disk center and at the accretion disk cusp follows a geodesic on the equatorial plane.
Due to the axisymmetry of the BS spacetime, the geodesic corresponds to a Keplerian circular orbit. 
The specific angular momentum $\ell$ at the center and cusp is therefore identical to the Keplerian specific angular momentum $\ell_K$. Considering the von Zeipel theorem, the specific angular momentum $\ell$ can be expressed as,
\begin{align}
    \ell^\pm_K(r) = - \frac{g_{t\varphi} + g_{\varphi\varphi}\Omega^\pm_K}{g_{tt} + g_{t\varphi}\Omega^\pm_K},
\end{align}
with $\Omega^\pm_K$ being the Keplerian angular velocity. 
The positive and negative sign refer to prograde and retrograde motion, respectively. 
In Fig.~\ref{fig:KeplerEll_Kerr_BS} examples of Keplerian specific angular momenta are shown for a set of Kerr black holes (a) and BSs (b).

The specific angular momentum is taken as a free parameter in $\mathcal{W}$, the chosen value $\ell_0$ determines the position of the minima and maxima of $\mathcal{W}$ with $r_c = \{ r: \ell_0 = \ell^\pm_K(r) , \frac{\partial^2 \mathcal{W}}{\partial r^2} > 0\}$ and $r_{cusp} = \{ r: \ell_0 = \ell^\pm_K(r), \frac{\partial^2 \mathcal{W}}{\partial r^2} < 0 \}$. Tab. \ref{tab:conditions} displays all possible thick disk morphologies for the various BS solutions in dependence of $\ell_0$.

\begin{table}[H]
\centering
\begin{tabular}{c|c|c|c|c}
\toprule
      & Centers & Cusp & $\ell_0$ condition & BS models \\ \hline
           &       &       &  $|\ell_0| \notin (\ell_K^{min},\ell_{mb}^{in})$  &  $0.665 \leq \omega \leq 0.806$  \\ \cline{4-5}
 Type $1$  &  $1$  &  $0$  &  $|\ell_0| \notin (\ell_K^{min},\ell_K^{max})$    &  $0.806 < \omega \leq 0.853$  \\ \cline{4-5}
           &       &       &  no condition                     &  $0.853 <  \omega < 1.000$  \\ \hline
 Type $2$  &  $2$  &  $1$  &  $|\ell_0| \in (\ell_K^{min},\ell_{mb}^{out})$    &  $0.665 \leq \omega \leq 0.806$  \\ \cline{4-5}
           &       &       &  $|\ell_0| \in (\ell_K^{min},\ell_K^{max})$       &  $0.806 < \omega \leq 0.853$  \\ \hline
 Type $3$  &  $2$  &  $0$  &  $|\ell_0| \in (\ell_{mb}^{out},\ell_{mb}^{in})$  &  $0.665 < \omega < 0.806$ \\ \hline
\end{tabular}
\caption{Conditions for the different types of non-magnetized thick disks around BSs \cite{Teodoro_2021b}. Since some BSs solutions contain two marginally bound orbits, $\ell_{mb}^{in}$ refers to the specific angular momentum corresponding to the inner marginally bound orbit, whereas $\ell_{mb}^{out}$ refers to the specific angular momentum at the outer marginally bound orbit. $\ell_K^{min}$ and $\ell_K^{max}$ are defined by the local minimum and local maximum of the specific angular momentum.}
\label{tab:conditions}
\end{table}

\begin{figure}[H]
  \centering
  \includegraphics[width=0.5\textwidth]{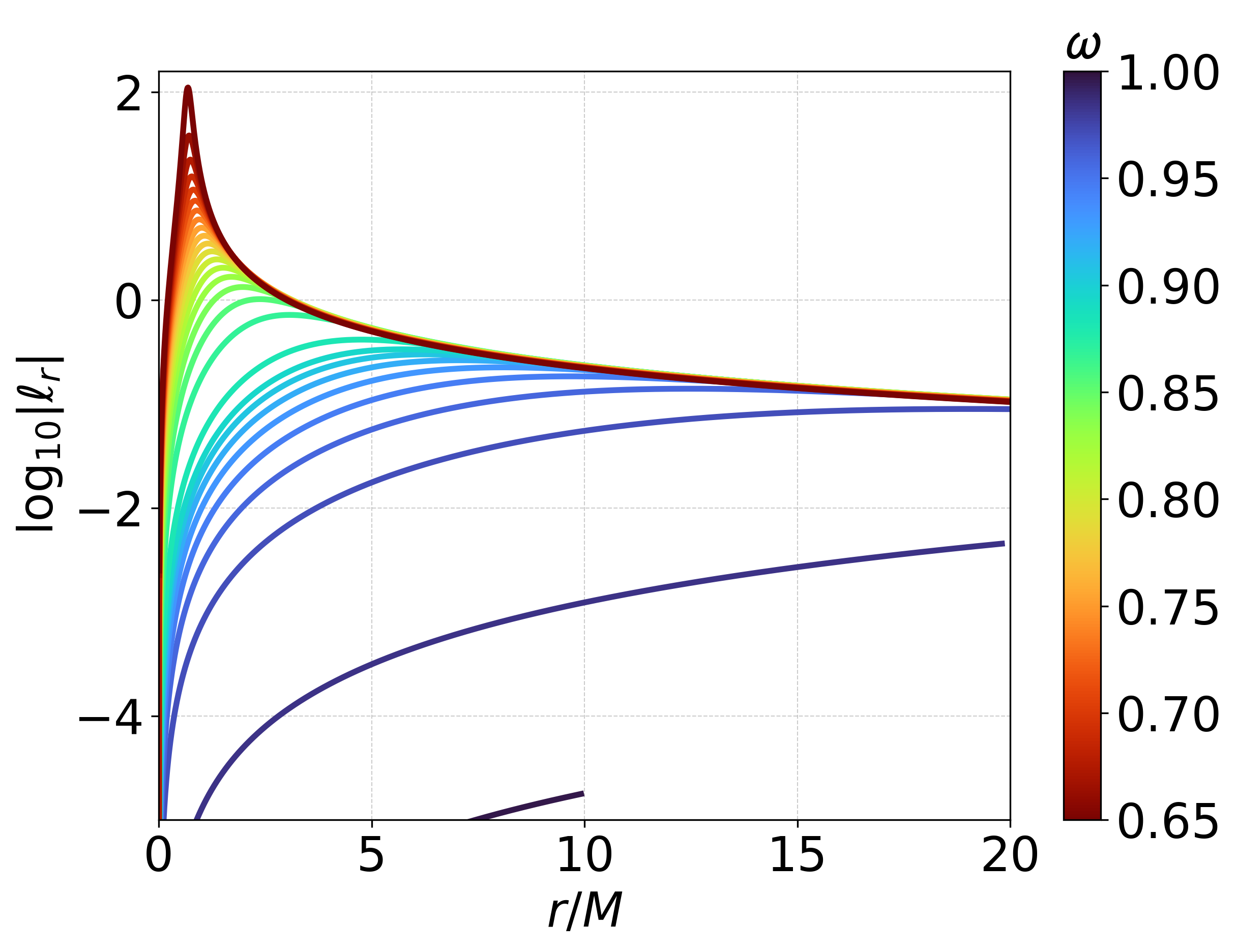}
  \caption{$\log_{10}$ of the absolute value of the rest specific angular momentum distribution $\ell_r$ in the equatorial plane for the set of BS solutions.}
  \label{fig:BS_rest_momentum}
\end{figure}

Some of the retrograde accretion disk solutions, possess locations where the fluid stays at rest for a ZAMO at infinity. Those locations are called static surfaces and are defined by the three dimensional generalization of the so called static rings, which represent orbits remaining at rest \cite{Collodel_2017a}. They are realised in the shape of toroidal surfaces located inside the accretion disk. Outside of the surfaces the fluid flows in a retrograde motion, while inside these surfaces the fluid flows in a prograde motion. Since the fluid stays at rest at these surfaces, the angular velocity $\Omega$ is zero and therefore the specific angular momentum at the surface is given by the rest specific angular momentum $\ell_r \coloneqq - \frac{g_{t\phi}}{g_{tt}}$. Disk solutions containing static surfaces occur when $\exists \ r: \ell_0 = \ell_r(r)$. Fig.~\ref{fig:BS_rest_momentum} shows the equatorial rest specific angular momentum distribution for various BS solutions.

The rest specific angular momentum is significantly smaller for the less relativistic BS solutions, therefore static surfaces are more likely to appear for the more relativistic BSs.

\section{Magnetized Thick Disks}
In order to construct magnetized torus solutions we follow the procedure employed in \cite{Komissarov:2006nz, Font_2019}. 
We extend the hydrodynamical thick disk model by adding a toroidal magnetic field. 
The fundamental conservation laws of relativistic magnetohydrodynamics need to be solved under the assumption,
\begin{align}
    \nabla_\mu (\rho u^\mu) &= 0  \label{eq:continuity_m}, \\
    \nabla_\mu T^{\mu \nu} &= 0   \label{eq:stress-energy-conserv}, \\
    \nabla_\mu (^*F^{\mu \nu}) &= 0, \label{eq:maxwell}
\end{align}
where $\rho$ is the rest-mass density, $u^\mu$ is the four-velocity of a co-moving observer, $T^{\mu\nu}$ is the stress-energy tensor and $^*F^{\mu\nu}$ the dual Faraday tensor,
\begin{align}
    T^{\mu\nu} &\equiv (\rho h + b^2)u^\mu u^\nu + \left(p + \frac{b^2}{2} \right) g^{\mu \nu} - b^\mu b ^\nu , \label{eq:T}\\
    ^*F^{\mu\nu} &\equiv u^\mu b^\nu - u^\nu b^\mu.
\end{align}
Here $b^\mu$ is the magnetic field with $b^2 \equiv b^\mu b_\mu$. In the fluid frame the magnetic field can be represented as $b^\mu = (0,\mathbf{B})$, where $\mathbf{B}$ is the three-dimensional magnetic field measured by a co-moving observer. Due to the axisymmetry, stationarity and azimuthal magnetic field distribution eqs.~(\ref{eq:continuity_m}) and (\ref{eq:maxwell}) are always satisfied. 
Contracting eq.~(\ref{eq:stress-energy-conserv}) with the orthogonal projection tensor $h^\nu_\alpha$ one gets,
\begin{align}
     (\rho h + b^2)u_\nu \partial_\alpha u^\nu + \partial_\alpha \left(p + \frac{b^2}{2} \right) - b_\nu \partial_\alpha b^\nu = 0,
\end{align}
where $\alpha = r,\theta$ is non-trivial. 
Expressing the equation in terms of the angular velocity and specific angular momentum leads to \cite{Komissarov:2006nz}
 \begin{align}
     \partial_\alpha(\ln|u_t|) - \frac{\Omega \partial_\alpha \ell}{1- \ell \Omega} + \frac{\partial_\alpha p}{\rho h} + \frac{\partial_\alpha(\mathcal{L} p_m)}{\mathcal{L}\rho h} = 0,
     \label{eq:potential_magnetized}
 \end{align}
with $\mathcal{L} \equiv g_{t\varphi}^2 - g_{tt}g_{\varphi\varphi}$ and the magnetic pressure $p_m \equiv \frac{b^2}{2}$. 

By assuming a polytropic equation of state also for the magnetic part, we can write $p_m = K_m \mathcal{L}^{q-1}(\rho h)^q$, with $K_m$ and $q$ being the polytropic magnetic constant and exponent. Using the definitions $\widetilde{p}_m = \mathcal{L} p_m$ and $\widetilde{\omega} = \mathcal{L} \rho h$ integration of eq.~(\ref{eq:potential_magnetized}) yields
\begin{align}
     \ln|u_t| - \int^\ell_0 \frac{\Omega}{1- \ell' \Omega}\mathrm{d}\ell' + \int^p_0 \frac{1}{\rho h} \mathrm{d}p' + \int^{\widetilde{p}_m}_0 \frac{1}{\widetilde{\omega}} \mathrm{d} \tilde{p}_m' = C. \label{eq:magnet_integral}
\end{align}
The integration constant is defined by the boundary conditions at the edge of the disk and therefore given by $C = \ln|(u_t)_{in}| = \mathcal{W}_{in}$. 
Since we suppose a constant specific angular momentum distribution the integral term regarding $\ell$ vanishes and integration leads to
\begin{align}
    \mathcal{W} - \mathcal{W}_{in} + \ln(h) + \frac{q}{q-1} K_m (\mathcal{L}\rho h)^{q-1} = 0.
    \label{eq:magnet_integrated}
\end{align}
Since eq.~(\ref{eq:magnet_integrated}) is a transcendental equation for the rest-mass density $\rho$, it needs to be solved numerically at every point of the numerical grid. In order to fix the gauge, the rest-mass density will be normalized at the center of the non-magnetized disk to $\rho_c = 1$, while in the case of two-centered solutions the densest center will be chosen.
Rewriting eq.~(\ref{eq:magnet_integrated}) with respect to the torus center and expressing the specific enthalpy $h$ in terms of the rest-mass density, we obtain an expression where $K$ is the only unknown parameter,
\begin{align}
        \mathcal{W}_c - \mathcal{W}_{in} + \ln\left(1 + \frac{\Gamma K}{\Gamma - 1} \rho_c^{\Gamma-1} \right)
        + \frac{q}{q-1} \frac{K \rho_c^\Gamma}{\beta_{mc} \left(\rho_c + \frac{K \Gamma \rho_c^\Gamma}{\Gamma-1}\right)} = 0.
        \label{eq:magnet_integrated_c}
\end{align}
Here we have defined the magnetization parameter $\beta_{mc} = \frac{p_c}{{p_m}_c}$ as the ratio between the thermodynamic and magnetic pressure at the densest center of the disk. Therefore, a high magnetization parameter $\beta_{mc}$ corresponds to an essentially non-magnetized disk, $\beta_{mc}\sim 1$ describes a mildly magnetized  disk, and a low magnetization parameter $\beta_{mc}$ implies a strongly magnetized disk.
The polytropic exponents are chosen as for relativistic degenerate matter, $\Gamma = q = \frac{4}{3}$. 

With $\beta_{mc}$ as the only parameter, eq.~(\ref{eq:magnet_integrated_c}) can be solved for $K$ for different degrees of magnetization represented by $\beta_{mc}$. After computation of $K$ the general eq.~(\ref{eq:magnet_integrated}) for the rest-mass density can be solved at every point of the numerical grid. As a numerical solving algorithm the bisection method was chosen with an absolute convergence error of $\epsilon = 10^{-15}$ between the last and second to last iteration step. 

\newpage
\section{Results}
We now present our results for magnetized thick tori, discussing first the one-centered disks and then the two-centered ones. The selected BS solutions are based on \cite{Teodoro_2021b} and intended to be exemplary for the spectrum of possible disk types. For less relativistic BS solutions $\omega=0.960$ was chosen, for mildly and highly relativistic $\omega = 0.798$ and $\omega = 0.671$. To conduct a more comprehensible analysis of the magnetized disks, we define the density center and the density cusp of a disk solution as the radial location where the equatorial density has a local maximum and local minimum, respectively. The density and thermodynamic pressure gradients vanish at these locations. Since the effective potential is not affected by the magnetization parameter, density center and density cusp only coincide in non-magnetized solutions with the accretion disk center and disk cusp defined by the minima and maxima of the effective potential. Therefore we will from now on distinguish between them in the magnetized solutions and refer with $r_c$ to the density center and with $r_{cusp}$ to the density cusp. In pursuit of a more precise study of the one-centered solutions we define an effective equatorial columnar radius $\tilde{R}$ and the corresponding mean equatorial columnar density $\bar{\rho}$,
\begin{align}
    \tilde{R} = \left\{R : \frac{\int_0^R \rho(r,\theta=\frac{\pi}{2})dr}{\int_0^\infty \rho(r,\theta=\frac{\pi}{2})dr} = 0.99 \right\} \ \ ; \ \ \bar{\rho} = \frac{1}{\tilde{R}} \int_0^{\tilde{R}} \rho(r,\theta=\frac{\pi}{2})dr.
\end{align}
The volume integral of the density $\bar{\rho}$  can give an estimate for the disk mass, approximating it in the weak field limit \cite{Rezzolla}, \cite{Font_2019}. We will use this quantity in the following analysis as a measure for the disk compactness.

\subsection{One-centered Disks}
\begin{figure}[H]
\centering
\begin{floatrow}
  \subfloat[$\omega = 0.960, \ell_0 = -4.5M$]{\includegraphics[height=0.1915\textheight]{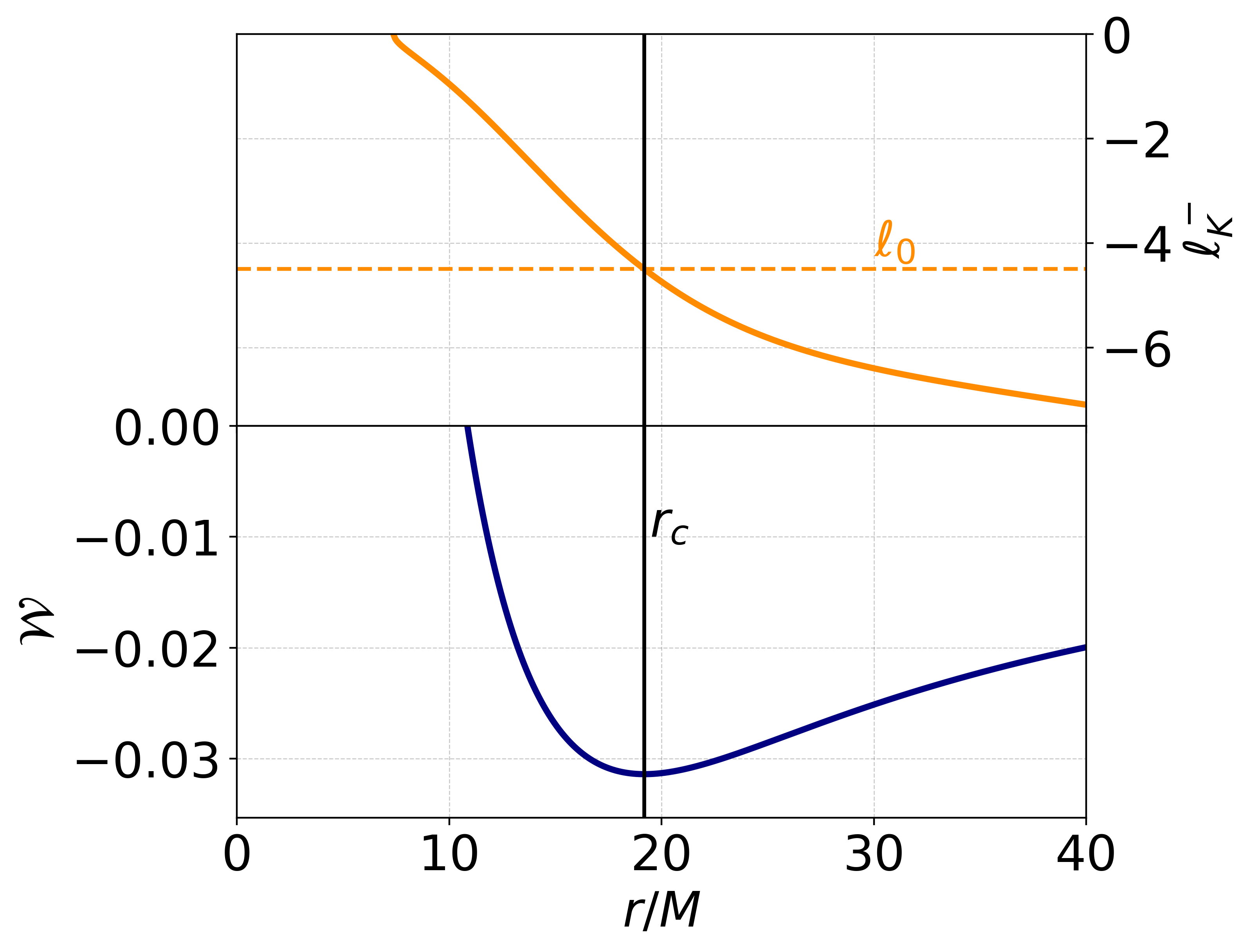}}
  \subfloat[$\omega = 0.960, \ell_0 = -0.1M$]{\includegraphics[height=0.1915\textheight]{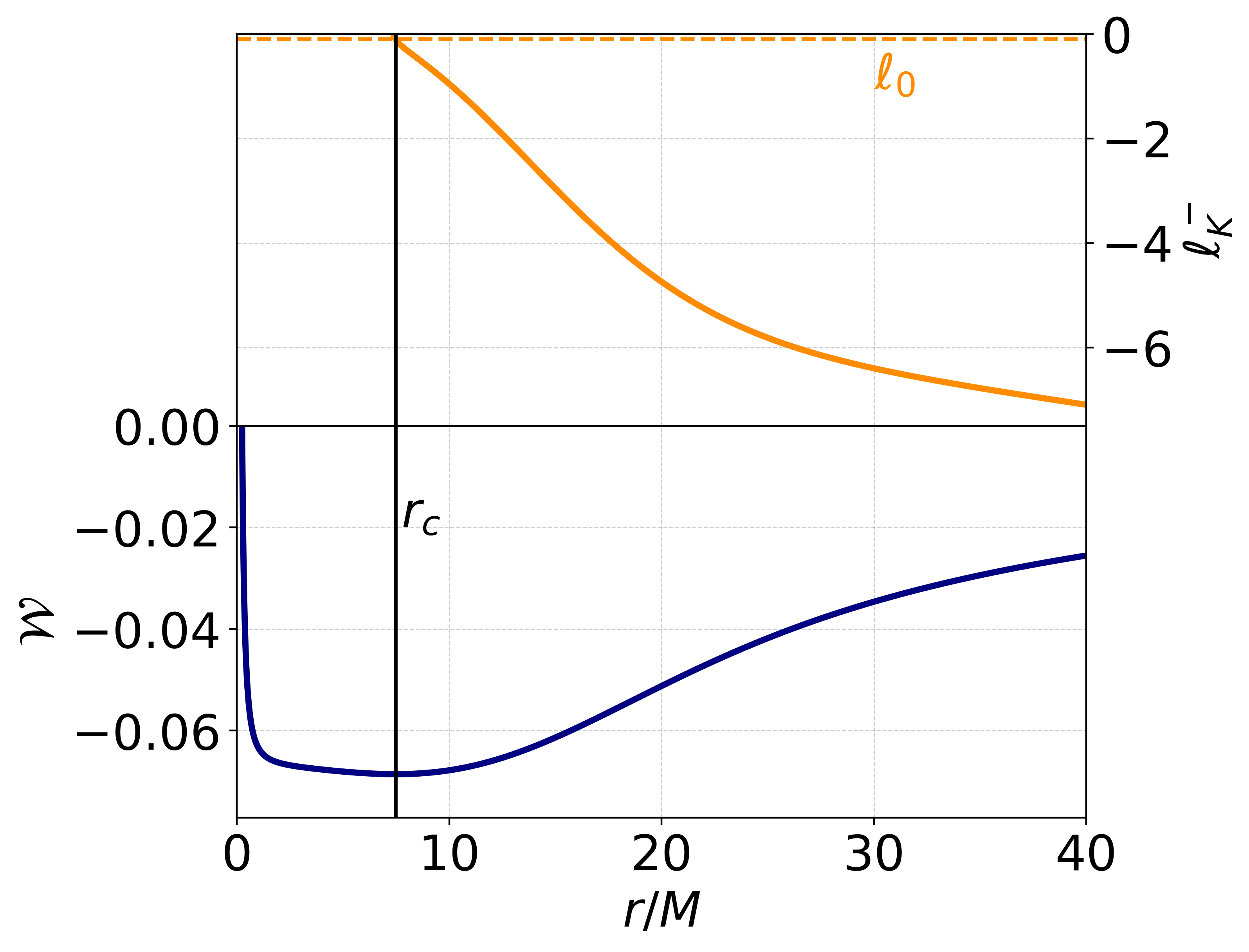}}
  \subfloat[$\omega = 0.798, \ell_0 = -0.4666M$]{\includegraphics[height=0.1915\textheight]{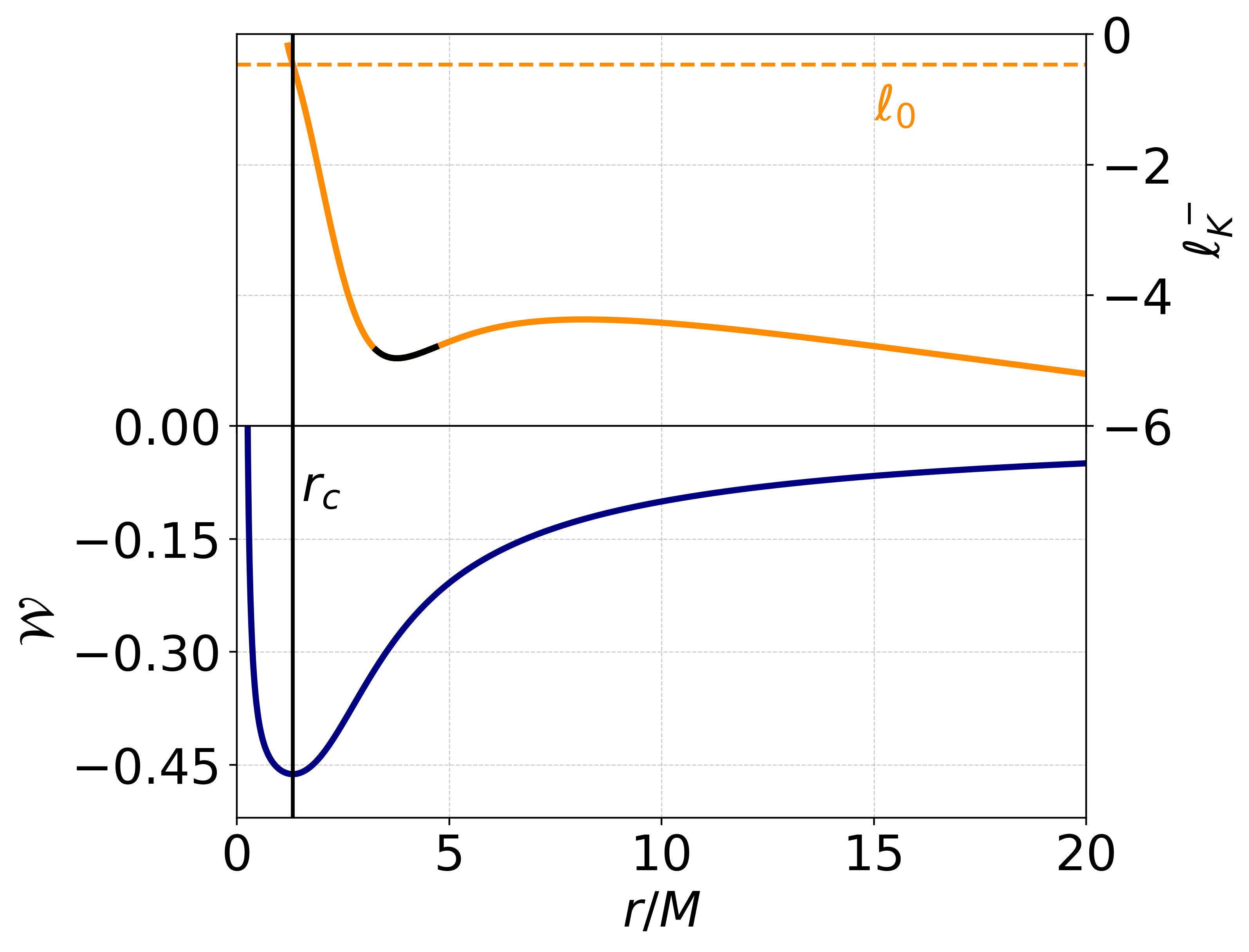}}
\end{floatrow}
\caption{The upper part of the figures presents the normalized Keplerian specific angular momentum $\ell_K^-$ of the BS solution. The dashed horizontal line illustrates the chosen $\ell_0$ value. The lower part of the figures shows the corresponding effective potential $\mathcal{W}$. Black vertical lines indicate the radial location of the disk and (non-magnetized) density center.}
\label{fig:Momentum_pot_one_centered}
\end{figure}

For the one-centered disks the selected solutions are: $A$.~$\omega=0.960, \ell_0 = -4.5M$, $B$.~$\omega=0.960, \ell_0 = -0.1M$, and $C$.~$\omega=0.798, \ell_0 = -0.4666M$. Fig. \ref{fig:Momentum_pot_one_centered} shows the normalized Keplerian angular momentum distribution $\ell_K^-$ of the BS solutions and the effective potential $\mathcal{W}$ of the chosen disk solutions. \\

\begin{figure}[H]
    \centering
    \subfloat[]{\includegraphics[width=1\linewidth]{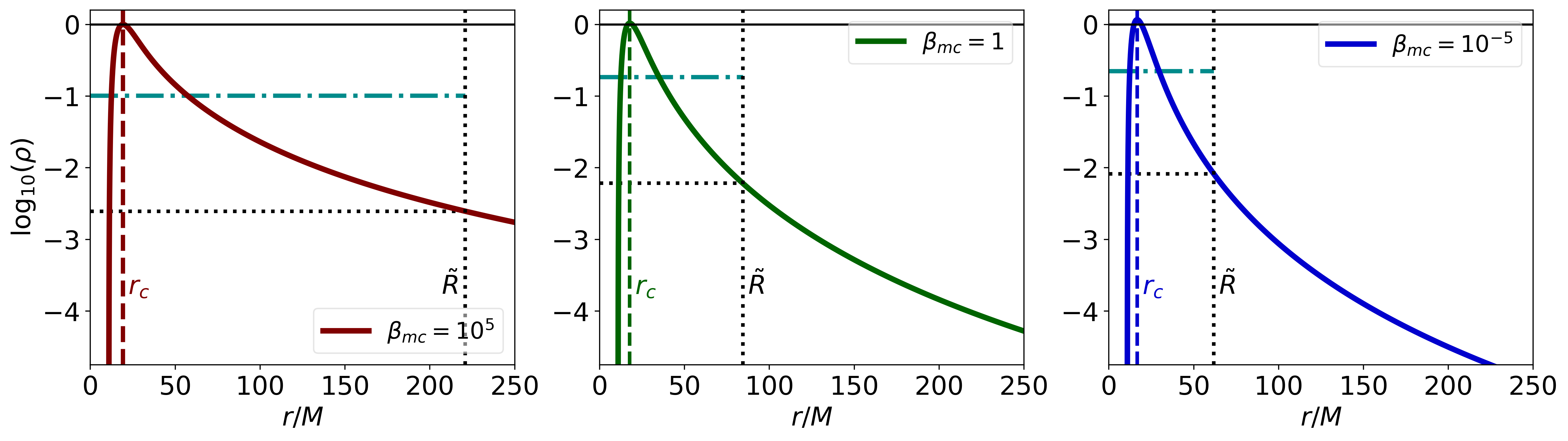}} \\
    \subfloat[]{\includegraphics[height=0.23\textheight]{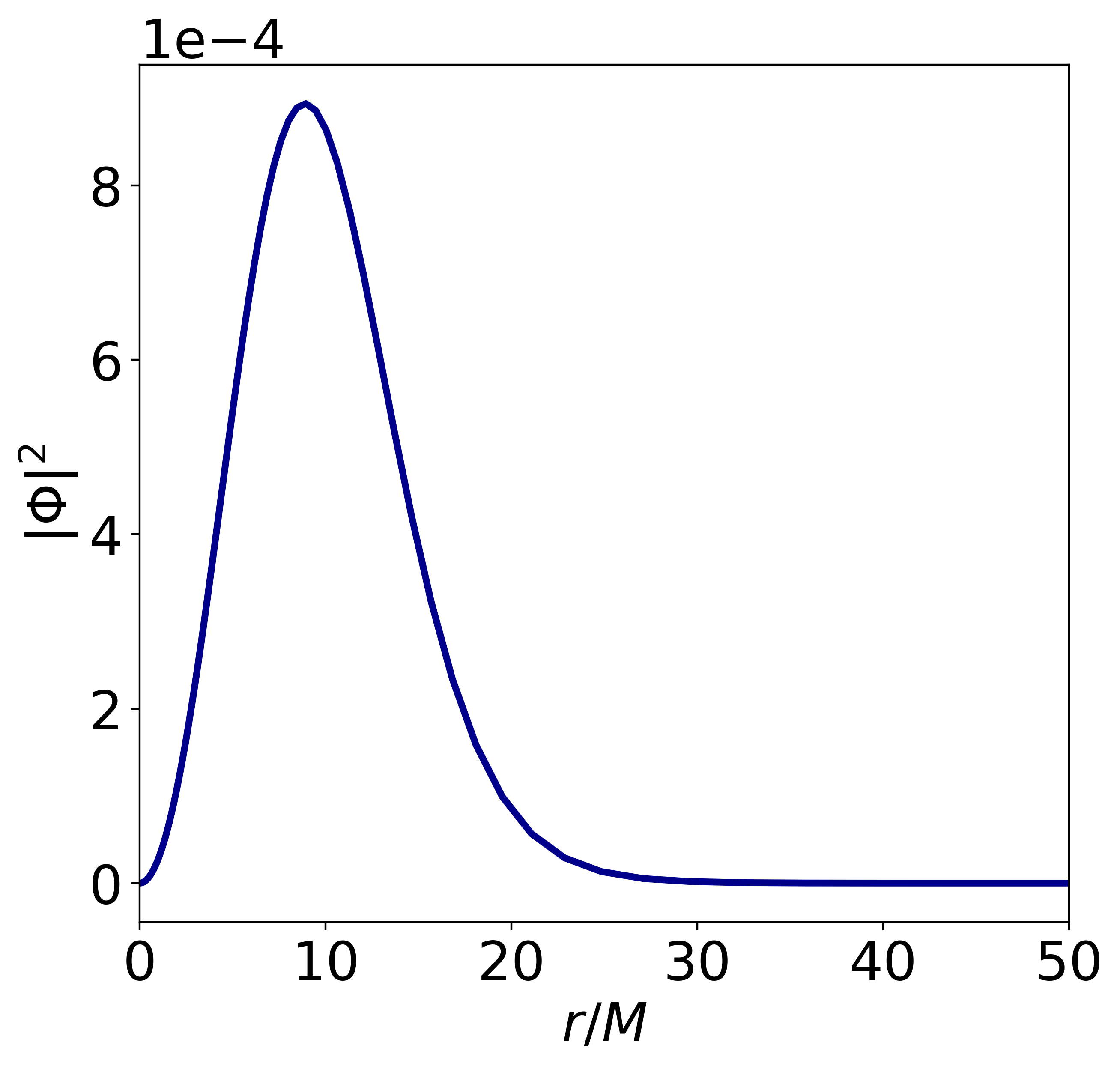}} \hspace{-0.85em}
    \bigskip
    
    \begin{tabular}{c|c|c|c|c|c|c|c|c}
    \toprule
    $\log_{10}\beta_{mc}$ & $\rho_{c}$ & $r_c$ & $\Delta r_c$ &  $\tilde{R}$ &  $\bar{\rho}$ &  $\log_{10}\left(p_{max}\right)$ & $\log_{10}\left(p_{m_{max}}\right)$ & $h_{max}$ \\
    \midrule
      5 &         1.000 &     19.190 &       - &      220.867 &         0.102 &                         -2.098 &                -7.091 &                            1.032 \\
      0 &         1.041  &     17.759 &       1.431 &       84.341 &         0.184 &                     -2.378 &                -2.395 &                            1.016 \\
      -5 &         1.165 &     16.678 &       2.511 &       61.774 &         0.224 &                    -7.017 &                -2.053 &                            1.000 \\
    \bottomrule
    \end{tabular}
    
    \caption{$A$. $\omega=0.960, \ell_0 = -4.5M$: (a) $\log_{10}$ of the density in the equatorial plane for the different magnetization parameters $\beta_{mc}$. Vertical dashed lines represent the position of the density center. Vertical dotted lines represent the effective columnar radius $\tilde{R}$. Horizontal dotted lines correspond to the density value $\rho(\tilde{R},\frac{\theta}{2})$. Dashed dotted horizontal blue lines represent the mean equatorial columnar density $\bar{\rho}$. (b) Scalar profile of the BS solution, the maximum is located at $r=8.936$. The table presents properties of the accretion disks, where $\rho_c$ is the maximum of the density, $r_c$ is the corresponding radial location of the maximum, $\Delta r_c$ is the distance between the density center locations of magnetized and non-magnetized tori, $p_{max}$ and $p_{m_{max}}$ are the maximum of the thermodynamic and magnetic pressure and $h_{max}$ is the maximum of the specific enthalpy.}
    \label{fig:Density_0.960_-4.5}
\end{figure}

\begin{figure}[H]
\centering
\begin{floatrow}
  \subfloat[$\beta_{mc}=10^5$]{\includegraphics[width=0.3283\columnwidth]{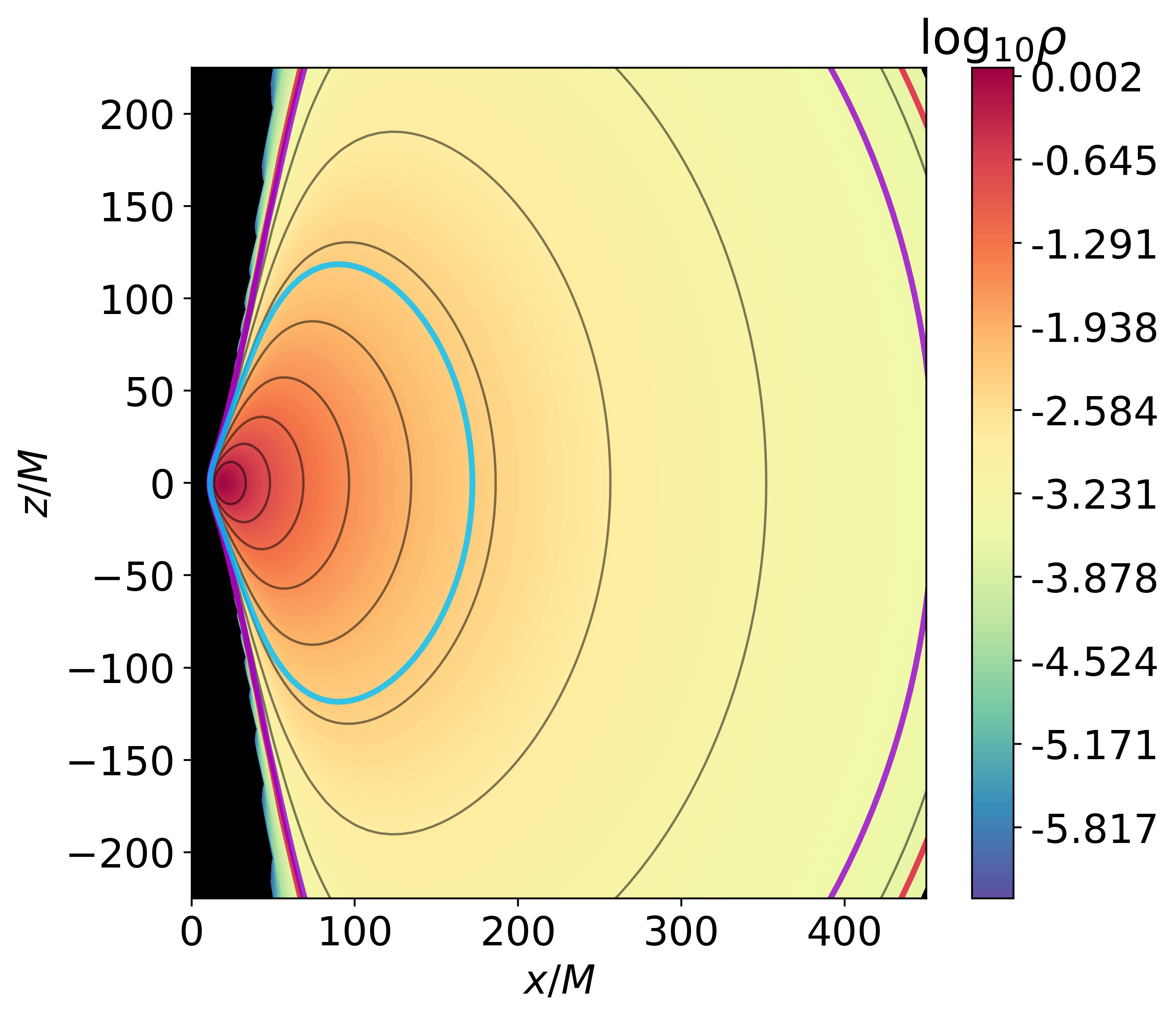}}
  \subfloat[$\beta_{mc}=1$]{\includegraphics[width=0.3283\columnwidth]{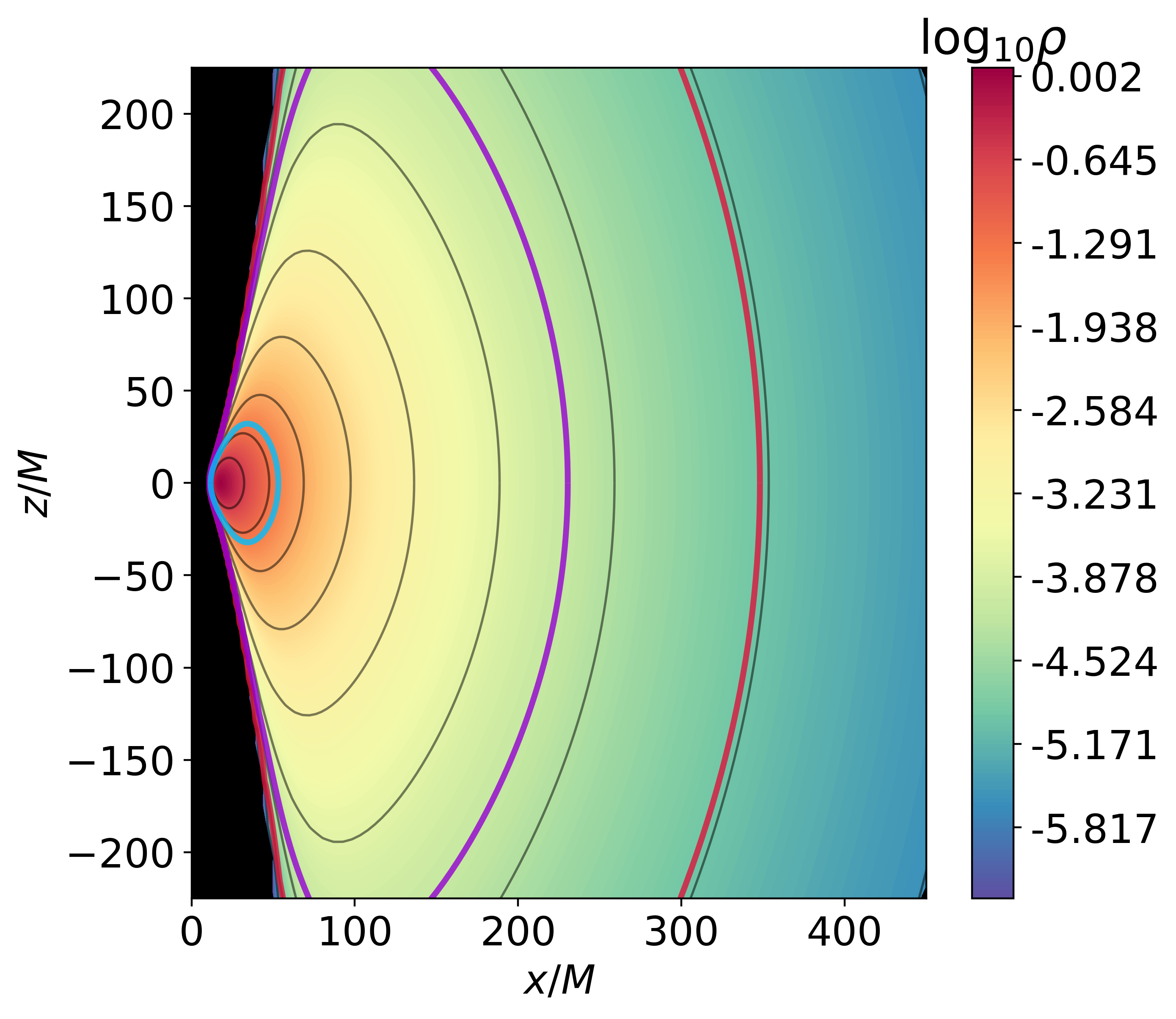}}
  \subfloat[$\beta_{mc}=10^{-5}$]{\includegraphics[width=0.3283\columnwidth]{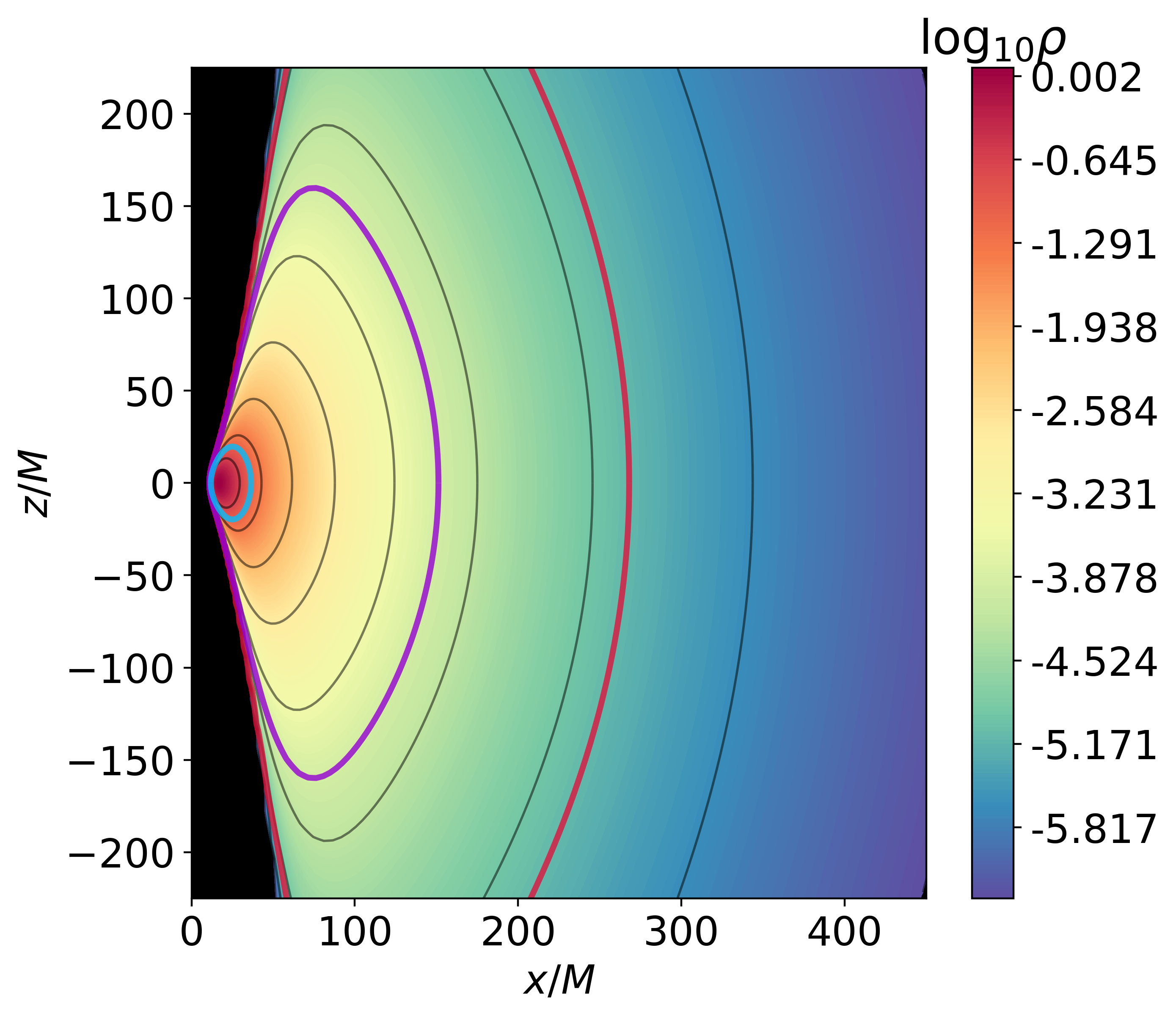}}
\end{floatrow}

\bigskip
\begin{floatrow}
\begin{tabular}{c|c|c|c}
    \toprule
     & $R$ & $\log_{10}\rho$ \\
    \midrule
    $\frac{\int_0^R \rho \mathrm{d}V}{\int_0^\infty \rho \mathrm{d}V} \approx 0.5$ & 172.05 & -2.30 \\ \hline
    $\frac{\int_0^R \rho \mathrm{d}V}{\int_0^\infty \rho \mathrm{d}V} \approx 0.95$ & 459.35 & -3.52 \\ \hline
    $\frac{\int_0^R \rho \mathrm{d}V}{\int_0^\infty \rho \mathrm{d}V} \approx 0.99$ & 483.68 & -3.62 \\ \hline
    \bottomrule
    \end{tabular}\hspace{6em}

\begin{tabular}{c|c|c|c}
    \toprule
     & $R$ & $\log_{10}\rho$ \\
    \midrule
    $\frac{\int_0^R \rho \mathrm{d}V}{\int_0^\infty \rho \mathrm{d}V} \approx 0.5$ & 53.13 & -1.41 \\ \hline
    $\frac{\int_0^R \rho \mathrm{d}V}{\int_0^\infty \rho \mathrm{d}V} \approx 0.95$ & 230.47 & -4.12 \\ \hline
    $\frac{\int_0^R \rho \mathrm{d}V}{\int_0^\infty \rho \mathrm{d}V} \approx 0.99$ & 348.11 & -4.94 \\ \hline
    \bottomrule
    \end{tabular}\hspace{5em}

\begin{tabular}{c|c|c|c}
    \toprule
     & $R$ & $\log_{10}\rho$ \\
    \midrule
    $\frac{\int_0^R \rho \mathrm{d}V}{\int_0^\infty \rho \mathrm{d}V} \approx 0.5$ & 36.49 & -0.16 \\ \hline
    $\frac{\int_0^R \rho \mathrm{d}V}{\int_0^\infty \rho \mathrm{d}V} \approx 0.95$ & 151.24 & -3.92 \\ \hline
    $\frac{\int_0^R \rho \mathrm{d}V}{\int_0^\infty \rho \mathrm{d}V} \approx 0.99$ & 268.08 & -5.12 \\ \hline
    \bottomrule
    \end{tabular}
\end{floatrow}

\caption{$A$. $\omega=0.960, \ell_0 = -4.5M$: Density distribution visualized for the different magnetization parameters. The black solid contour lines represent equidensity surfaces. The minimum of the density is set to $\log_{10}\rho = -6.37$ for all figures and the maximum is set to the density maximum of the highly magnetized solution. The light blue, violet and red contour lines represent each the equidensity surface within which $50\%, \ 95\%$ and $99\%$ of the approximated mass lies. The tables below each figure contain the corresponding radial value in the equatorial plane of these equidensity surfaces and their density. It should be noted, that owing to the numerical nature of the solutions (and all further solutions), the computed values of these surfaces are only representing approximations.}
\label{fig:Torus_0.960_-4.5}
\end{figure}

\paragraph{$\omega=0.960, \ell_0 = -4.5M:$}\label{par:a}
This solution is representative for the less relativistic BS solutions, the accretion disk consists of an one-centered far-reaching torus, since the slope of the density curve flattens with increasing radial value (Fig.~\ref{fig:Density_0.960_-4.5}). The density center is represented by the maximum of the density curves. With higher magnetization the maximum density increases and the location of the density center shifts closer to the center of the BS. For all radial values smaller than the disk center for the non-magnetized case, the density of the magnetized disks is higher compared to the non-magnetized solution. In contrast to this, the density in the magnetized solutions becomes smaller for larger values of $r$ compared to the non-magnetized case, scaling up to several orders of magnitude difference for large values of $r$. The effective equatorial columnar radius $\tilde{R}$ decreases with a higher magnetisation, implying a more compact mass distribution, whereas the mean equatorial columnar density $\bar{\rho}$ remains similar, which is a consequence of the denser center and steeper slope of the density curve. As seen in Fig.~\ref{fig:Torus_0.960_-4.5}, the disk gets compressed for a strong magnetization, leading to a more elliptic shape of the equidensity surfaces. Most of the accretion disk mass is located in a smaller volume around the disk center, since the mass distribution decreases even more rapidly with increasing magnetization. We conclude that strong magnetic fields lead to a compactification of the torus, while the general shape and geometry of the torus is preserved for the less relativistic BS solutions. \\

\begin{figure}[H]
    \centering
    \subfloat[]{\includegraphics[width=1\linewidth]{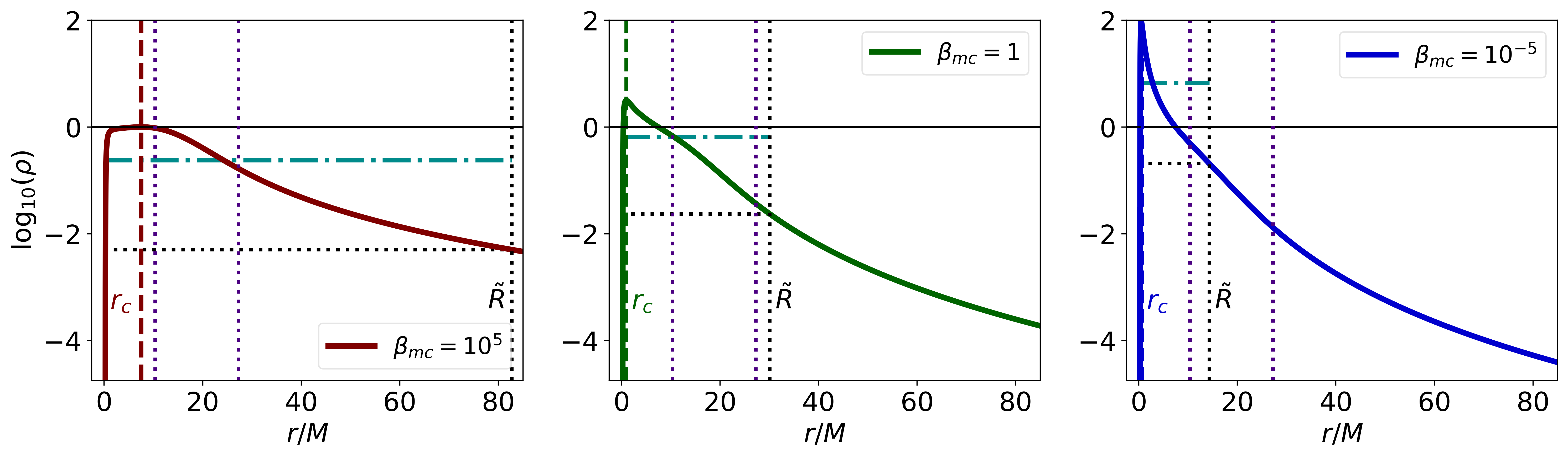}} \\
    \subfloat[]{\includegraphics[height=0.23\textheight]{Figures/Results/scalar_profile_0.960.png}} \hspace{-0.85em}
    \bigskip
    
    \begin{tabular}{c|c|c|c|c|c|c|c|c}
    \toprule
    $\log_{10}\beta_{mc}$ &  $\rho_{c}$ & $r_c$ & $\Delta r_c$ &  $\tilde{R}$ &  $\bar{\rho}$ &  $\log_{10}\left(p_{max}\right)$ & $\log_{10}\left(p_{m_{max}}\right)$ & $h_{max}$ \\
    \midrule
      5 &         1.000 &      7.474 &       0.000 &       82.741 &         0.240 &                     -1.751 &                -6.681 &                            1.071 \\
      0 &         3.135 &      0.960 &       6.513 &       30.083 &         0.651 &                     -1.394 &                -1.887&                            1.051 \\
     -5 &        94.798 &      0.490 &       6.983 &       14.307 &         6.705 &                     -4.130 &                0.094&                            1.000 \\
    \bottomrule
    \end{tabular}
    
    \caption{$B$. $\omega=0.960, \ell_0 = -0.1M$: (a) $\log_{10}$ of the density in the equatorial plane for the different magnetization parameters. Vertical dotted indigo lines represent the radial value corresponding to the intersection of the static surface with the equatorial plane. Vertical dashed lines represent the position of the density center. Vertical dotted lines represent the effective columnar radius $\tilde{R}$. Horizontal dotted lines correspond to the density value $\rho(\tilde{R},\frac{\theta}{2})$. The dashed dotted horizontal blue lines represent the mean equatorial density $\tilde{\rho}$. (b) Scalar profile of the BS solution, the maximum is located at $r=8.936$.}
    \label{fig:Density_0.960_-0.1}
\end{figure}

\begin{figure}[ht!]
\centering
\begin{floatrow}
  \subfloat[$\beta_{mc}=10^5$]{\includegraphics[width=0.3283\columnwidth]{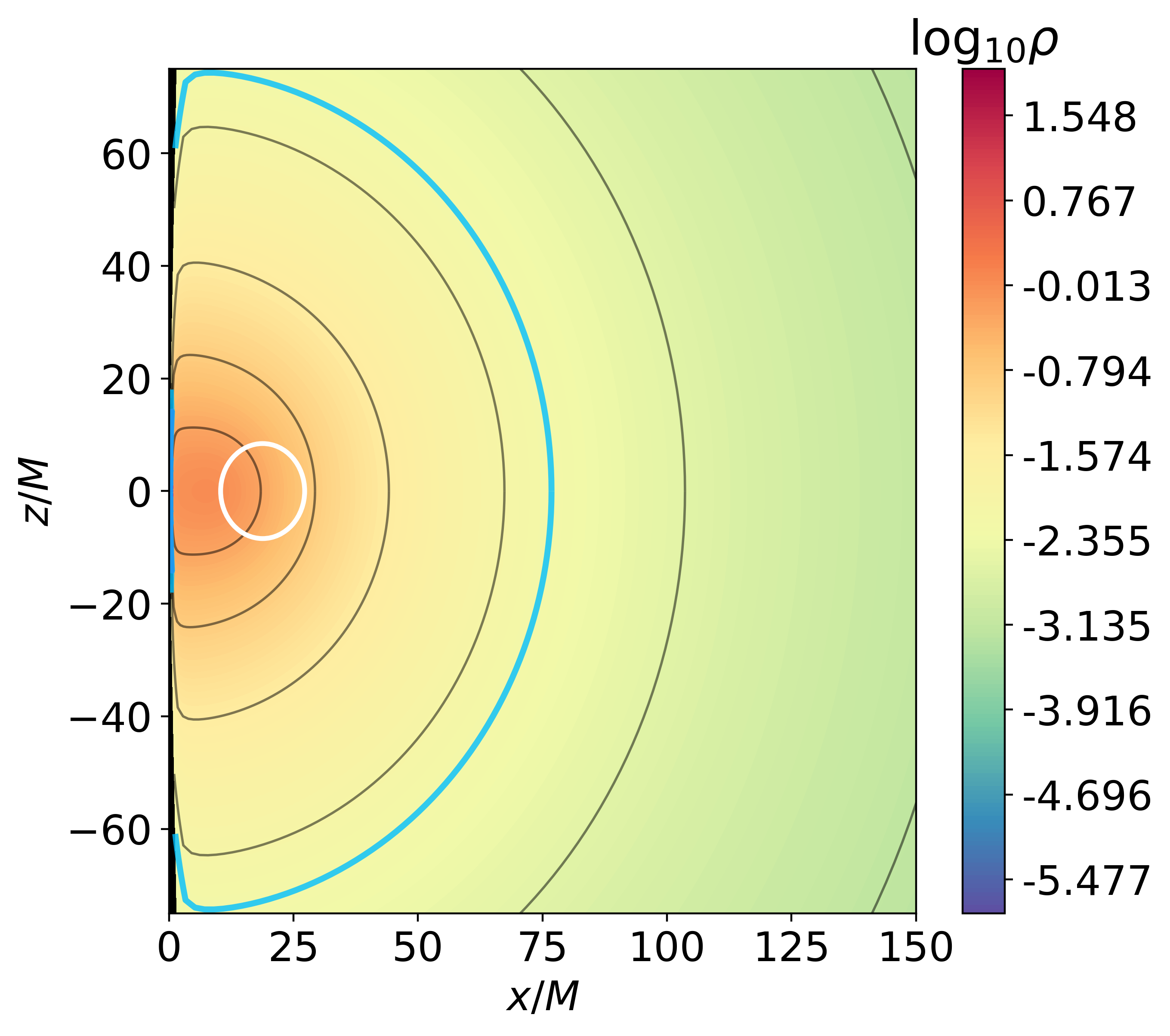}}
  \subfloat[$\beta_{mc}=1$]{\includegraphics[width=0.3283\columnwidth]{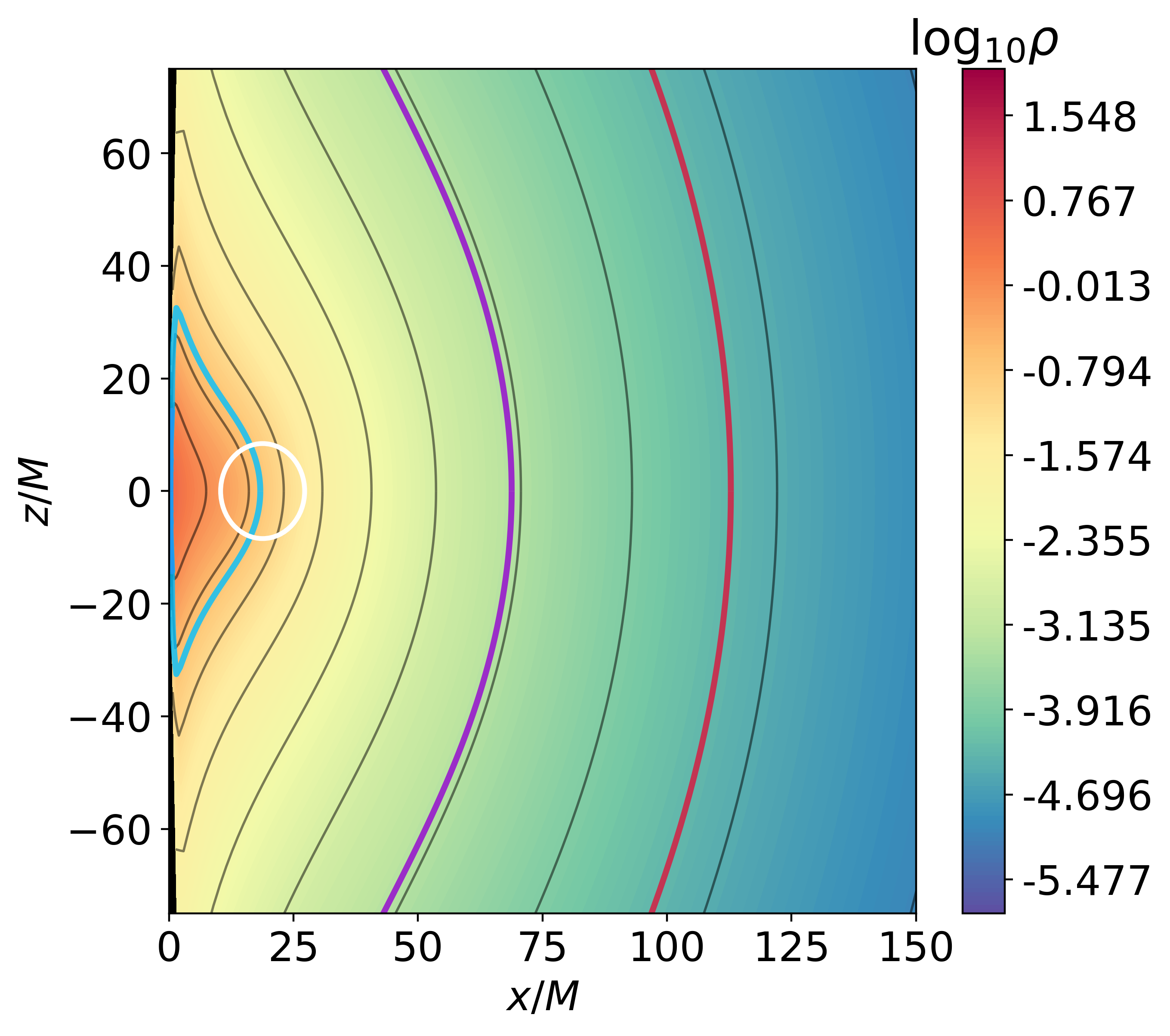}}
  \subfloat[$\beta_{mc}=10^{-5}$]{\includegraphics[width=0.3283\columnwidth]{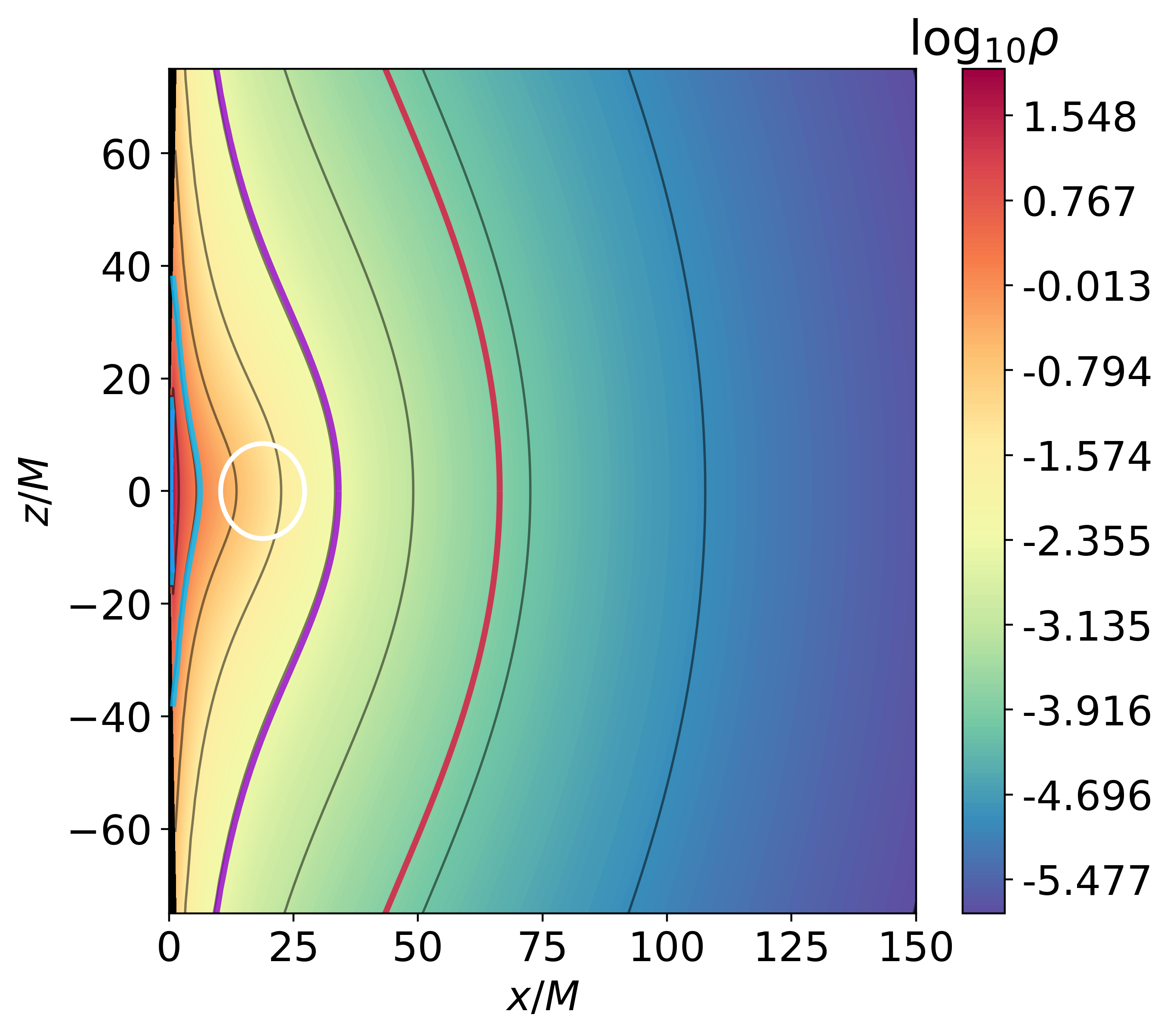}}
\end{floatrow}

\bigskip
\begin{floatrow}
    \begin{tabular}{c|c|c|c}
    \toprule
     & $R$ & $\log_{10}\rho$ \\
    \midrule
    $\frac{\int_0^R \rho \mathrm{d}V}{\int_0^\infty \rho \mathrm{d}V} \approx 0.5$ & 76.82 & -2.20 \\ \hline
    $\frac{\int_0^R \rho \mathrm{d}V}{\int_0^\infty \rho \mathrm{d}V} \approx 0.95$ & 421.09 & -4.42 \\ \hline
    $\frac{\int_0^R \rho \mathrm{d}V}{\int_0^\infty \rho \mathrm{d}V} \approx 0.99$ & 478.55 & -4.61 \\ \hline
    \bottomrule
\end{tabular} \hspace{6em}

\begin{tabular}{c|c|c|c}
    \toprule
     & $R$ & $\log_{10}\rho$ \\
    \midrule
    $\frac{\int_0^R \rho \mathrm{d}V}{\int_0^\infty \rho \mathrm{d}V} \approx 0.5$ & 18.34 & -0.74 \\ \hline
    $\frac{\int_0^R \rho \mathrm{d}V}{\int_0^\infty \rho \mathrm{d}V} \approx 0.95$ & 68.82 & -3.30 \\ \hline
    $\frac{\int_0^R \rho \mathrm{d}V}{\int_0^\infty \rho \mathrm{d}V} \approx 0.99$ & 112.83 & -4.31 \\ \hline
    \bottomrule
\end{tabular} \hspace{5em}

\begin{tabular}{c|c|c|c}
    \toprule
     & $R$ & $\log_{10}\rho$ \\
    \midrule
    $\frac{\int_0^R \rho \mathrm{d}V}{\int_0^\infty \rho \mathrm{d}V} \approx 0.5$ & 6.23 & 0.16 \\ \hline
    $\frac{\int_0^R \rho \mathrm{d}V}{\int_0^\infty \rho \mathrm{d}V} \approx 0.95$ & 34.09 & -2.39 \\ \hline
    $\frac{\int_0^R \rho \mathrm{d}V}{\int_0^\infty \rho \mathrm{d}V} \approx 0.99$ & 66.42 & -3.87 \\ \hline
    \bottomrule
\end{tabular}
\end{floatrow}

\caption{$B$. $\omega=0.960, \ell_0 = -0.1M$: Density distribution visualized for different magnetization parameters. Black solid contour lines represent equidensity surfaces, the white circle represents the static surface. The minimum of the density is set to $\log_{10}\rho = -5.79$ in all figures and the maximum is set to the density maximum of the highly magnetized solution. The light blue, violet and red lines represent each the equidensity surface within which $50\%, \ 95\%$ and $99\%$ of the approximated mass lies. The tables below each figure contain the corresponding radial value in the equatorial plane of these equidensity surfaces and their density value.} 
\label{fig:Torus_0.960_-0.1}
\end{figure}

\paragraph{$\omega=0.960, \ell_0 = -0.1M:$}
This example represents a special set of solutions, which are also known as \textit{fat tori}, they occur for small values of the specific angular momentum. Since the specific angular momentum is sufficiently small in these solutions, the tori are capable of possessing static surfaces. As seen in Fig.~\ref{fig:Density_0.960_-0.1} the density center for strongly magnetized disks is located closer to the center of the BS and the maximum density is approximately two orders of magnitude higher compared to the maximum density of the non-magnetized disk. The mean columnar density $\bar{\rho}$ and effective columnar radius $\tilde{R}$ are significantly higher/lower for the strong magnetized case, indicating a very compressed disk and a more longitudinal mass distribution. Furthermore the density at the static surface decreases, resulting in less matter located inside and at the static surface. In the magnetized solution the static surface lies completely outside the volume in which approximately $50\%$ of the mass is contained (Fig.~\ref{fig:Torus_0.960_-0.1}). As seen in Fig.~\ref{fig:Torus_0.960_-0.1} strong magnetic fields compress the equidensity surfaces parallel to the equatorial plane in the direction of the BS center, resulting in extremal tori with a cylindrical shape centered around the rotational axis. The geometry mimics a sharp ellipsoid centered around the BS center, where most of the mass is located close to the BS center and alongside the rotational axis. We conclude that strong magnetic fields have a significant effect on fat tori, highly compactifying and elongating them alongside the rotational axis. \\

\begin{figure}[H]
    \centering
    \subfloat[]{\includegraphics[width=\linewidth]{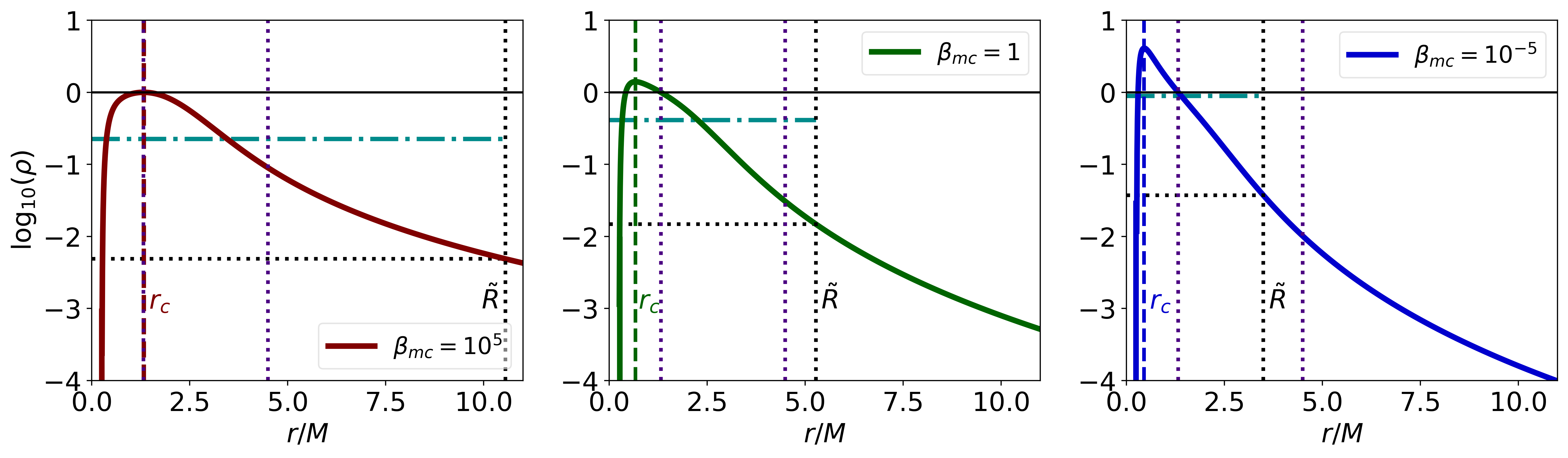}} \\
    \subfloat[]{\includegraphics[height=0.23\textheight]{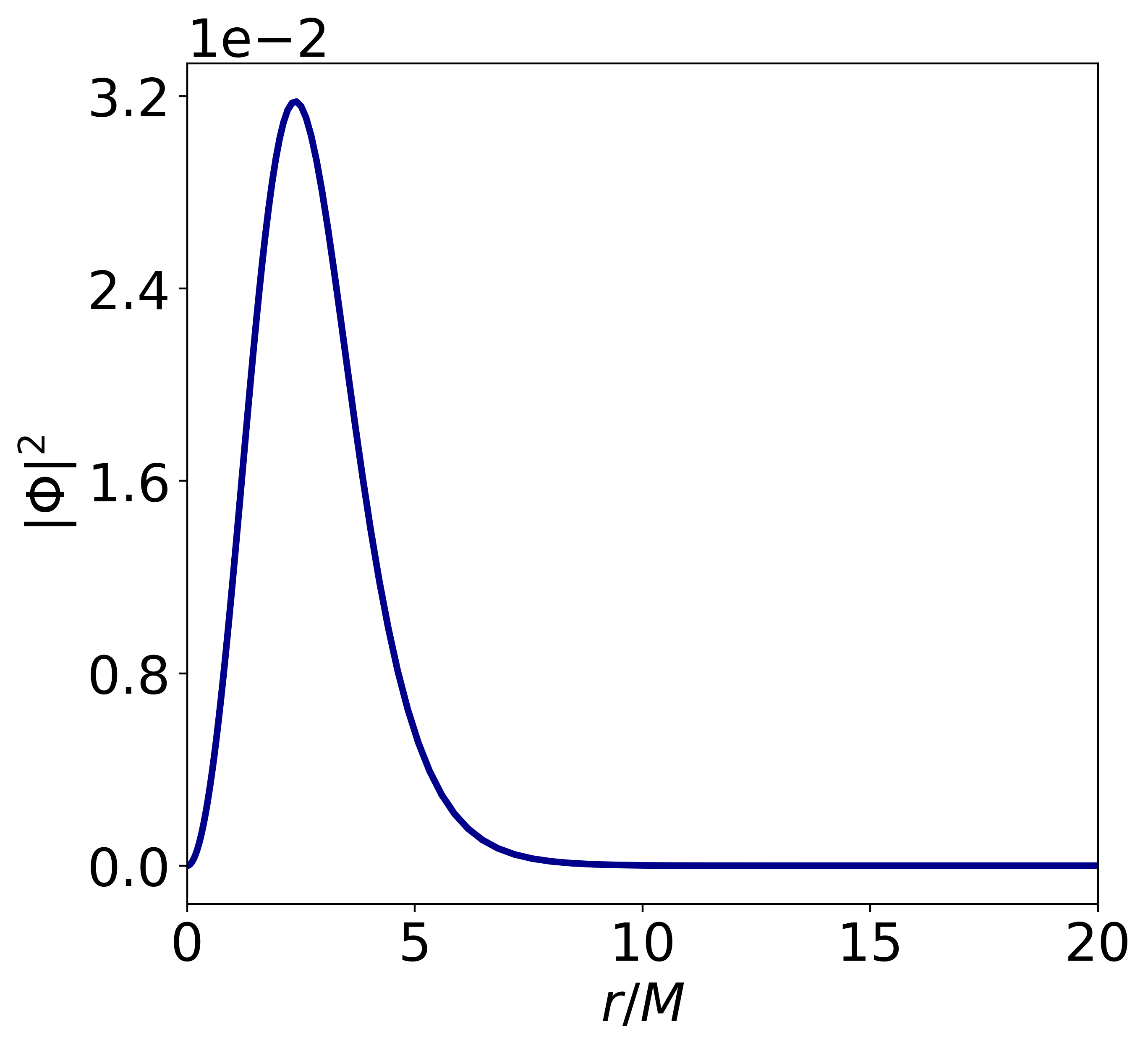}} \hspace{-0.95em}
    \bigskip

    \begin{tabular}{c|c|c|c|c|c|c|c|c}
    \toprule
    $\log_{10}\beta_{mc}$ &  $\rho_{c}$ & $r_c$ & $\Delta r_c$ &  $\tilde{R}$ &  $\bar{\rho}$ &  $\log_{10}\left(p_{max}\right)$ & $\log_{10}\left(p_{m_{max}}\right)$ & $h_{max}$ \\
    \midrule
    5 &         1.000 &      1.321 &        0.00 &       10.555 &         0.226 &                       -0.833 &                     -5.801 &                            1.588 \\
    0 &         1.403 &      0.670 &        0.65 &        5.273 &         0.413 &                       -0.962 &                    -1.103&                            1.311 \\
    -5 &         4.101 &      0.450 &        0.87 &        3.492 &         0.891 &                      -5.120 &                    -0.415&                            1.000 \\
    \bottomrule
    \end{tabular}
    
    \caption{$C$. $\omega=0.798, \ell_0 = -0.4666M$: (a) $\log_{10}$ of the density in the equatorial plane for the different magnetization parameters. Vertical dotted indigo lines represent the radial value corresponding to the intersection of the static surface with the equatorial plane. Vertical dashed lines represent the position of the density center. Vertical dotted lines represent the effective columnar radius $\tilde{R}$. Horizontal dotted lines correspond to the density value $\rho(\tilde{R},\frac{\theta}{2})$. The dashed dotted horizontal blue lines represent the mean equatorial density $\tilde{\rho}$. (b) Scalar profile of the BS solution, the maximum is located at $r = 2.402$.}
    \label{fig:Density_0.798_-0.4666}
\end{figure}

\begin{figure}[H]
\centering
\begin{floatrow}
  \subfloat[$\beta_{mc}=10^5$]{\includegraphics[width=0.3283\columnwidth]{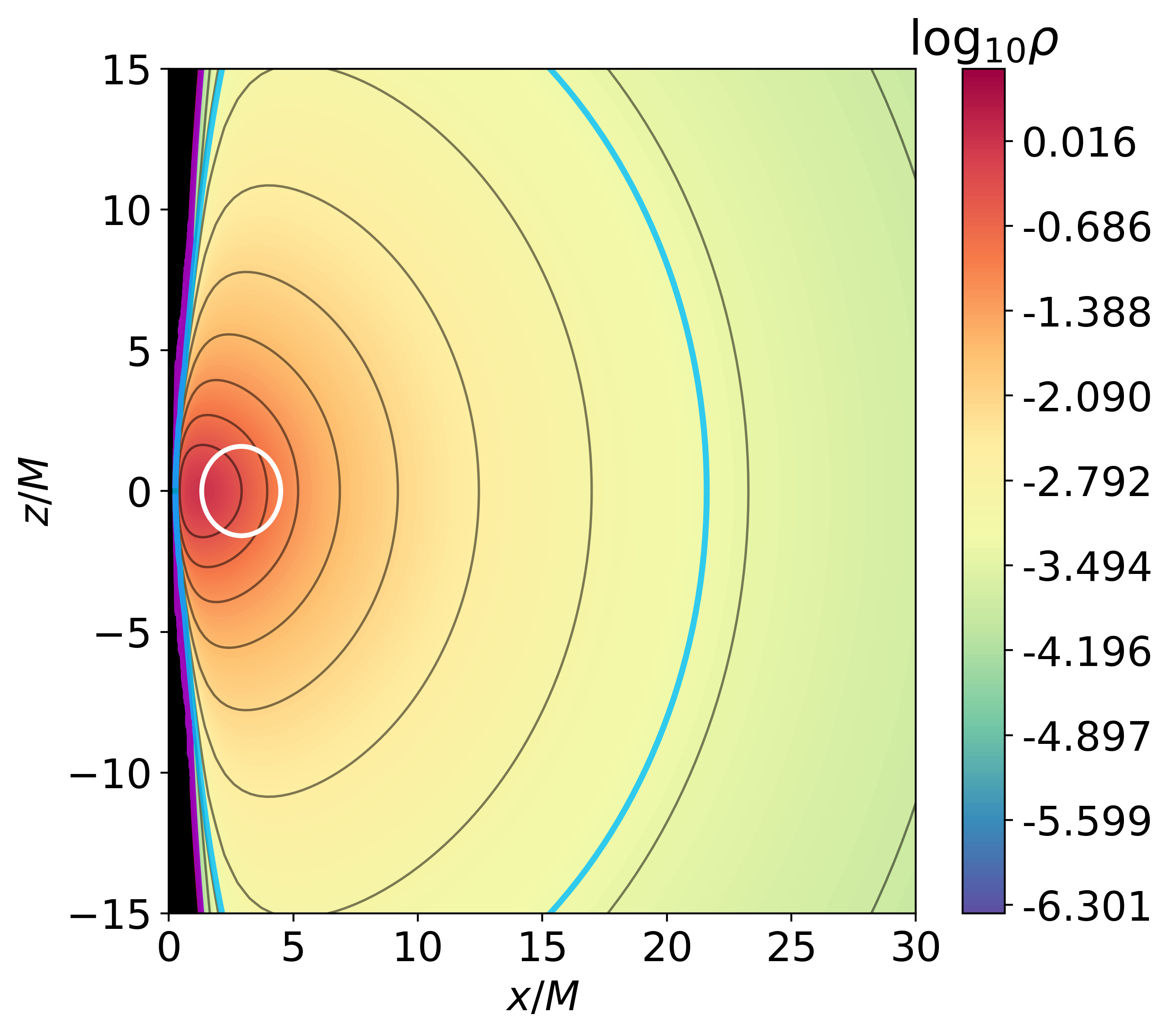}}
  \subfloat[$\beta_{mc}=1$]{\includegraphics[width=0.3283\columnwidth]{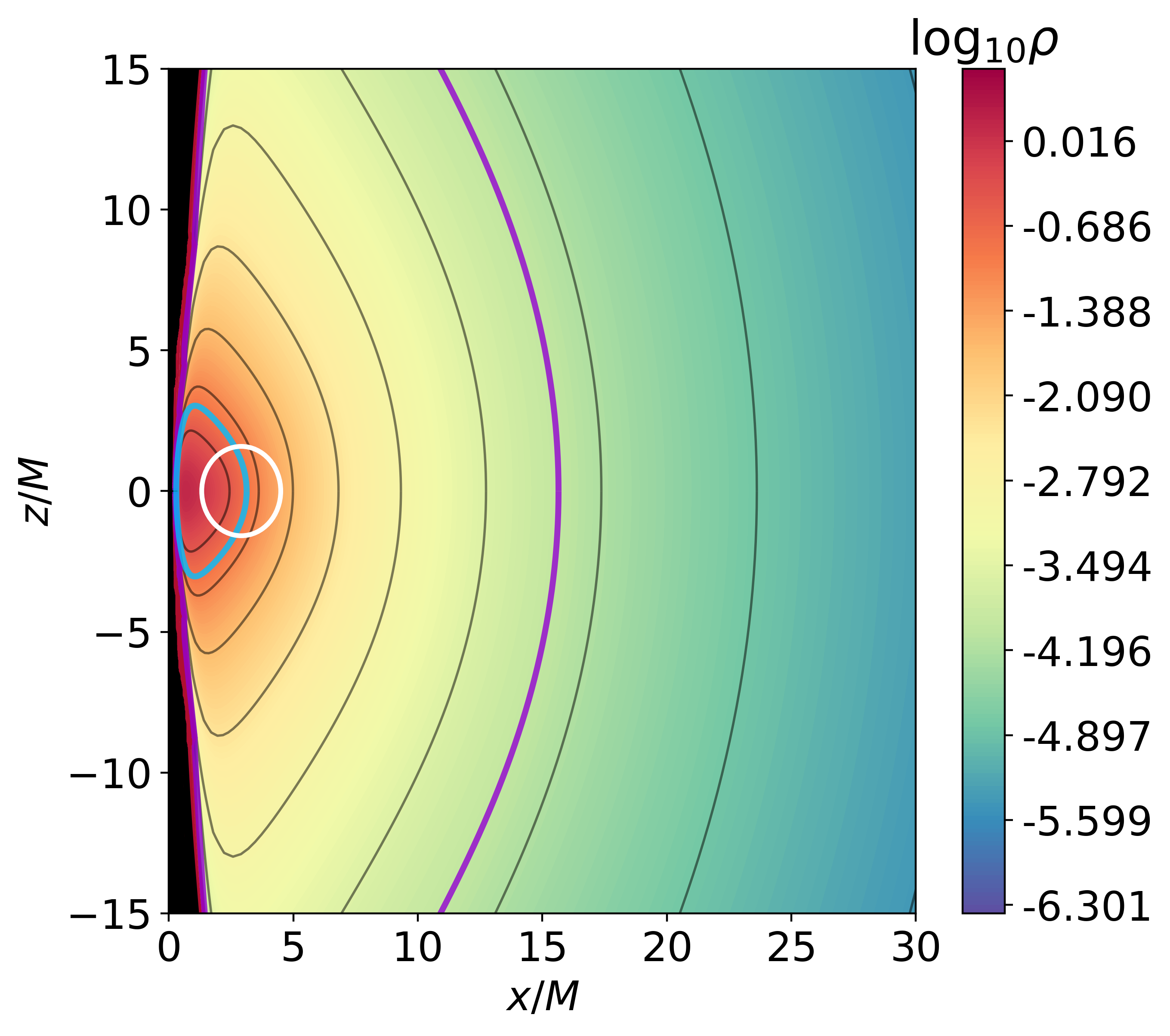}}
  \subfloat[$\beta_{mc}=10^{-5}$]{\includegraphics[width=0.3283\columnwidth]{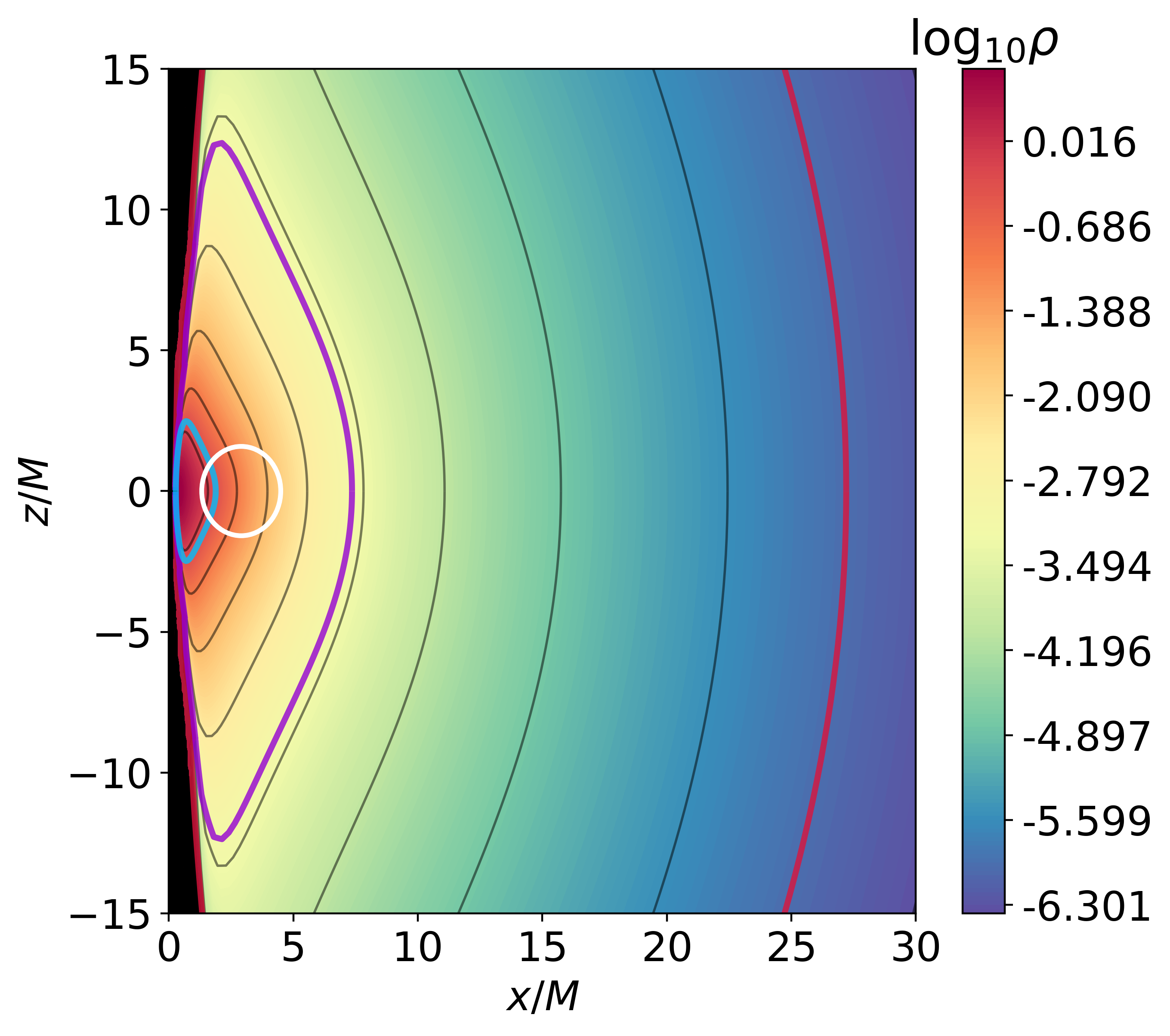}}
\end{floatrow}

\bigskip
\begin{floatrow}
\begin{tabular}{c|c|c|c}
    \toprule
     & $R$ & $\log_{10}\rho$ \\
    \midrule
    $\frac{\int_0^R \rho \mathrm{d}V}{\int_0^\infty \rho \mathrm{d}V} \approx 0.5$ & 21.60 & -3.29  \\ \hline
    $\frac{\int_0^R \rho \mathrm{d}V}{\int_0^\infty \rho \mathrm{d}V} \approx 0.95$ & 376.12 & -7.00 \\ \hline
    $\frac{\int_0^R \rho \mathrm{d}V}{\int_0^\infty \rho \mathrm{d}V} \approx 0.99$ & 413.57 & -7.34 \\ \hline
    \bottomrule
\end{tabular} \hspace{6em}

\begin{tabular}{c|c|c|c}
    \toprule
     & $R$ & $\log_{10}\rho$ \\
    \midrule
    $\frac{\int_0^R \rho \mathrm{d}V}{\int_0^\infty \rho \mathrm{d}V} \approx 0.5$ & 3.12 & -0.84 \\ \hline
    $\frac{\int_0^R \rho \mathrm{d}V}{\int_0^\infty \rho \mathrm{d}V} \approx 0.95$ & 15.65 & -3.99\\ \hline
    $\frac{\int_0^R \rho \mathrm{d}V}{\int_0^\infty \rho \mathrm{d}V} \approx 0.99$ & 44.01 & -6.10 \\ \hline
    \bottomrule
\end{tabular} \hspace{5em}

\begin{tabular}{c|c|c|c}
    \toprule
     & $R$ & $\log_{10}\rho$ \\
    \midrule
    $\frac{\int_0^R \rho \mathrm{d}V}{\int_0^\infty \rho \mathrm{d}V} \approx 0.5$ & 1.88 & -0.36 \\ \hline
    $\frac{\int_0^R \rho \mathrm{d}V}{\int_0^\infty \rho \mathrm{d}V} \approx 0.95$ & 7.35 & -3.11 \\ \hline
    $\frac{\int_0^R \rho \mathrm{d}V}{\int_0^\infty \rho \mathrm{d}V} \approx 0.99$ & 27.20 & -5.99 \\ \hline
    \bottomrule
\end{tabular}
\end{floatrow}
  
\caption{$C$. $\omega=0.798, \ell_0 = -0.4666M$: Torus solutions visualized for different magnetization parameters. Black solid contour lines represent equidensity surfaces and the white circle represents the static surface. The minimum of the density is set to $\log_{10}\rho = -6.37$ in all figures and the maximum is set to the density maximum of the highly magnetized solution. The light blue, violet and red lines represent each the equidensity surface within which $50\%, \ 95\%$ and $99\%$ of the approximated mass lies. The tables below each figure contain the corresponding radial value in the equatorial plane of these equidensity surfaces and their density value.}
\label{fig:Torus_0.798_-0.4666}
\end{figure}

\paragraph{$\omega=0.798, \ell_0 = -0.4666M:$}
For the mildly relativistic BS solutions the non-magnetized disk is already relatively compact compared to the previous solutions and located close to the center of the BS. The static surface is located close to the disk center, with the inner intersection with the equatorial plane being located at the disk center. As in the other solutions, magnetization compresses the disk leading to a higher density around the torus center and a shift towards the BS center. This shift is relatively small, since the non-magnetized center location is already close to the BS center. The decrease of $\tilde{R}$ and increase of $\bar{\rho}$ indicate again a further compactification and elongated density distribution (Fig.~\ref{fig:Density_0.798_-0.4666}).
The density at the outer intersection of the static surface with the equatorial plane is approximately one order of magnitude lower compared to the non-magnetized solution (Fig.~\ref{fig:Density_0.798_-0.4666}), in the high magnetized case it lies only partially in the volume which contains half the accretion disk mass, as shown in Fig.~\ref{fig:Torus_0.798_-0.4666}. In general the disk gets characteristically more compressed, with sharp edged contour lines forming equidensity surfaces with a smaller diameter, resulting in a torus geometry similar to a sharp ellipsoid which is centered around the BS center with most of the mass located close to the center (Fig.~\ref{fig:Torus_0.798_-0.4666}).

\newpage

\newpage
\subsection{Two-centered Disks}
\begin{figure}[H]
\centering
\begin{floatrow}
  \subfloat[$\omega=0.798, \ell_0 = -4.5M$]{\includegraphics[height=0.1915\textheight]{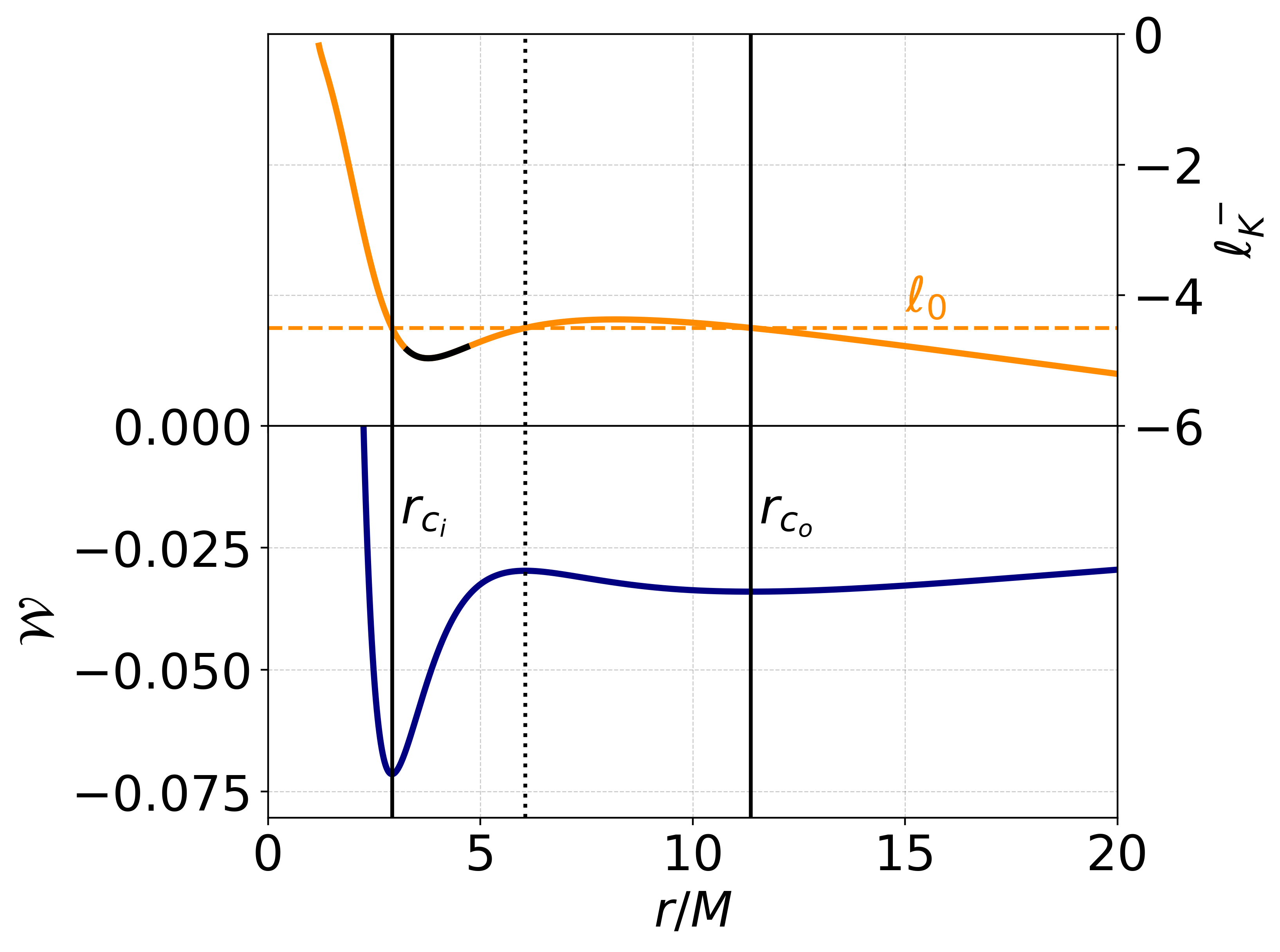}}
  \subfloat[$\omega=0.671, \ell_0 = -4.5M$]{\includegraphics[height=0.1915\textheight]{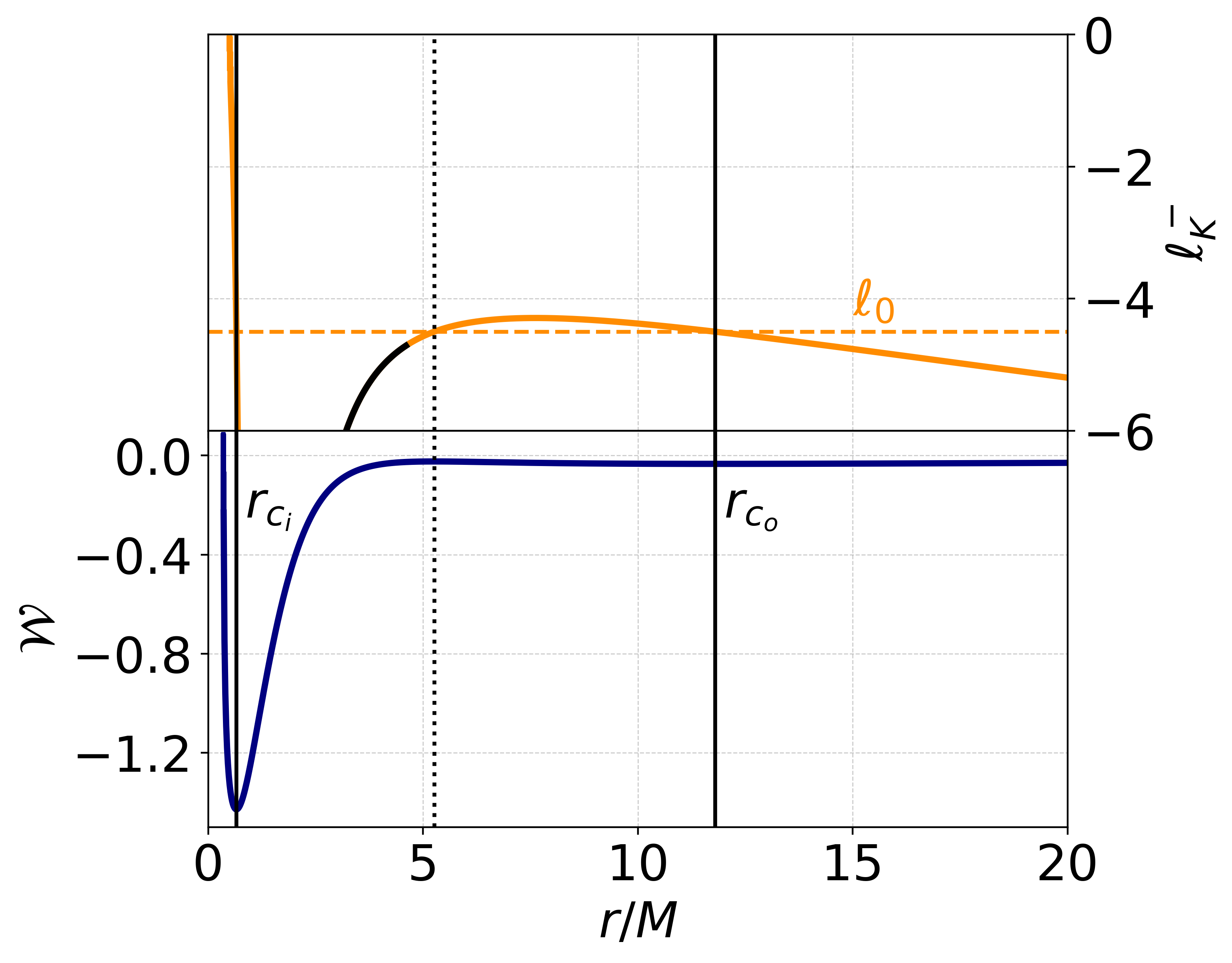}}
  \subfloat[$\omega=0.671, \ell_0 = -5M$]{\includegraphics[height=0.1915\textheight]{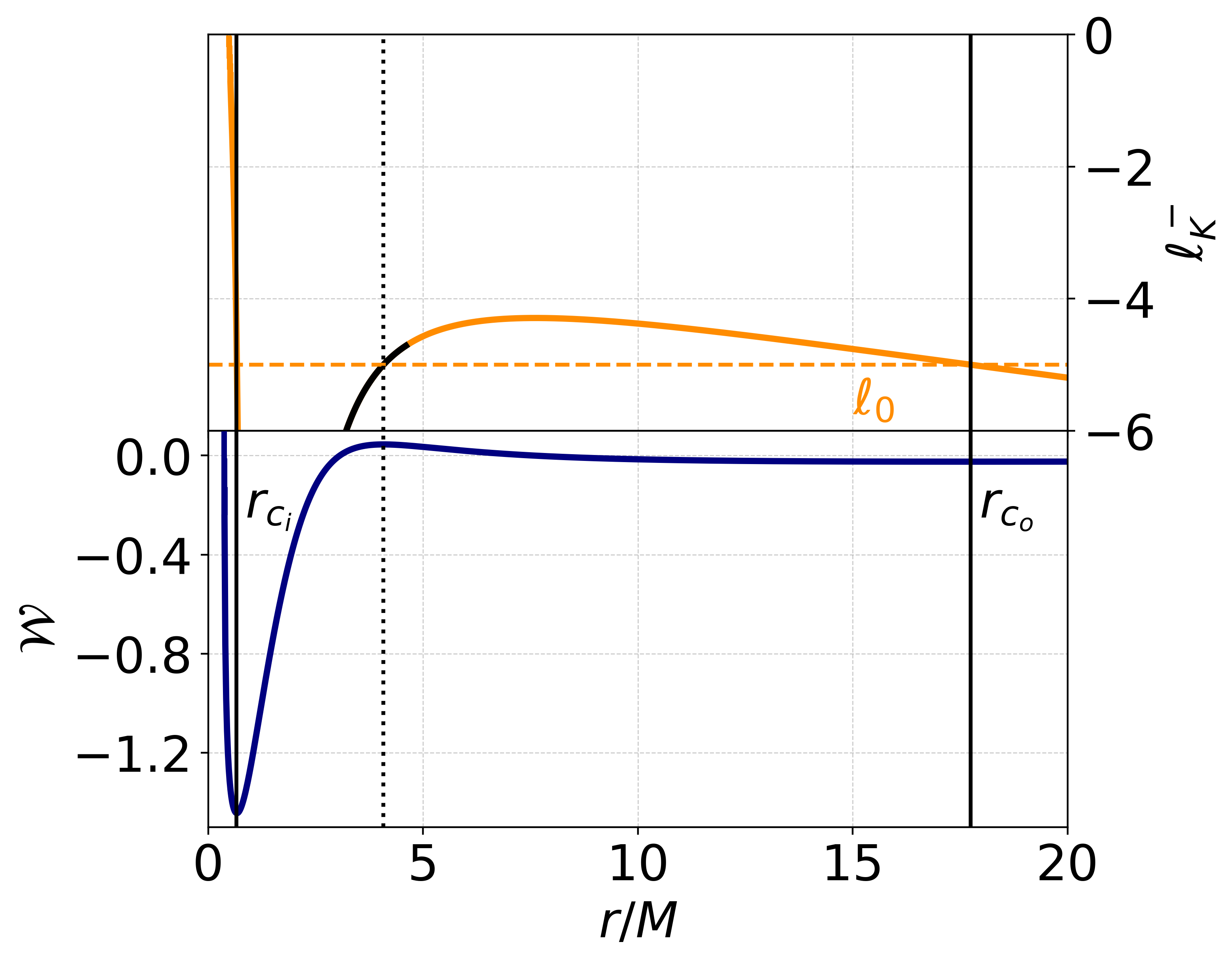}}
\end{floatrow}

\begin{floatrow}
  \subfloat[$\omega=0.798, \ell_0 = -4.75M$]{\includegraphics[width=0.3283\columnwidth]{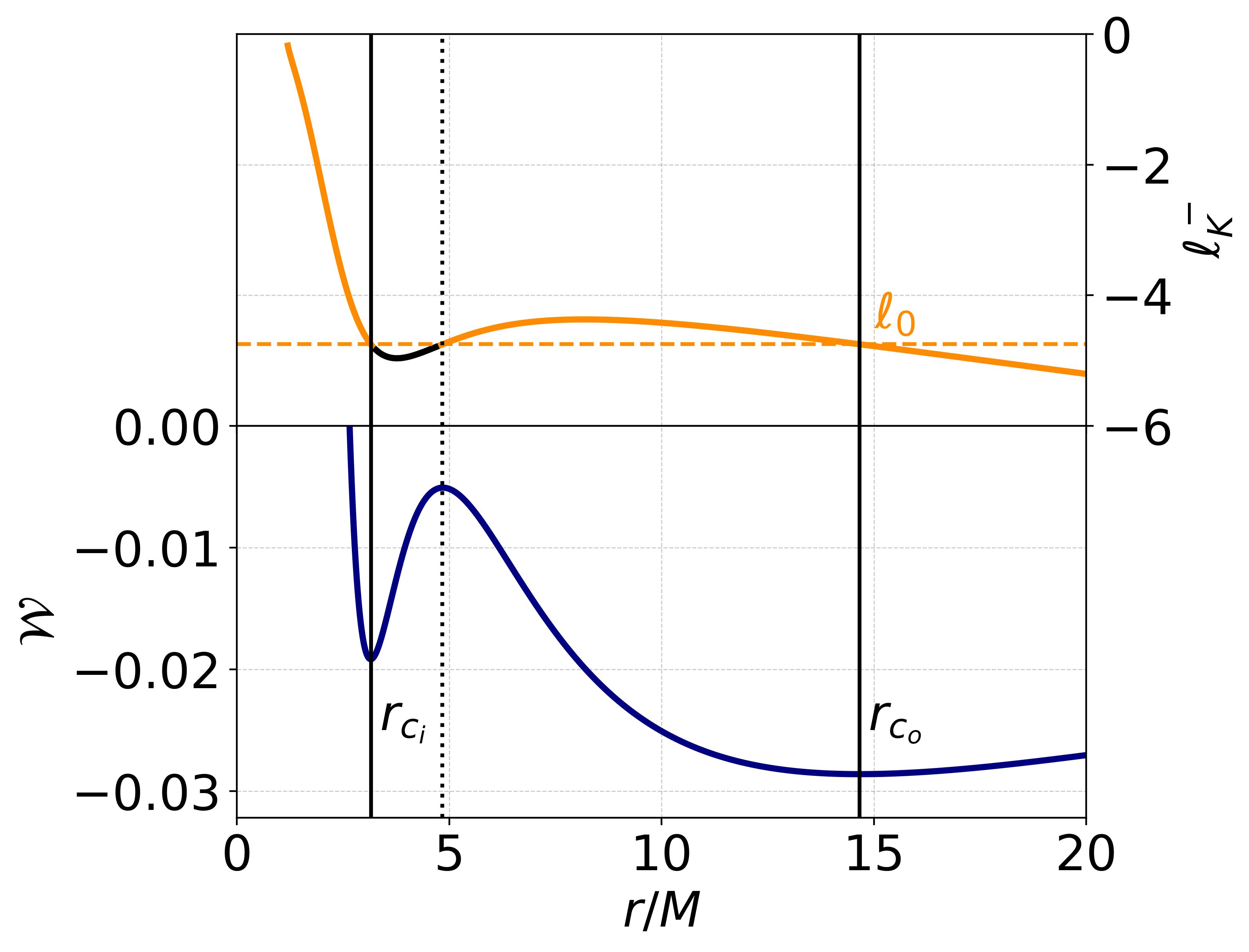}}\hspace{-4.75em}
\end{floatrow}
\caption{The upper part of the figures shows the normalized Keplerian specific angular momentum $\ell_K^-$, black curve sections represent the radial coordinates for which only unbound orbits are possible. Dashed horizontal lines illustrate the chosen $\ell_0$ value. The lower part of the figures shows the corresponding effective potential $\mathcal{W}$. Solid vertical lines indicate the disk and (non-magnetized) density center locations, with $r_{c_i}$ as the inner disk and density center and $r_{c_o}$ as the outer disk and density center. Dotted vertical lines indicate the position of the cusp.}
\label{fig:Momentum_pot_two_centered}
\end{figure}

In contrast to the Kerr black hole, it is possible for mildly and highly relativistic BSs to shelter accretion disk solutions with more than one center (albeit solutions with two centers are shown to exist for Kerr black holes with scalar hair \cite{Teodoro_2021, Gimeno-Soler}). These two-centered solutions differ quite strongly from those analyzed so far, they either possess a cusp, which connects the two centers of the disk or they have no cusp. In the latter case the accretion disk consists of two separated tori, one inner torus and one outer torus. For the two-centered disks we have selected the following 4 representative examples:
$A$.~$\omega=0.798, \ell_0 = -4.5M$, $B$.~$\omega=0.671, \ell_0 = -4.5M$, $C$.~$\omega=0.671, \ell_0 = -5M$ and $D$.~$\omega=0.798, \ell_0 = -4.75M$. Fig. \ref{fig:Momentum_pot_two_centered} illustrates the normalized Keplerian specific angular momentum and effective potential of these solutions.\\

\begin{figure}[H]
    \centering
    \subfloat[]{\includegraphics[width=0.725\linewidth]{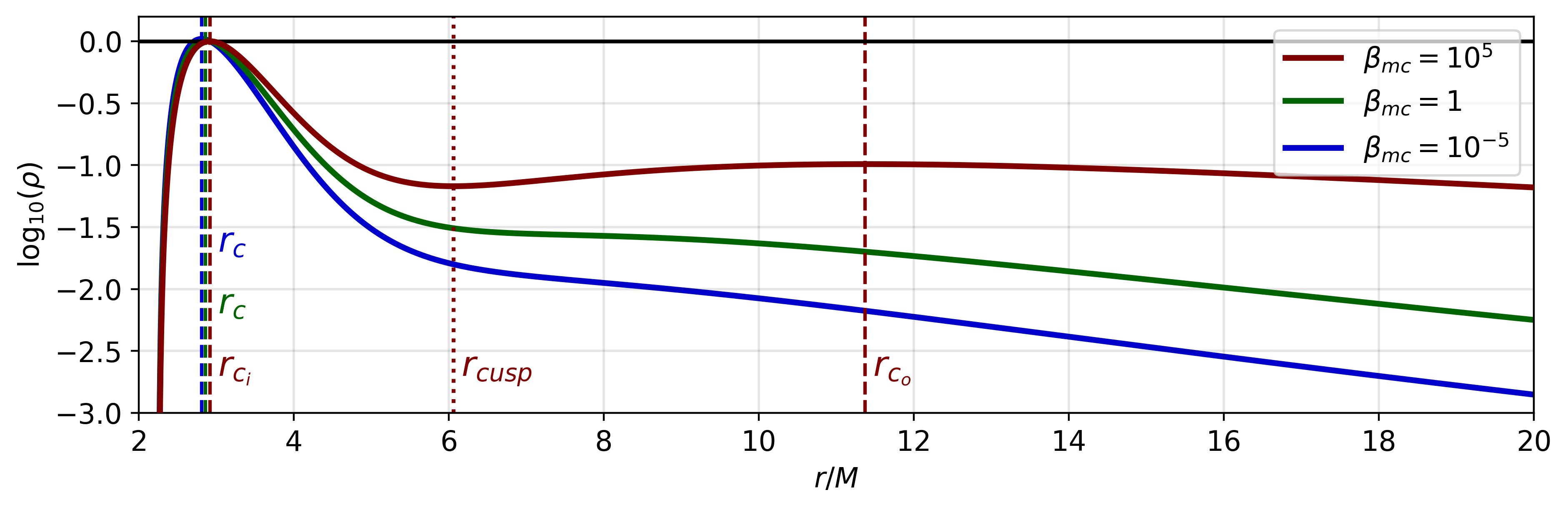}}
    \subfloat[]{\includegraphics[width=0.265\linewidth]{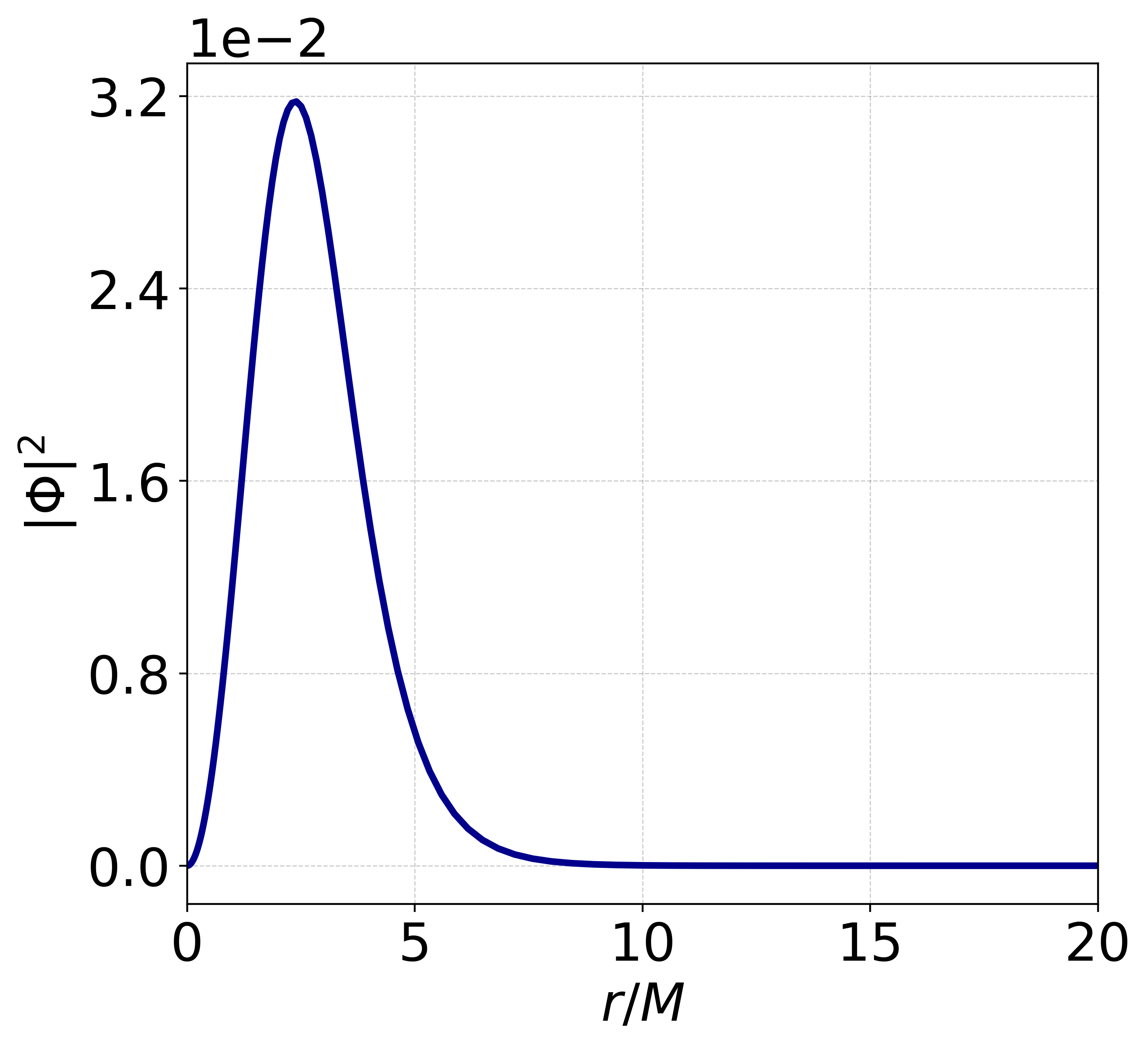}}
    \bigskip
    
    \begin{tabular}{c|c|c|c|c|c|c|c|c|c|c}
    \toprule
    $\log_{10}\beta_{mc}$ &  $\rho_{c_i}$ & $r_{c_i}$ & $\rho_{c_o}$ &  $r_{c_o}$ & $\rho_{cusp}$ & $r_{cusp}$ & $\frac{\rho_{c_i}}{\rho_{c_o}}$ &  $\log_{10}\left(p_{max}\right)$ & $\log_{10}\left(p_{m_{max}}\right)$ & $h_{max}$ \\
    \midrule
     5 &         1.000 &      2.921 &    0.102 &      11.372 &        0.068 &      6.061 &       9.813 &              -1.733 &                               -6.731 &                            1.074 \\
            0 &         1.011 &      2.861 &    - &       - &        - &      - &       - &                           -2.031 &                              -2.036 &                            1.037 \\
           -5 &         1.044 &      2.811 &    - &       - &        - &      - &       - &                           -6.724 &                               -1.734 &                            1.000 \\
    \bottomrule
    \end{tabular}
    
    \caption{$A$. $\omega=0.798, \ell_0 = -4.5M$: (a) $\log_{10}$ of the density in the equatorial plane for different magnetization parameters. Dashed vertical lines represent center positions, with $r_{c_i}$ referring to the inner density center and $r_{c_o}$ to the outer density center. The dotted vertical line represents the cusp. The outer center (and therefore also the cusp) vanishes in the magnetized solutions. (b) Scalar profile of the BS solution, the maximum is located at $r = 2.402$.}
    \label{fig:Density_0.798_-4.5}
\end{figure}

\begin{figure}[H]
\centering
   \subfloat[$1 \leq \beta_{mc} \leq 100$]{\includegraphics[width=0.42\columnwidth]{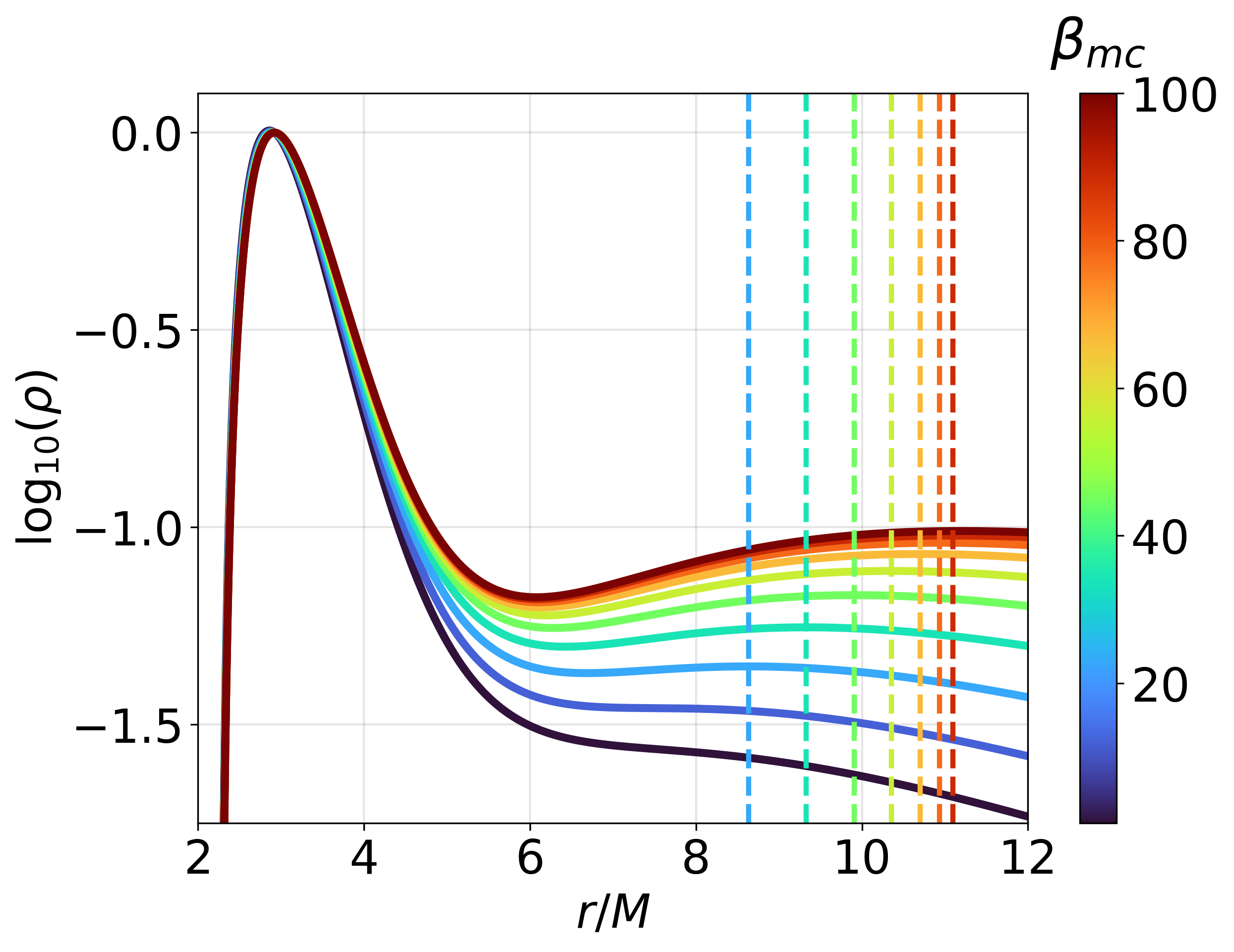}}\hfill
   \subfloat[$1.72 \leq \beta_{mc} \leq 1.78$]{\includegraphics[width=0.44\columnwidth]{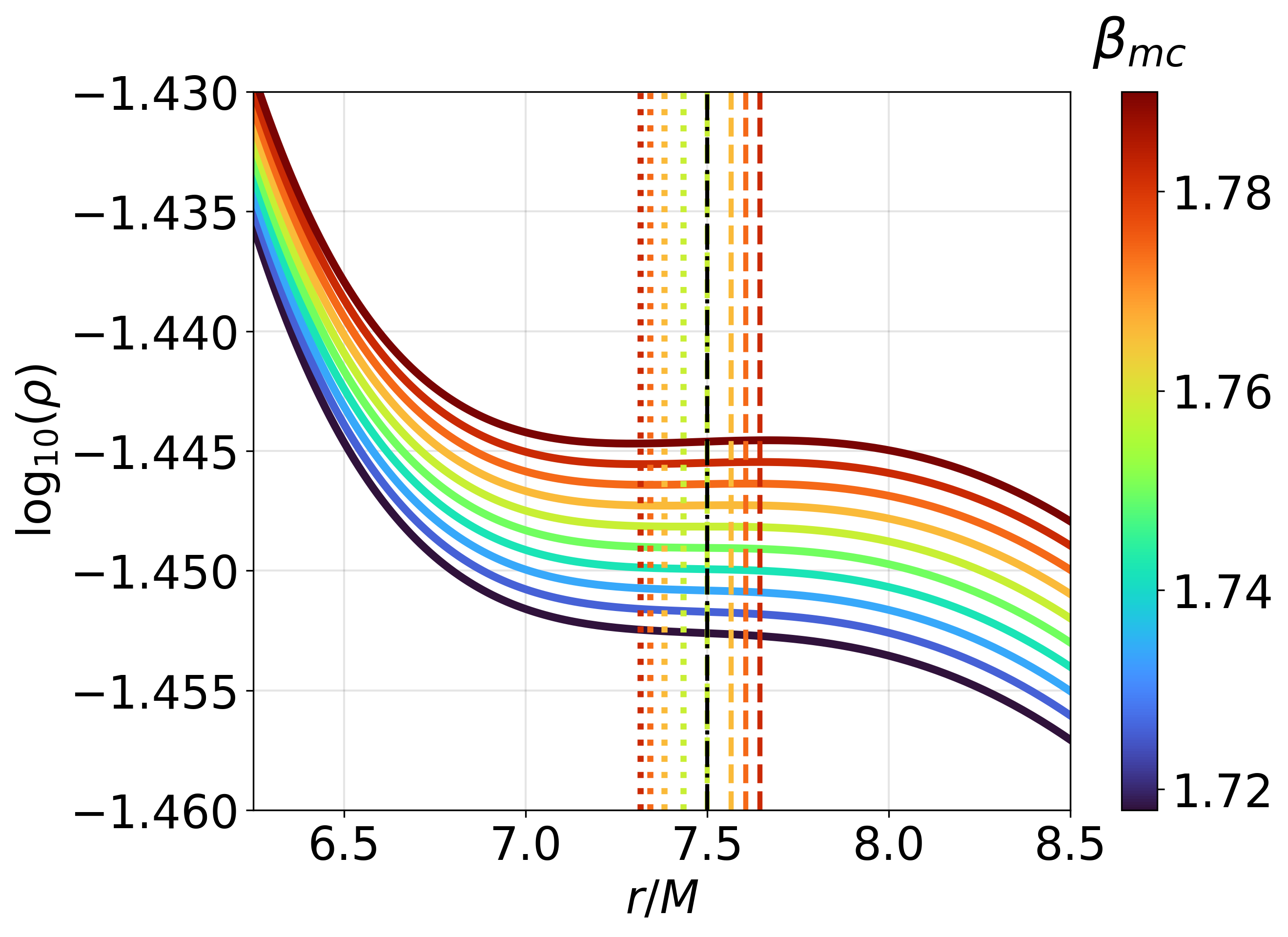}}
   
\caption{$A$. $\omega=0.798, \ell_0 = -4.5M$: (a) Solutions for $\beta_{mc}$ in the range of 1 to 100, representing mildly magnetized disks. 
(b) Closeup of solutions for $\beta_{mc}$ around the threshold value $\beta_0$. 
Dashed lines represent the location of the outer density center $r_{c_0}$, dotted lines represent the location of the density cusp. The dashed-dotted black line in (b) shows the location of the saddle point, which exists for the density curve corresponding to the threshold value of $\beta_0 \approx 1.757$. For the lowest two curves in (a) and for the lowest five curves in (b) $\beta_{mc}$ is below $\beta_0$, therefore there exists no density cusp and outer density center.}
\label{fig:threshold_0.798_-4.5}
\end{figure}

\begin{figure}[H]
\centering
\begin{floatrow}
  \subfloat[$\beta_{mc}=10^5$]{\includegraphics[width=0.3283\columnwidth]{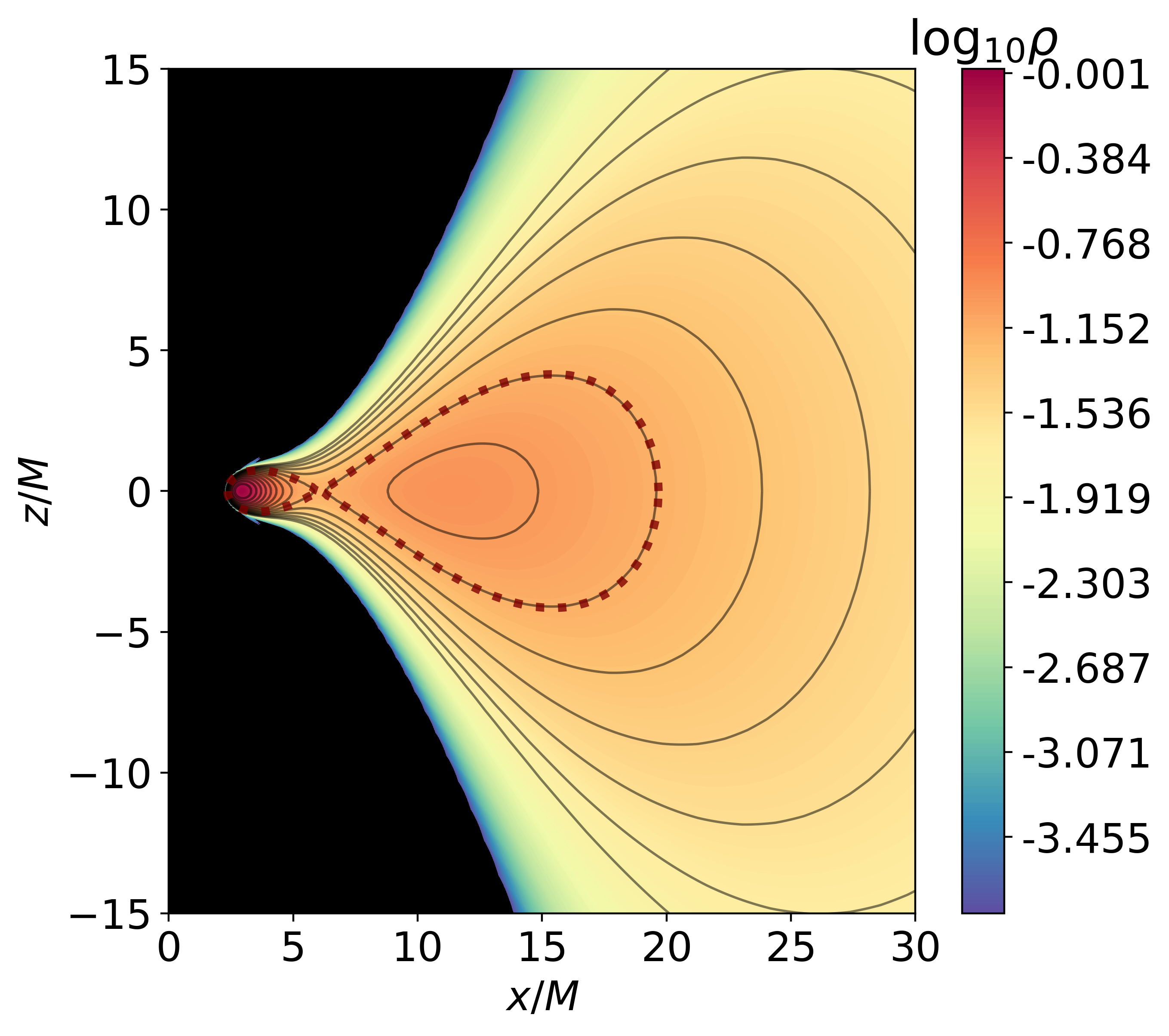}}
  \subfloat[$\beta_{mc}= 1$]{\includegraphics[width=0.3283\columnwidth]{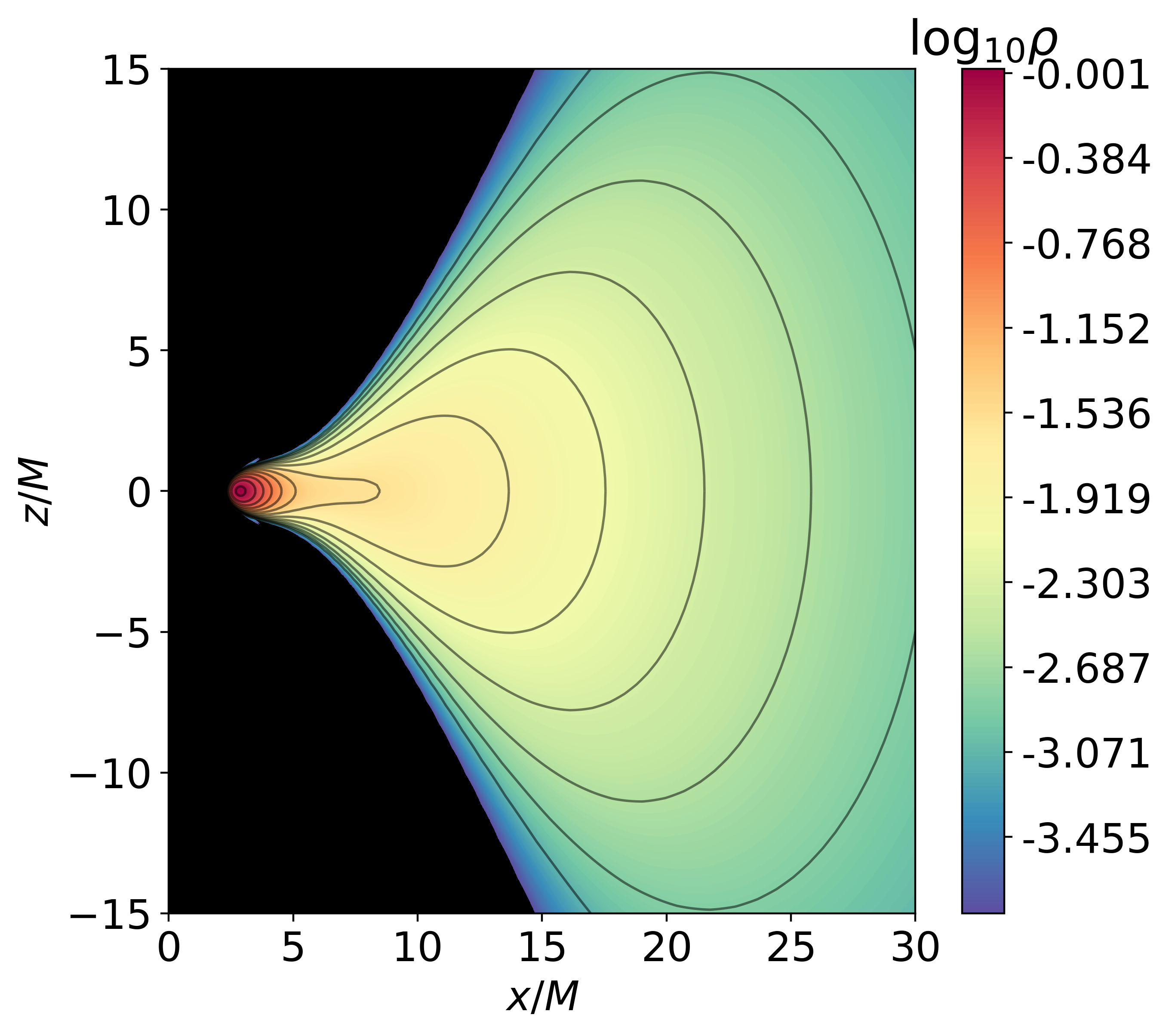}}
  \subfloat[$\beta_{mc}=10^{-5}$]{\includegraphics[width=0.3283\columnwidth]{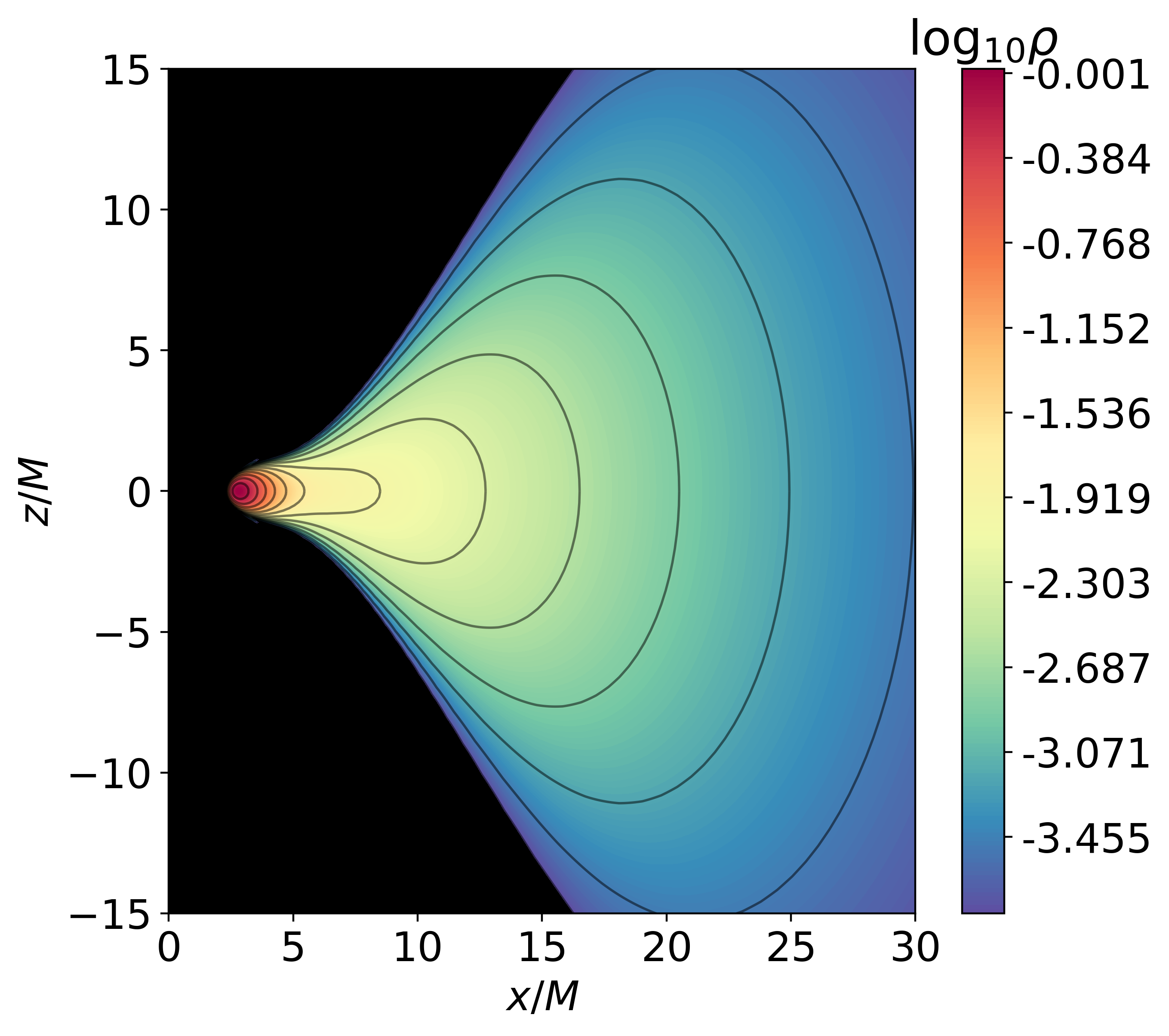}}
\end{floatrow}

\begin{floatrow}
  \subfloat[$\beta_{mc}= 1.757$]{\includegraphics[width=0.3283\columnwidth]{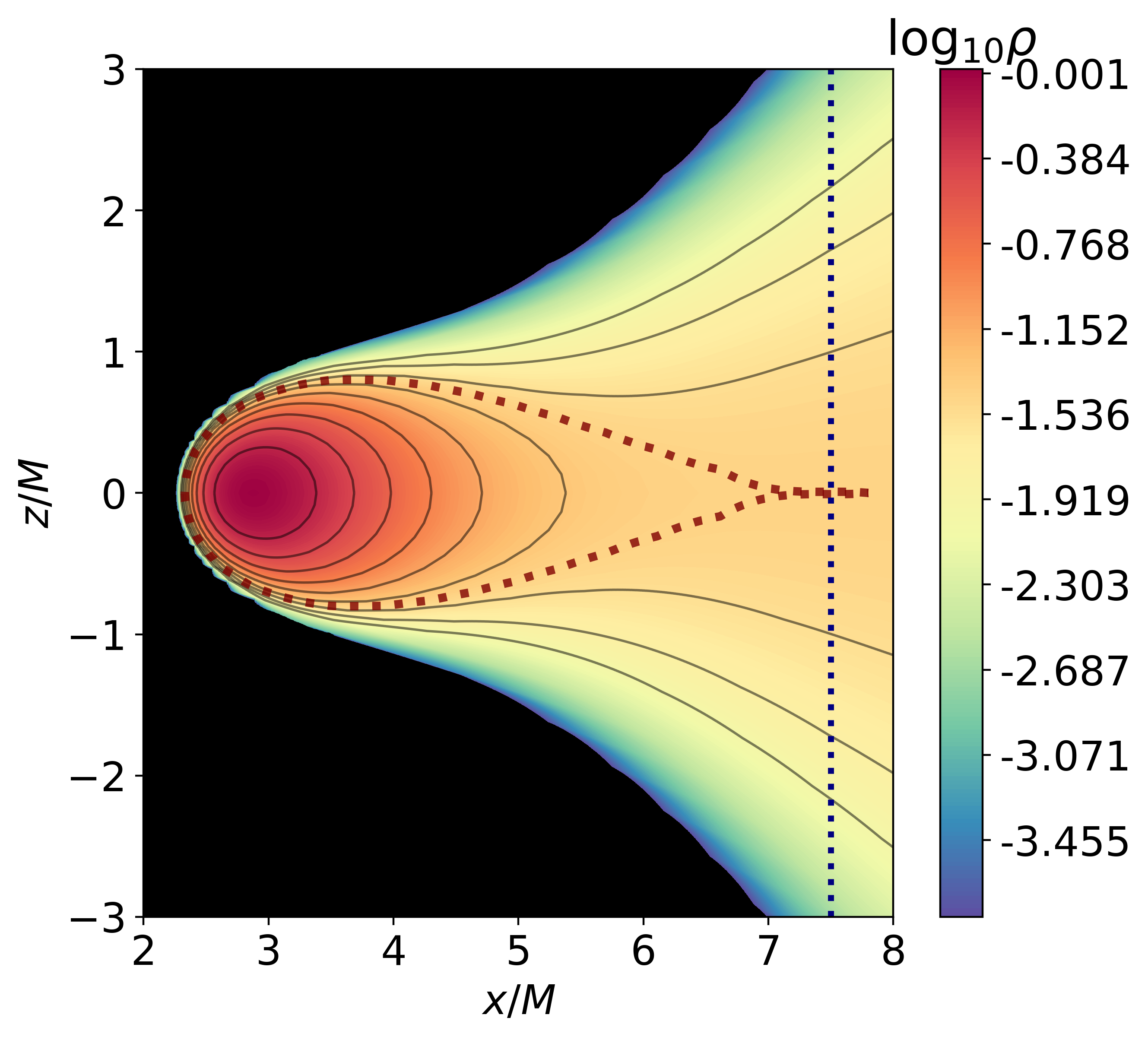}}\hspace{-5.2em}
\end{floatrow}

\caption{$A$. $\omega=0.798, \ell_0 = -4.5M$: Density distribution visualized for different magnetization parameters. The minimum of the density is set to $\log_{10}\rho = -3.8$ for (a) - (d) and the maximum to the inner density maximum of the strong magnetized solution. Black solid lines represent equidensity surfaces, dotted red lines represent the equidensity surface corresponding to the cusp. The dashed-dotted blue line in (d) represents the location of the saddle point for the solution with the threshold value $\beta_0 = 1.757$.}
\label{fig:Torus_0.798_-4.5}
\end{figure}

\paragraph{$\omega=0.798, \ell_0 = -4.5M:$}
This solution is in the non-magnetized case composed of a two-centered disk connected by a cusp. As shown in Fig.~\ref{fig:Density_0.798_-4.5}, the equatorial density of the mildly and strong magnetized solution only possesses one maximum and no minimum. Therefore the density cusp and outer density center vanish, resulting in a change of disk morphology from a two-torus solution to a one-torus solution. For all radial values greater than the inner disk center of the non-magnetized solution, the density is monotonically decreasing in correspondence to the magnetization parameter. Since the local extreme points vanish for strongly magnetized disks, there must be a threshold value $\beta_0$, below which only one-torus solutions exist, the disk topology is therefore dependent on the magnetization parameter. Fig.~\ref{fig:threshold_0.798_-4.5} shows an analysis of density curves for magnetization parameters close to this threshold value. The density cusp and outer density center converge to one location for the threshold value $\beta_0 \approx 1.757$, which marks a saddle point of the density curve (Fig.~\ref{fig:threshold_0.798_-4.5} (b)). As seen in Fig.~\ref{fig:Torus_0.798_-4.5} the disk gets compactified towards the inner density center and the contour lines of the equidensity surfaces close to the center are of circular shape and smaller in diameter compared to the contour lines of lower density, which have a greater extent and possess a teardrop-like shape. \\

\begin{figure}[H]
    \centering
    \subfloat[]{\includegraphics[width=0.725\linewidth]{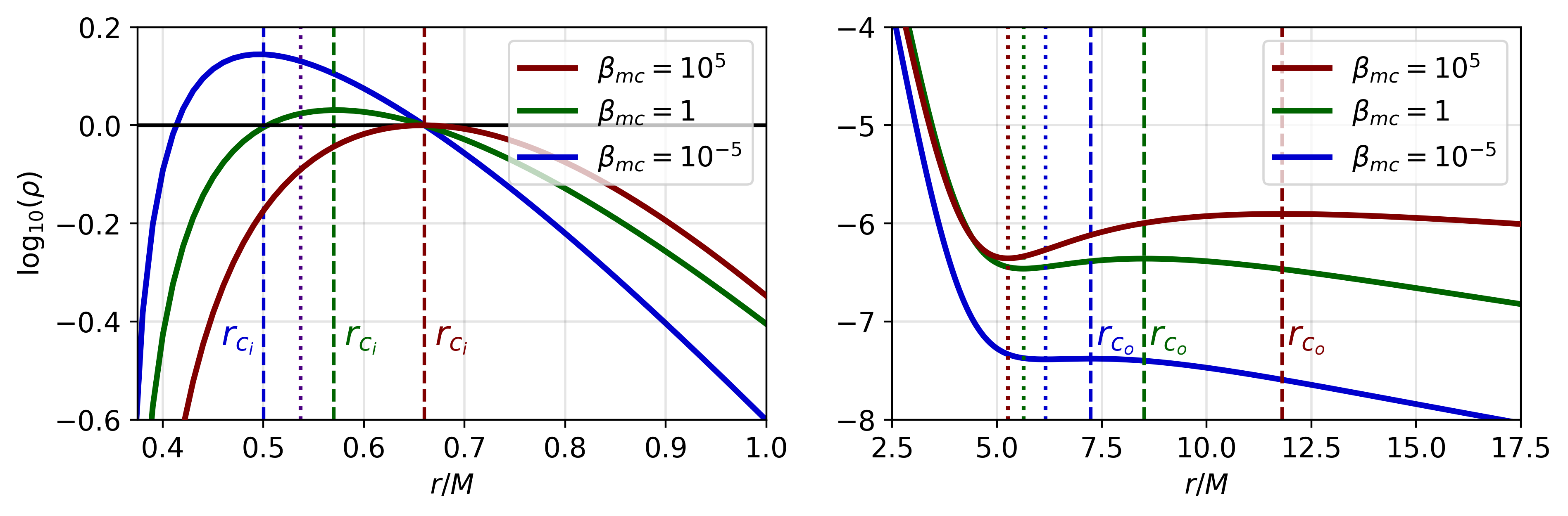}}
    \subfloat[]{\includegraphics[width=0.265\linewidth]{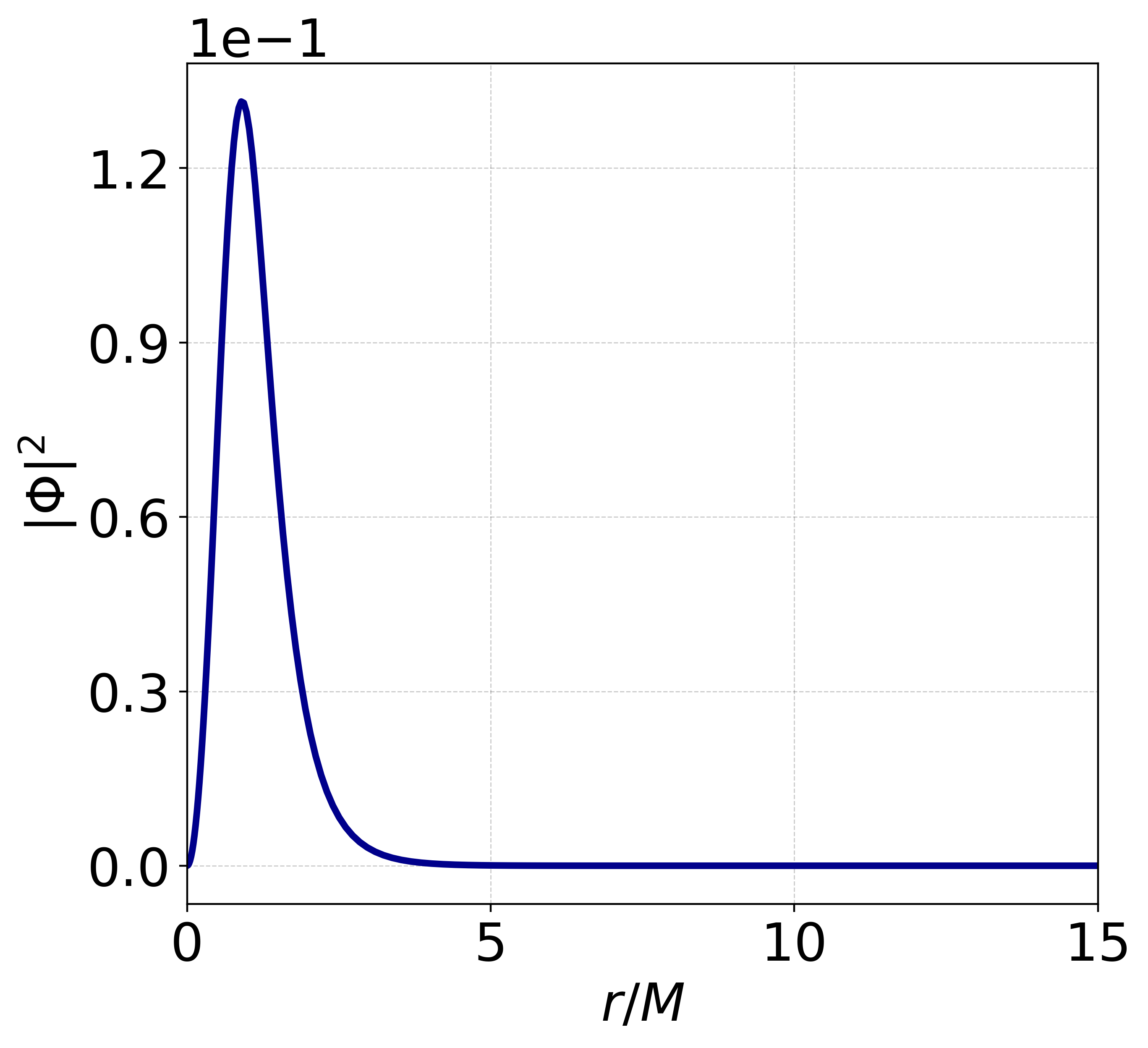}}
    \bigskip
    
    \begin{tabular}{c|c|c|c|c|c|c|c|c|c|c}
    \toprule
    $\log_{10}\beta_{mc}$ &  $\log_{10}\rho_{c_i}$ & $r_{c_i}$ & $\log_{10}\rho_{c_o}$ &  $r_{c_o}$ & $\log_{10}\rho_{cusp}$ & $r_{cusp}$ & $\log_{10}\left(\frac{\rho_{c_i}}{\rho_{c_o}}\right)$ &  $\log_{10}\left(p_{max}\right)$ & $\log_{10}\left(p_{m_{max}}\right)$ & $h_{max}$ \\
    \midrule
    5 &         0.000 &       0.66 &   -5.903 &      11.802 &       -6.356 &      5.261 &       5.903 &                           -0.101 &                               -5.095  &                            4.172 \\
            0 &         0.031 &       0.57 &   -6.359 &       8.512 &       -6.461 &      5.631 &       6.389 &                           -0.431 &                               -0.457&                            2.382 \\
           -5 &         0.144 &       0.50 &   -7.377 &       7.241 &       -7.384 &      6.151 &       7.522 &                           -5.255 &                               -0.330 &                            1.000 \\
    \bottomrule
    \end{tabular}
    
    \caption{$B$. $\omega=0.671, \ell_0 = -4.5M$: (a) The left panel shows the $\log_{10}$ of the density in the equatorial plane around the inner density center of the disk, with the dashed lines marking the position of the inner center and the indigo dotted line the position of the inner intersection of the static surface with the equatorial plane. The right panel shows the density in the equatorial plane around the outer density center, with the dotted lines representing the location of the density cusp and the dashed lines the location of the outer center. (b) Scalar profile of the BS solution, the maximum is located at $r = 0.893$.}
    \label{fig:density_0.671_-4.5}
\end{figure}

\begin{figure}[H]
\centering
   \subfloat[\label{Loc_Cente_Cusp_0.671_-4.5}]{\includegraphics[width=0.42\columnwidth]{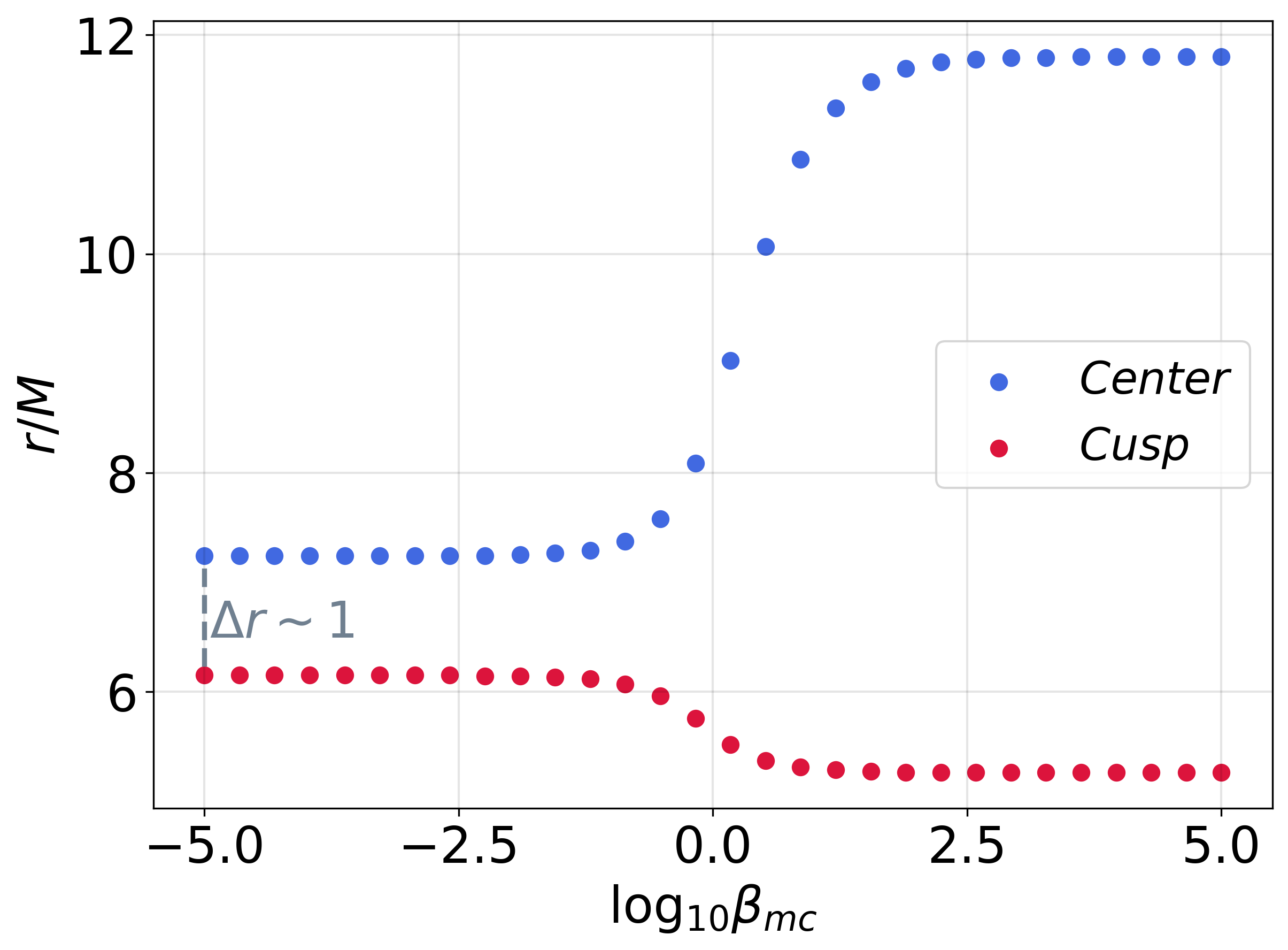}}\hfill
   \subfloat[\label{Density_Center_Cusp0.671_-4.5}]{\includegraphics[width=0.4485\columnwidth]{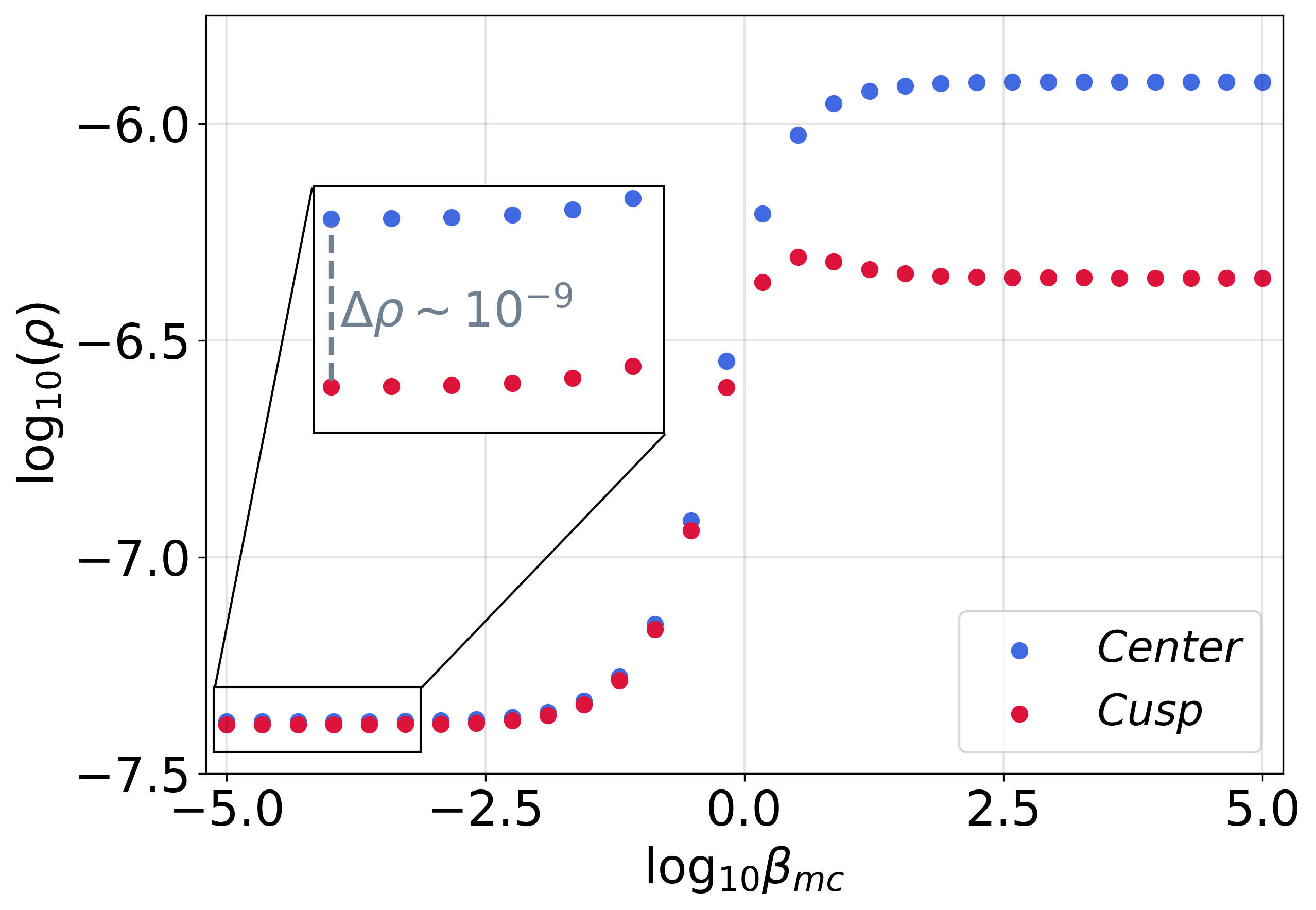}}
   
\caption{$B$. $\omega=0.671, \ell_0 = -4.5M$: (a) Locations of the outer density center and density cusp versus $\beta_{mc}$. 
(b) Density values at the outer density center and density cusp versus $\beta_{mc}$.}
\label{fig:Cusp_center_0.671_-4.5}
\end{figure}

\begin{figure}[H]
\centering
\begin{floatrow}
  \subfloat[$\beta_{mc}=10^5$]{\includegraphics[width=0.3283\columnwidth]{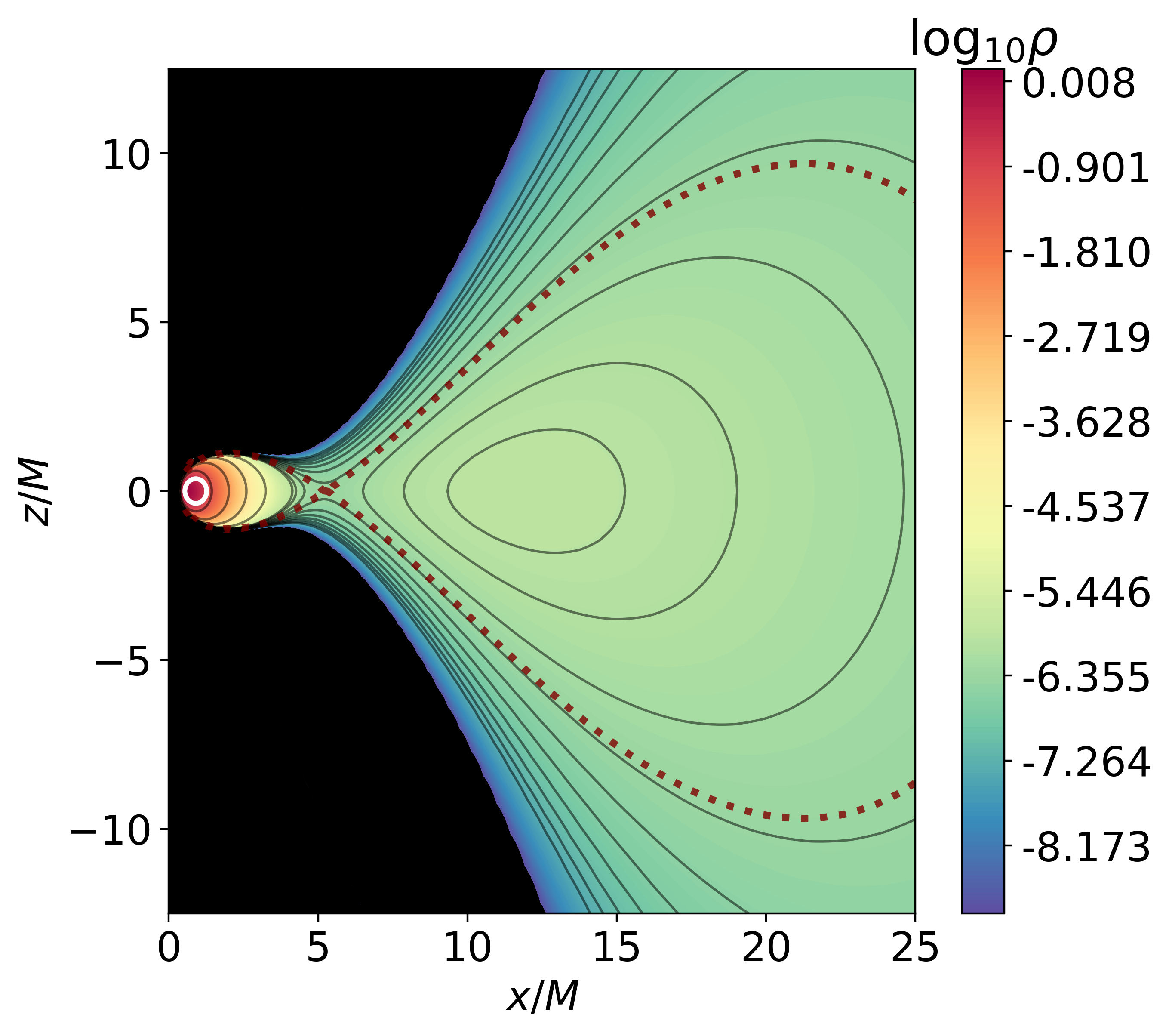}}
  \subfloat[$\beta_{mc}= 1$]{\includegraphics[width=0.3283\columnwidth]{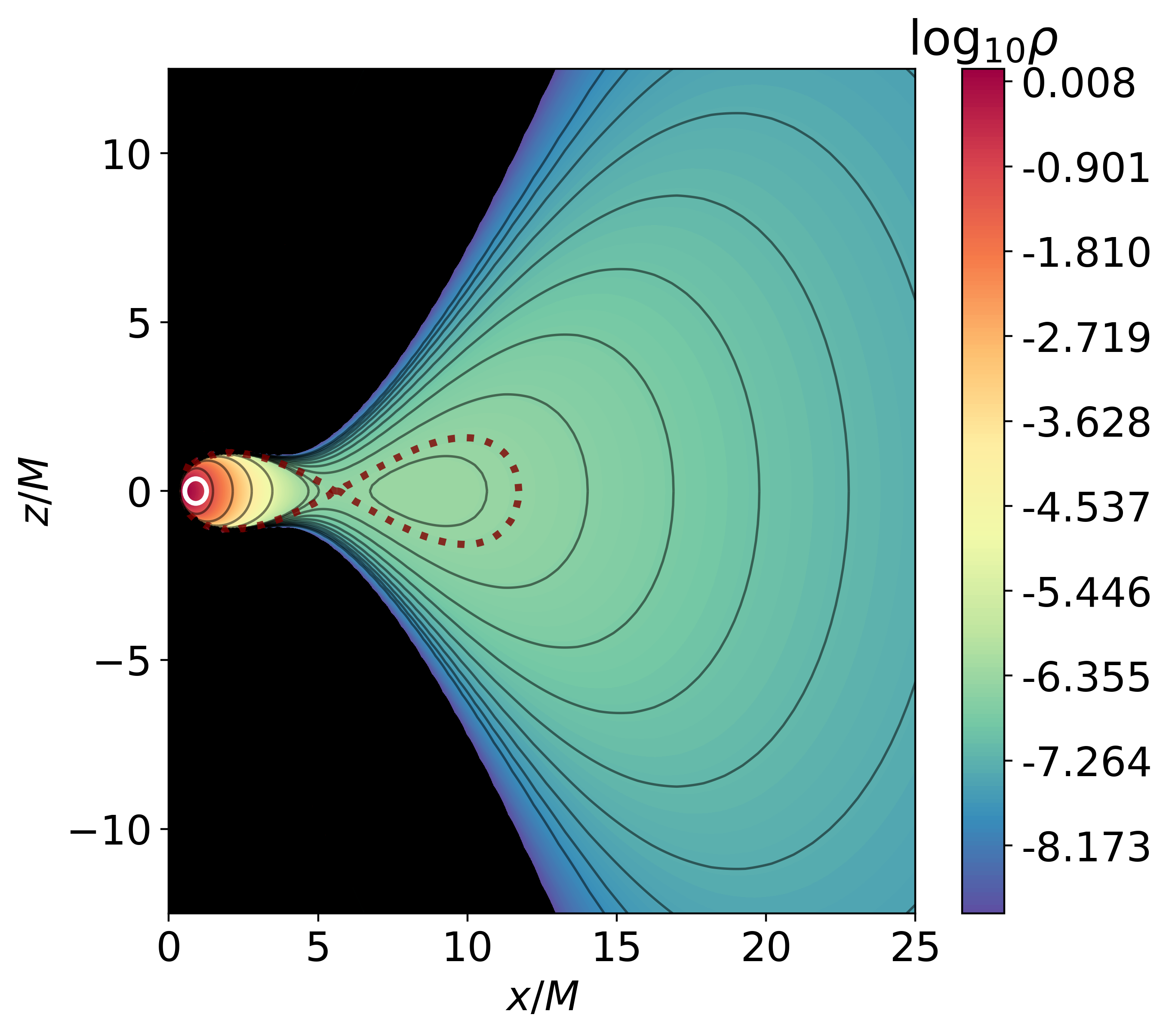}}
  \subfloat[$\beta_{mc}=10^{-5}$]{\includegraphics[width=0.3283\columnwidth]{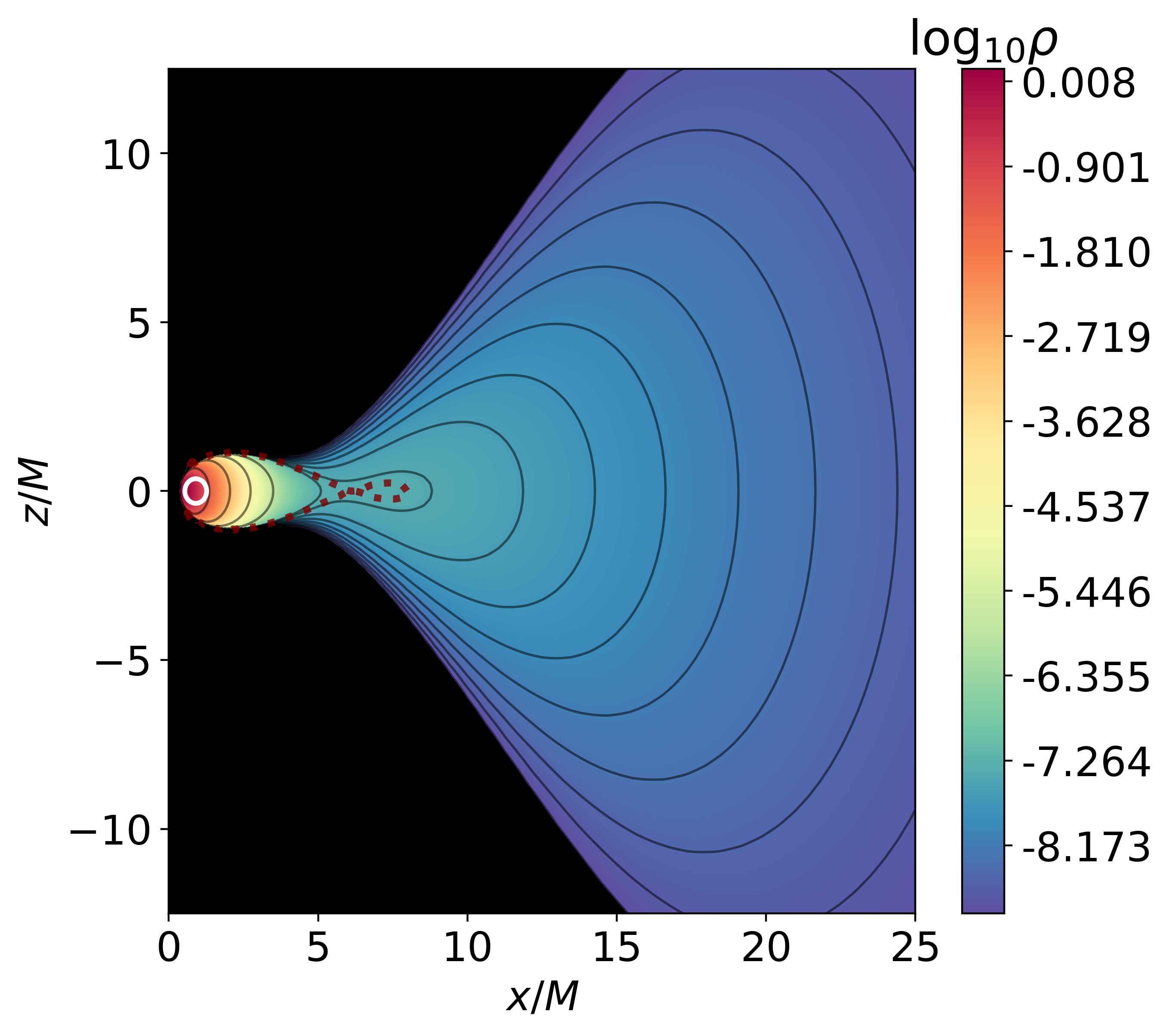}}
\end{floatrow}
\caption{$B$. $\omega=0.671, \ell_0 = -4.5M$: Density distribution visualized for different magnetization parameters, the minimum of the density is set to $\log_{10}\rho = -8.9$ and the maximum to the inner density maximum of the high magnetized solution. The black solid lines represent equidensity surfaces, the red dotted lines represent the cusp.}
\label{fig:Torus_0.671_-4.5}
\end{figure}

\paragraph{$\omega=0.671, \ell_0 = -4.5M$:}
The solution presented in Fig.~\ref{fig:density_0.671_-4.5} is highly relativistic and composed of a two-centered disk connected by a cusp with a static surface. A high magnetization does not affect the torus geometry around the inner center considerably. Similar to the other solutions it gets denser and more compressed, thus the density center is located closer to the BS center as compared to the non-magnetized case. In contrast, the geometry around the cusp and outer center is more affected by strong magnetic fields. The density cusp is located slightly further away from the BS center, whereas the outer density center moves closer to it, therefore the distance between density cusp and outer center decreases significantly, as shown in Fig.~\ref{fig:Cusp_center_0.671_-4.5}. The distance between them converges for low magnetization parameters to $\Delta r \equiv r_{c_o} - r_{cusp} = 1.09$. Furthermore the corresponding density values are similar for a high magnetization. The difference between them converges to $\Delta \rho \equiv \rho_{c_0} - \rho_{cusp} = 6.93 \cdot 10^{-10}$. Since the difference of the densities is significantly small, the physical properties of the disk between density cusp and outer center would be similar, leaving it hard to distinguish between them. It should be noted that the density in general is very small around the outer torus center, having a magnitude around $10^{-7}$ in the strong magnetized case. Equidensity surfaces around the outer center become smaller in diameter for a low magnetization parameter, as seen in Fig.~\ref{fig:Torus_0.671_-4.5}. In general we conclude that strong magnetic fields are suppressing the outer density center. Since the static surface is located close to the inner density center, the properties in and at the static surface are only slightly influenced by strong magnetization. \\

\begin{figure}[H]
    \centering
    \subfloat[]{\includegraphics[width=0.725\linewidth]{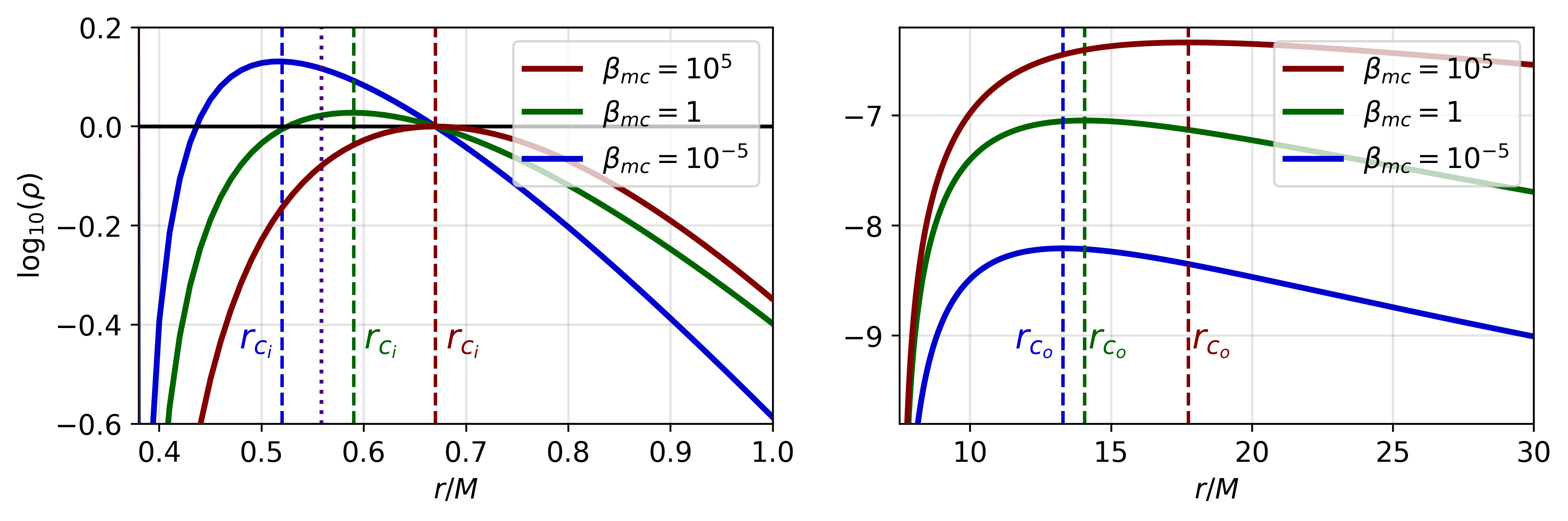}}
    \subfloat[]{\includegraphics[width=0.265\linewidth]{Figures/Results/Two-centered/scalar_profile_0.671.png}}
    
    \bigskip
    
    \begin{tabular}{c|c|c|c|c|c|c|c|c|c|c}
    \toprule
    $\log_{10}\beta_{mc}$ &  $\log_{10}\rho_{c_i}$ & $r_{c_i}$ & $\log_{10}\rho_{c_o}$ &  $r_{c_o}$ & $\log_{10}\rho_{cusp}$ & $r_{cusp}$ & $\log_{10}\left(\frac{\rho_{c_i}}{\rho_{c_o}}\right)$ &  $\log_{10}\left(p_{max}\right)$ & $\log_{10}\left(p_{m_{max}}\right)$ & $h_{max}$ \\
    \midrule
     5 &        0.000 &       0.67 &   -6.337 &      17.744 &       - &       -  &       6.337 &                           -0.092 &                               -5.081 &                            4.237 \\
            0 &         0.028 &       0.59 &   -7.046 &      14.053 &       - &       - &       7.074 &                           -0.427 &                               -0.455 &                            2.404 \\
           -5 &         0.131 &       0.52 &   -8.207 &      13.293 &       -- &       - &       8.338 &                           -5.268 &                               -0.350&                            1.000 \\
    \bottomrule
    \end{tabular}
    
    \caption{$C$. $\omega=0.671, \ell_0 = -5M$: (a) The left panel shows the equatorial density around the inner density center of the disk, the right panel shows the equatorial density around the outer density center of the disk. Dotted vertical lines are marking the positions of the inner and outer center. The indigo dotted line represents the location of the inner intersection of the static surfaces. (b) Scalar profile of the BS solution, the maximum is located at $r = 0.893$.}
    \label{fig:density_0.671_-5}
\end{figure}

\paragraph{$\omega=0.671, \ell_0 = -5M$:}
Considering the same BS solution as in example B. and setting the specific angular momentum to $\ell_0 = -5M$ leads to a two-centered solution without a cusp, as presented in Fig.~\ref{fig:density_0.671_-5}. The outer density center moves closer to the BS center for a low magnetization parameter and the density values of the outer torus are for the most part more than two orders of magnitude lower compared to the non-magnetized case, meaning there is almost no matter in the outer torus (in comparison to the inner torus), with density levels around $\sim 10^{-9}$ (Fig.~\ref{fig:Torus_0.671_-4.5}). The inner torus with the static surface behaves similarly to the solution with the cusp. \\

\begin{figure}[H]
    \centering
    \subfloat[]{\includegraphics[width=0.725\linewidth]{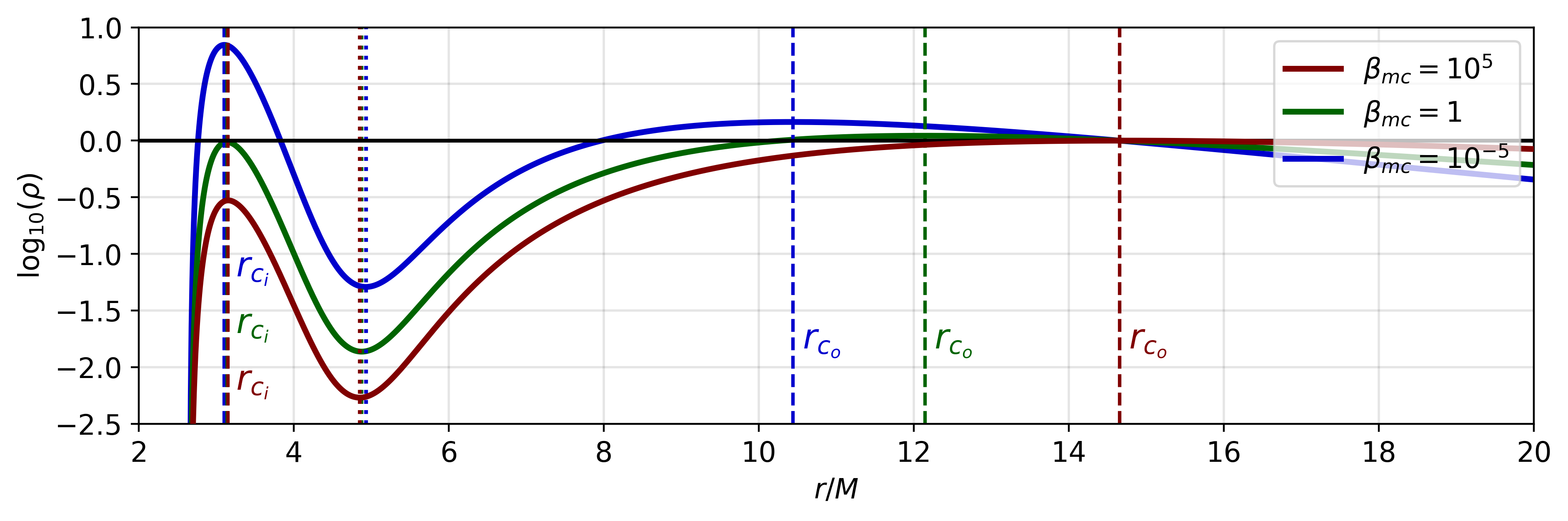}}
    \subfloat[]{\includegraphics[width=0.265\linewidth]{Figures/Results/Two-centered/scalar_profile_0.798_-4.5.png}}
    
    \bigskip
    
    \begin{tabular}{c|c|c|c|c|c|c|c|c|c|c}
    \toprule
    $\log_{10}\beta_{mc}$ & $\rho_{c_i}$ & $r_{c_i}$ & $\rho_{c_o}$ &  $r_{c_o}$ & $\rho_{cusp}$ & $r_{cusp}$ & $\frac{\rho_{c_i}}{\rho_{c_o}}$ &  $\log_{10}\left(p_{max}\right)$ & $\log_{10}\left(p_{m_{max}}\right)$ & $h_{max}$ \\
    \midrule
    5 &         0.297 &      3.151 &    1.000 &      14.653 &        0.005 &      4.851 &       0.297 &                          -2.139 &                               -7.121 &                            1.019 \\
            0 &         0.969 &      3.141 &    1.104 &      12.142 &        0.014 &      4.871 &       0.878 &                   -2.384 &                               -2.426 &                            1.014 \\
           -5 &         6.998 &      3.101 &    1.461 &      10.442 &        0.051 &      4.931 &       4.790 &                      -6.018 &                               -1.476 &                            1.000 \\
    \bottomrule
    \end{tabular}
    
    \caption{$D$. $\omega=0.798, \ell_0 = -4.75M$: (a) $\log_{10}$ of the density in the equatorial plane. Dashed lines show the locations of the inner and outer density center, the dotted lines show the position of the density cusp. (b) Scalar profile of the BS solution, the maximum is located at $r = 2.402$.}
    \label{fig:density_0.798_-4.75}
\end{figure}

\begin{figure}[H]
\centering
   \subfloat[\label{Center_0.798_-4.75}]{\includegraphics[width=0.42\columnwidth]{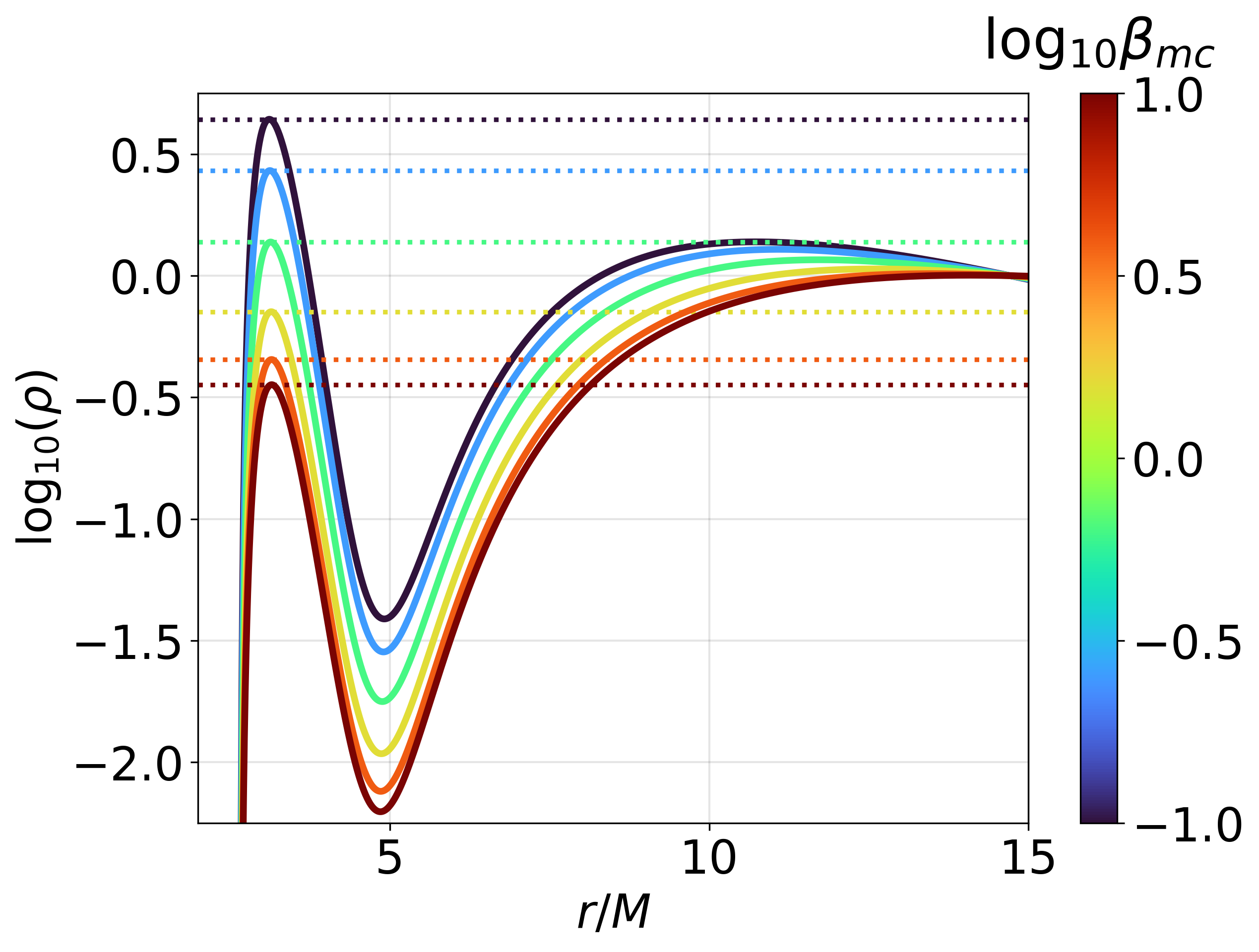}}\hfill
   \subfloat[\label{Threshold_0.798_-4.75}]{\includegraphics[width=0.43\columnwidth]{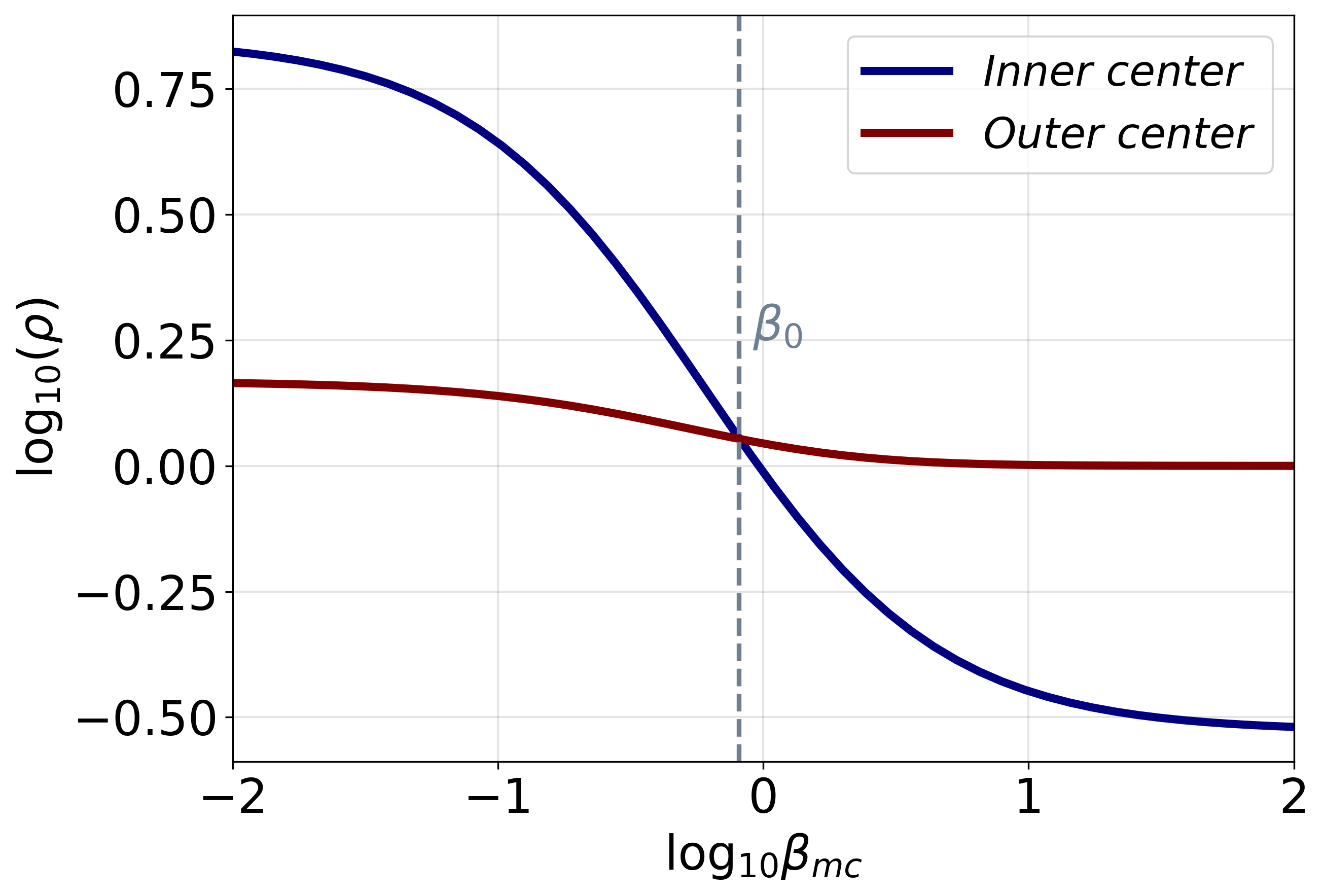}}
\caption{$D$. $\omega=0.798, \ell_0 = -4.75M$: (a) Equatorial density curves for $\beta_{mc}$ in the range of $0.1 \leq \beta_{mc} \leq 10$. Horizontal dotted lines mark the density value at the inner density center. 
(b) Density values of the inner density center and outer density center versus $\beta_{mc}$. 
The dashed vertical line represents the threshold value $\beta_0 = 0.814$.}
\label{fig:Inner_Outer_0.798_-4.5}
\end{figure}

\begin{figure}[H]
\centering
\begin{floatrow}
  \subfloat[$\beta_{mc}=10^5$]{\includegraphics[width=0.3283\columnwidth]{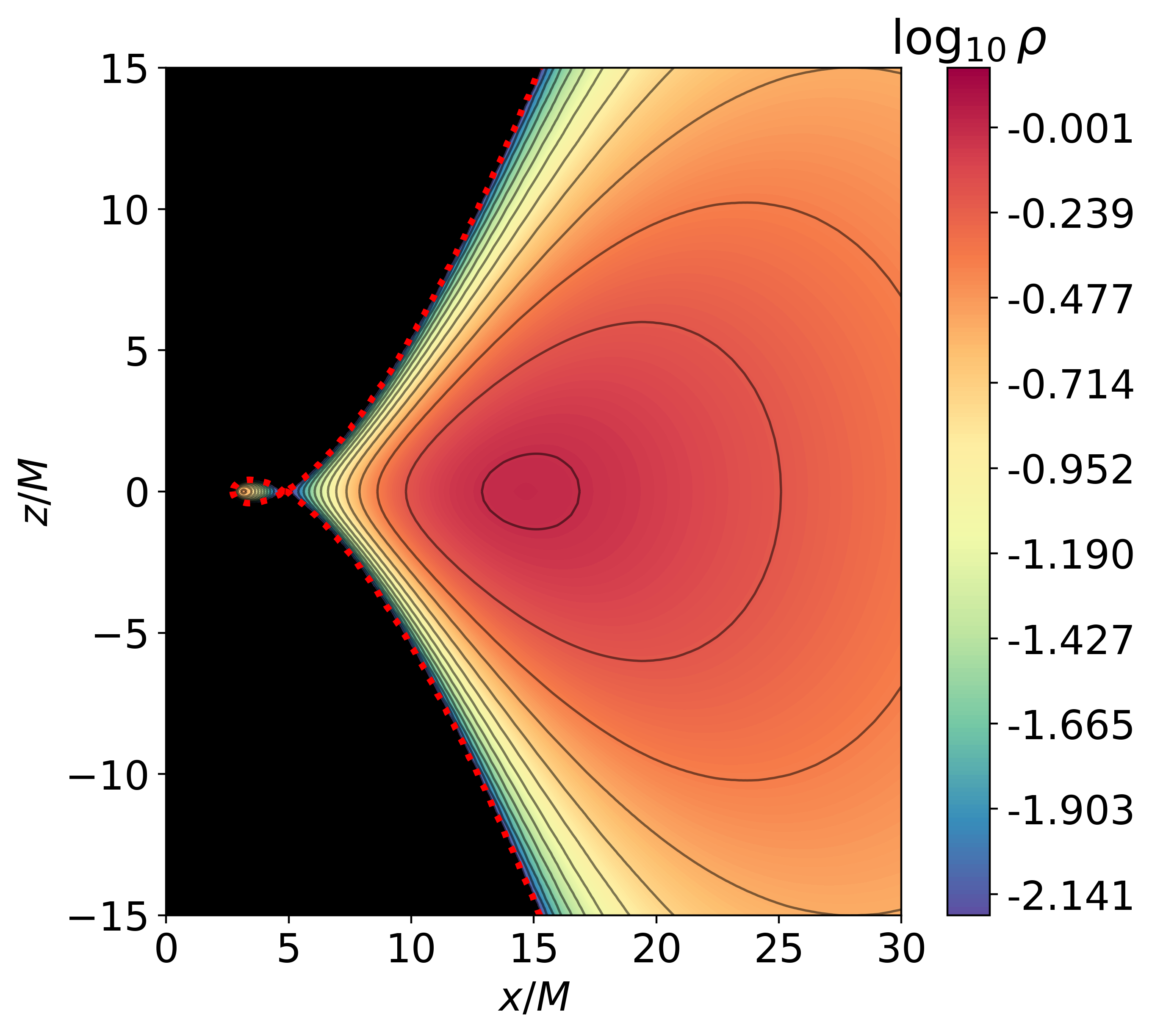}}
  \subfloat[$\beta_{mc}=1$]{\includegraphics[width=0.3283\columnwidth]{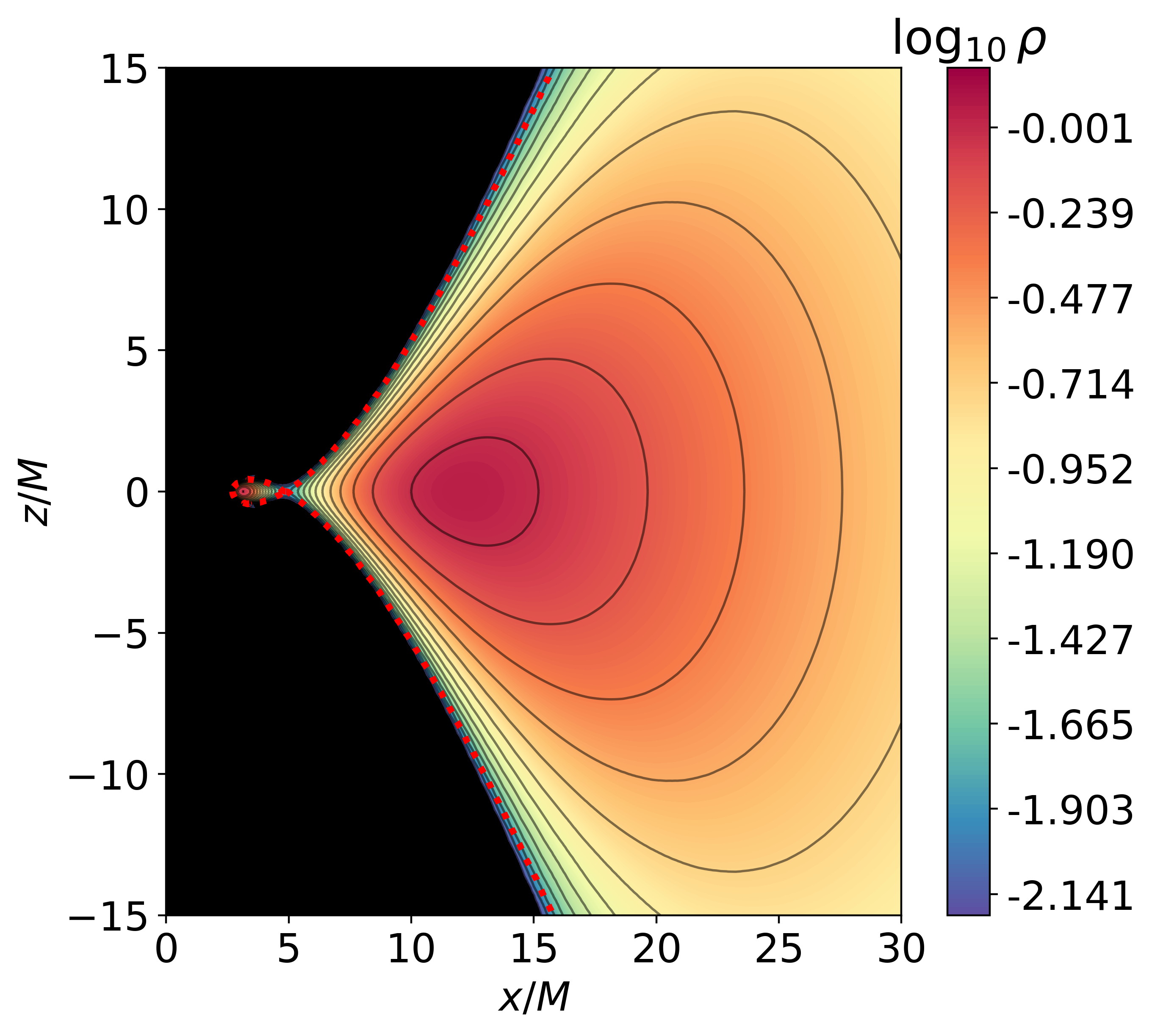}}
  \subfloat[$\beta_{mc}=10^{-5}$]{\includegraphics[width=0.3283\columnwidth]{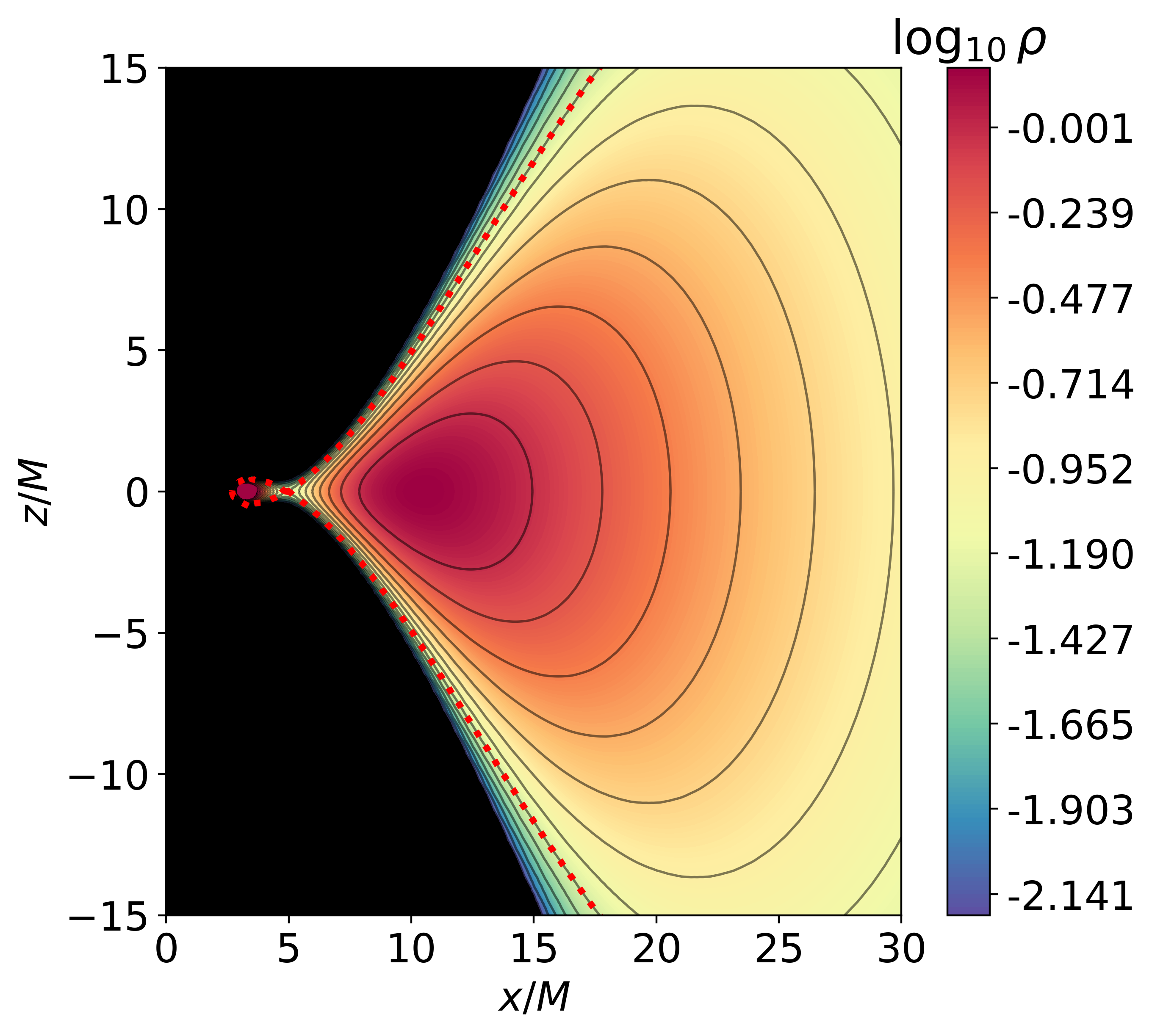}}
\end{floatrow}
\caption{$D$. $\omega=0.798, \ell_0 = -4.75M$: Density distribution visualized for different magnetization parameters. The minimum density is set to $\log_{10}\rho= -2.2$ and the maximum to the outer density maximum of the high magnetized solution. The black solid lines represent equidensity surfaces, the dotted red line represents the equidensity surface corresponding to the cusp.}
\label{fig:Torus_0.798_-4.75}
\end{figure}

\paragraph{$\omega=0.798, \ell_0 = -4.75M$:}
Looking at the $\omega = 0.798$ BS solution, there exist also two-centered solutions with a cusp, which are composed of a denser outer disk center, as seen in Fig.~\ref{fig:density_0.798_-4.75}. In the high magnetized solution the inner density center as well as the outer density center become denser compared to the non-magnetized solution, with the inner one being denser than the outer one. For all radial values smaller than the outer disk center of the non-magnetized solution, the equatorial density of the magnetized disks is higher compared to the non-magnetized solution. As a consequence there are higher density values at and around the density cusp. Fig.~\ref{fig:Torus_0.798_-4.75} shows the two-dimensional density distribution. The locations of the inner density center and density cusp are not significantly influenced by strong magnetization. The outer density center moves closer to the BS center and the cusp. Considering that the inner center becomes denser for a strong magnetization, there must be a threshold value $\beta_0$, below which the outer center has lower density. Fig.~\ref{fig:Inner_Outer_0.798_-4.5} presents solutions close to this threshold value. Since $\beta_0$ marks the intersection between the density of the inner and outer center, a two-centered magnetized disk solution with the same density at both centers is possible for this magnetization parameter. 

\begin{figure}[H]
\centering
\begin{floatrow}
  \subfloat[$\beta_{mc}=10^5$]{\includegraphics[width=0.3283\columnwidth]{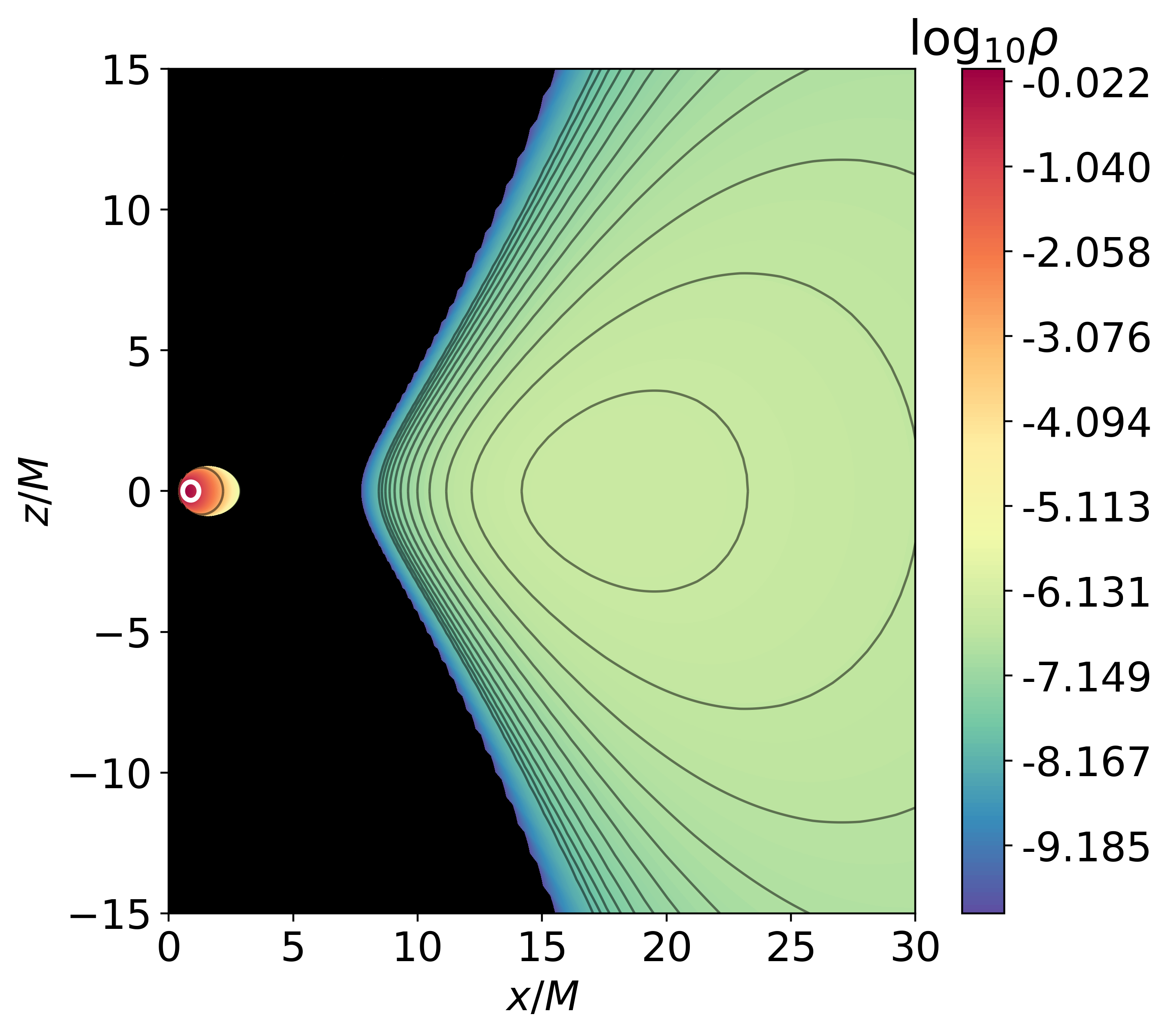}}
  \subfloat[$\beta_{mc}= 1$]{\includegraphics[width=0.3283\columnwidth]{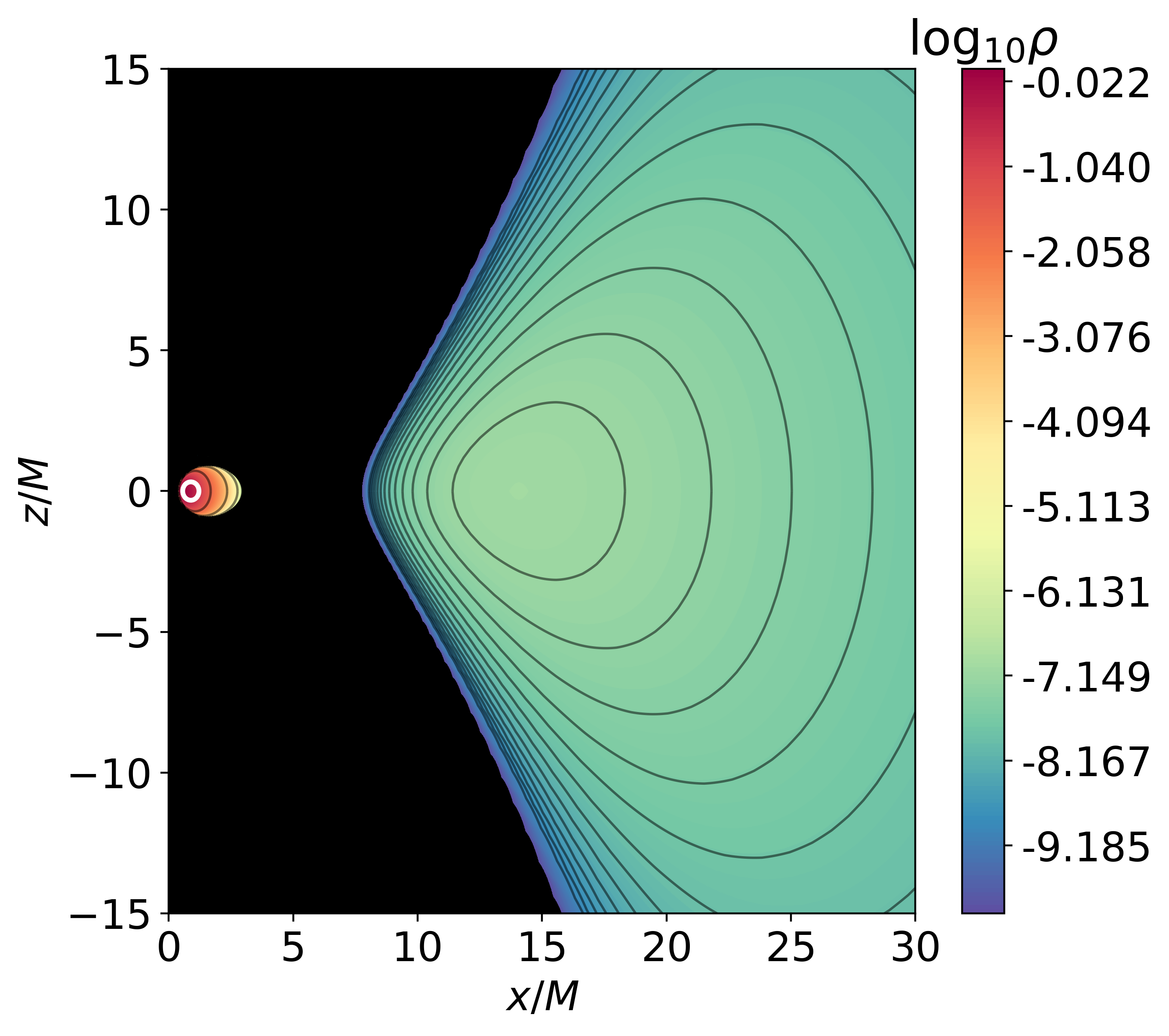}}
  \subfloat[$\beta_{mc}=10^{-5}$]{\includegraphics[width=0.3283\columnwidth]{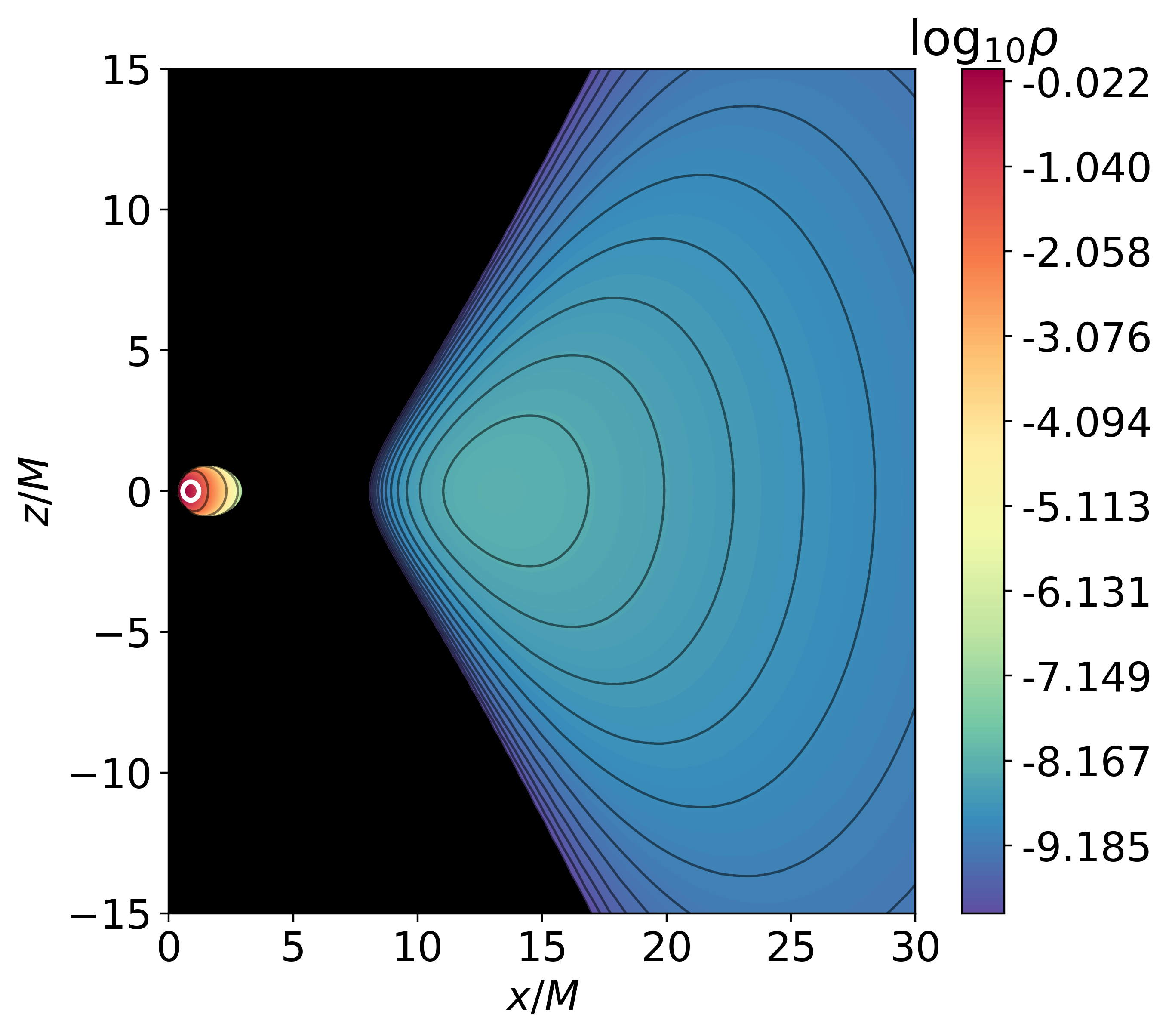}}
\end{floatrow}
\caption{$C$. $\omega=0.671, \ell_0 = -5M$: Density distribution visualized for different magnetization parameters, the minimum of the density is set to $\log_{10}\rho = -10$ and the maximum to the inner density maximum of the high magnetized solution. The black solid lines represent equidensity surfaces.}
\label{fig:torus_0.671_-5}
\end{figure}

\newpage
\section{Conclusion}
In this work we investigate the properties of non-selfgravitating magnetized thick disks in rotating boson star spacetimes. 
For the boson stars we here employ a massive complex scalar field without self-interactions, minimally coupled to gravity.
For such rotating boson stars a nonaxisymmetric dynamical instability has been found in \cite{Sanchis}. 
However, when the boson star frequency $\omega$ approaches its upper limit given by the boson mass $m$, the growth rate of this instability scales like $\propto \left(1 - \omega / m \right)$, and thus the growth rate approaches zero in the Newtonian limit \cite{Siemonsen:2020hcg}.
Furthermore, when appropriate nonlinear scalar self-interactions of the scalar field are included also the instability of the boson stars on the relativistic branch is quenched \cite{Siemonsen:2020hcg}.
Thus also highly compact rotating boson stars may be dynamically stable on relevant astrophysical scales, and possibly exhibit similar features for magnetized accretion disks as the ones observed here.

We construct the equilibrium configurations as solutions to the relativistic Euler equations assuming the presence of a toroidal magnetic field and constant angular momentum of the disk. We further describe the accreting plasma by a perfect fluid with polytropic equation of state. The influence of the magnetic field is evaluated by considering models with different degree of magnetization
quantified by the ratio of the thermal and magnetic pressure, which serves as a magnetization parameter. In order to compare to the purely hydrodynamical case we choose several representative boson star solutions ranging from less to highly relativistic, which are characterized by a qualitatively different morphology of the equilibrium tori. Then we analyse the impact of the magnetic field by exploring certain distinctive features such as modifications in the disk compactness, location of the density cusps and density centers, distribution of the predominant fluid density, transitions in the geometry or topology of the equilibrium configurations.

We observe the following systematic behavior. Mildly and less relativistic boson stars, which in the purely hydrodynamical case support equilibrium tori with a single center and no cusp preserve the disk morphology when a magnetic field is present. However, when the magnetization increases the disks become more compact and the predominant matter distribution shifts inwards towards the boson star center. This is manifested quantitatively by a range of features. In particular, the location of the density center moves closer to the boson star, the maximum density of the torus increases, and the predominant fluid density concentrates in smaller regions of the disk. In addition, for strong magnetic fields the disk can become highly compressed towards the rotational axis and the geometry of the equidensity surfaces can change from oblate to prolate. Due to the steep density gradients, these solutions could be unstable when time-evolved. In general the described effects are consistent with the studies of magnetized equilibrium tori around black holes, where similar qualitative behavior is observed \cite{Font_2019, Font_2021}.

In the case of highly relativistic boson stars we observe some qualitatively new features induced by the magnetic field. In the purely hydrodynamical case these solutions are characterized by two-centered equilibrium tori which can be either disjoint or connected by a cusp. We demonstrate that a sufficiently strong magnetic field can cause a transition in the disk topology. After a certain critical value of the magnetization parameter, some configurations lose their outer density center and become one-torus solutions. For other configurations the outer density center does not disintegrate in the strict sense but it moves inwards and the fluid density decreases so much in this region that it becomes insignificant. In these cases the ratio of the maximum fluid density at the inner and outer density center reaches eight orders of magnitude. Another phenomenon which we observed is a shift of the predominant fluid density between the torus centers. Configurations in which the fluid density is concentrated at the outer center in the purely hydrodynamical case can turn into configurations with dominant matter distribution at the inner center under a sufficiently strong magnetic field. The described effects can be interpreted as a manifestation of the same physical behavior as for the mildly relativistic boson stars. They illustrate further the trend that the presence of magnetic field leads to denser and more strongly compactified equilibrium configurations which are concentrated closer to the  gravitational field center. 
The non-linear evolution of the described equilibrium tori and their impact as initial conditions for the GRMHD simulations may be subject to further investigations.

\begin{acknowledgments}
We would like to gratefully acknowledge support by the DFG Research Training Group 1620 \textit{Models of Gravity}. J.K. gratefully acknowledges DFG project Ku612/18-1, P. N. gratefully acknowledges support by the Bulgarian National Science Fund Grants KP-06-RUSSIA/13 and KP-06-H38/2.
\end{acknowledgments}

\bibliography{literature}

\begin{thebibliography}{58}%
\makeatletter
\providecommand \@ifxundefined [1]{%
 \@ifx{#1\undefined}
}%
\providecommand \@ifnum [1]{%
 \ifnum #1\expandafter \@firstoftwo
 \else \expandafter \@secondoftwo
 \fi
}%
\providecommand \@ifx [1]{%
 \ifx #1\expandafter \@firstoftwo
 \else \expandafter \@secondoftwo
 \fi
}%
\providecommand \natexlab [1]{#1}%
\providecommand \enquote  [1]{``#1''}%
\providecommand \bibnamefont  [1]{#1}%
\providecommand \bibfnamefont [1]{#1}%
\providecommand \citenamefont [1]{#1}%
\providecommand \href@noop [0]{\@secondoftwo}%
\providecommand \href [0]{\begingroup \@sanitize@url \@href}%
\providecommand \@href[1]{\@@startlink{#1}\@@href}%
\providecommand \@@href[1]{\endgroup#1\@@endlink}%
\providecommand \@sanitize@url [0]{\catcode `\\12\catcode `\$12\catcode
  `\&12\catcode `\#12\catcode `\^12\catcode `\_12\catcode `\%12\relax}%
\providecommand \@@startlink[1]{}%
\providecommand \@@endlink[0]{}%
\providecommand \url  [0]{\begingroup\@sanitize@url \@url }%
\providecommand \@url [1]{\endgroup\@href {#1}{\urlprefix }}%
\providecommand \urlprefix  [0]{URL }%
\providecommand \Eprint [0]{\href }%
\providecommand \doibase [0]{https://doi.org/}%
\providecommand \selectlanguage [0]{\@gobble}%
\providecommand \bibinfo  [0]{\@secondoftwo}%
\providecommand \bibfield  [0]{\@secondoftwo}%
\providecommand \translation [1]{[#1]}%
\providecommand \BibitemOpen [0]{}%
\providecommand \bibitemStop [0]{}%
\providecommand \bibitemNoStop [0]{.\EOS\space}%
\providecommand \EOS [0]{\spacefactor3000\relax}%
\providecommand \BibitemShut  [1]{\csname bibitem#1\endcsname}%
\let\auto@bib@innerbib\@empty
\bibitem [{\citenamefont {Akiyama~et al.}(2019)}]{EHT_1}%
  \BibitemOpen
  \bibfield  {author} {\bibinfo {author} {\bibfnamefont {K.}~\bibnamefont
  {Akiyama~et al.}} (\bibinfo {collaboration} {Event Horizon Telescope}),\
  }\href@noop {} {\bibfield  {journal} {\bibinfo  {journal} {Astrophys. J.
  Lett.}\ }\textbf {\bibinfo {volume} {875}},\ \bibinfo {pages} {L1–L6}
  (\bibinfo {year} {2019})}\BibitemShut {NoStop}%
\bibitem [{\citenamefont {Akiyama~et al.}(2021)}]{EHT_10}%
  \BibitemOpen
  \bibfield  {author} {\bibinfo {author} {\bibfnamefont {K.}~\bibnamefont
  {Akiyama~et al.}} (\bibinfo {collaboration} {Event Horizon Telescope}),\
  }\href@noop {} {\bibfield  {journal} {\bibinfo  {journal} {Astrophys. J.
  Lett.}\ }\textbf {\bibinfo {volume} {910}},\ \bibinfo {pages} {L12–L13}
  (\bibinfo {year} {2021})}\BibitemShut {NoStop}%
\bibitem [{\citenamefont {Akiyama~et al.}(2022{\natexlab{a}})}]{EHT_2}%
  \BibitemOpen
  \bibfield  {author} {\bibinfo {author} {\bibfnamefont {K.}~\bibnamefont
  {Akiyama~et al.}} (\bibinfo {collaboration} {Event Horizon Telescope}),\
  }\href@noop {} {\bibfield  {journal} {\bibinfo  {journal} {Astrophys. J.
  Lett.}\ }\textbf {\bibinfo {volume} {930}},\ \bibinfo {pages} {L12–L17}
  (\bibinfo {year} {2022}{\natexlab{a}})}\BibitemShut {NoStop}%
\bibitem [{\citenamefont {Fishbone}\ and\ \citenamefont
  {Moncrief}(1976)}]{Fishbone1}%
  \BibitemOpen
  \bibfield  {author} {\bibinfo {author} {\bibfnamefont {L.~G.}\ \bibnamefont
  {Fishbone}}\ and\ \bibinfo {author} {\bibfnamefont {V.}~\bibnamefont
  {Moncrief}},\ }\href@noop {} {\bibfield  {journal} {\bibinfo  {journal}
  {Astrophys. J.}\ }\textbf {\bibinfo {volume} {207}},\ \bibinfo {pages} {962}
  (\bibinfo {year} {1976})}\BibitemShut {NoStop}%
\bibitem [{\citenamefont {{Abramowicz}}\ \emph {et~al.}(1978)\citenamefont
  {{Abramowicz}}, \citenamefont {{Jaroszynski}},\ and\ \citenamefont
  {{Sikora}}}]{Abramowicz}%
  \BibitemOpen
  \bibfield  {author} {\bibinfo {author} {\bibfnamefont {M.}~\bibnamefont
  {{Abramowicz}}}, \bibinfo {author} {\bibfnamefont {M.}~\bibnamefont
  {{Jaroszynski}}},\ and\ \bibinfo {author} {\bibfnamefont {M.}~\bibnamefont
  {{Sikora}}},\ }\href@noop {} {\bibfield  {journal} {\bibinfo  {journal}
  {A\&A}\ }\textbf {\bibinfo {volume} {63}},\ \bibinfo {pages} {221} (\bibinfo
  {year} {1978})}\BibitemShut {NoStop}%
\bibitem [{\citenamefont {{Kozlowski}}\ \emph {et~al.}(1978)\citenamefont
  {{Kozlowski}}, \citenamefont {{Jaroszynski}},\ and\ \citenamefont
  {{Abramowicz}}}]{Kozlowski}%
  \BibitemOpen
  \bibfield  {author} {\bibinfo {author} {\bibfnamefont {M.}~\bibnamefont
  {{Kozlowski}}}, \bibinfo {author} {\bibfnamefont {M.}~\bibnamefont
  {{Jaroszynski}}},\ and\ \bibinfo {author} {\bibfnamefont {M.~A.}\
  \bibnamefont {{Abramowicz}}},\ }\href@noop {} {\bibfield  {journal} {\bibinfo
   {journal} {A\&A}\ }\textbf {\bibinfo {volume} {63}},\ \bibinfo {pages} {209}
  (\bibinfo {year} {1978})}\BibitemShut {NoStop}%
\bibitem [{\citenamefont {Qian}\ \emph {et~al.}(2009)\citenamefont {Qian},
  \citenamefont {Abramowicz}, \citenamefont {Fragile}, \citenamefont
  {Hor{\'a}k}, \citenamefont {Machida},\ and\ \citenamefont {Straub}}]{Qian}%
  \BibitemOpen
  \bibfield  {author} {\bibinfo {author} {\bibfnamefont {L.}~\bibnamefont
  {Qian}}, \bibinfo {author} {\bibfnamefont {M.~A.}\ \bibnamefont
  {Abramowicz}}, \bibinfo {author} {\bibfnamefont {P.~C.}\ \bibnamefont
  {Fragile}}, \bibinfo {author} {\bibfnamefont {J.}~\bibnamefont {Hor{\'a}k}},
  \bibinfo {author} {\bibfnamefont {M.}~\bibnamefont {Machida}},\ and\ \bibinfo
  {author} {\bibfnamefont {O.}~\bibnamefont {Straub}},\ }\href@noop {}
  {\bibfield  {journal} {\bibinfo  {journal} {A\&A}\ }\textbf {\bibinfo
  {volume} {498}},\ \bibinfo {pages} {471} (\bibinfo {year}
  {2009})}\BibitemShut {NoStop}%
\bibitem [{\citenamefont {Daigne}\ and\ \citenamefont {Font}(2004)}]{Daigne}%
  \BibitemOpen
  \bibfield  {author} {\bibinfo {author} {\bibfnamefont {F.}~\bibnamefont
  {Daigne}}\ and\ \bibinfo {author} {\bibfnamefont {J.~A.}\ \bibnamefont
  {Font}},\ }\href@noop {} {\bibfield  {journal} {\bibinfo  {journal} {MNRAS}\
  }\textbf {\bibinfo {volume} {349}},\ \bibinfo {pages} {841} (\bibinfo {year}
  {2004})}\BibitemShut {NoStop}%
\bibitem [{\citenamefont {Komissarov}(2006)}]{Komissarov:2006nz}%
  \BibitemOpen
  \bibfield  {author} {\bibinfo {author} {\bibfnamefont {S.~S.}\ \bibnamefont
  {Komissarov}},\ }\href@noop {} {\bibfield  {journal} {\bibinfo  {journal}
  {MNRAS}\ }\textbf {\bibinfo {volume} {368}},\ \bibinfo {pages} {993}
  (\bibinfo {year} {2006})}\BibitemShut {NoStop}%
\bibitem [{\citenamefont {Gimeno-Soler}\ and\ \citenamefont
  {Font}(2017)}]{Font}%
  \BibitemOpen
  \bibfield  {author} {\bibinfo {author} {\bibfnamefont {S.}~\bibnamefont
  {Gimeno-Soler}}\ and\ \bibinfo {author} {\bibfnamefont {J.~A.}\ \bibnamefont
  {Font}},\ }\href@noop {} {\bibfield  {journal} {\bibinfo  {journal} {A\& A}\
  }\textbf {\bibinfo {volume} {607}},\ \bibinfo {pages} {A68} (\bibinfo {year}
  {2017})}\BibitemShut {NoStop}%
\bibitem [{\citenamefont {Wielgus}\ \emph {et~al.}(2015)\citenamefont
  {Wielgus}, \citenamefont {Fragile}, \citenamefont {Wang},\ and\ \citenamefont
  {Wilson}}]{Wielgus}%
  \BibitemOpen
  \bibfield  {author} {\bibinfo {author} {\bibfnamefont {M.}~\bibnamefont
  {Wielgus}}, \bibinfo {author} {\bibfnamefont {P.~C.}\ \bibnamefont
  {Fragile}}, \bibinfo {author} {\bibfnamefont {Z.}~\bibnamefont {Wang}},\ and\
  \bibinfo {author} {\bibfnamefont {J.}~\bibnamefont {Wilson}},\ }\href@noop {}
  {\bibfield  {journal} {\bibinfo  {journal} {MNRAS}\ }\textbf {\bibinfo
  {volume} {447}},\ \bibinfo {pages} {3593} (\bibinfo {year}
  {2015})}\BibitemShut {NoStop}%
\bibitem [{\citenamefont {Igumenshchev}\ \emph {et~al.}(2003)\citenamefont
  {Igumenshchev}, \citenamefont {Narayan},\ and\ \citenamefont
  {Abramowicz}}]{Igumenshchev}%
  \BibitemOpen
  \bibfield  {author} {\bibinfo {author} {\bibfnamefont {I.~V.}\ \bibnamefont
  {Igumenshchev}}, \bibinfo {author} {\bibfnamefont {R.}~\bibnamefont
  {Narayan}},\ and\ \bibinfo {author} {\bibfnamefont {M.~A.}\ \bibnamefont
  {Abramowicz}},\ }\href@noop {} {\bibfield  {journal} {\bibinfo  {journal}
  {The Astrophysical Journal}\ }\textbf {\bibinfo {volume} {592}},\ \bibinfo
  {pages} {1042} (\bibinfo {year} {2003})}\BibitemShut {NoStop}%
\bibitem [{\citenamefont {McKinney}\ \emph {et~al.}(2012)\citenamefont
  {McKinney}, \citenamefont {Tchekhovskoy},\ and\ \citenamefont
  {Blandford}}]{McKinney}%
  \BibitemOpen
  \bibfield  {author} {\bibinfo {author} {\bibfnamefont {J.~C.}\ \bibnamefont
  {McKinney}}, \bibinfo {author} {\bibfnamefont {A.}~\bibnamefont
  {Tchekhovskoy}},\ and\ \bibinfo {author} {\bibfnamefont {R.~D.}\ \bibnamefont
  {Blandford}},\ }\href@noop {} {\bibfield  {journal} {\bibinfo  {journal}
  {Monthly Notices of the Royal Astronomical Society}\ }\textbf {\bibinfo
  {volume} {423}},\ \bibinfo {pages} {3083} (\bibinfo {year}
  {2012})}\BibitemShut {NoStop}%
\bibitem [{\citenamefont {Lahiri}\ \emph {et~al.}(2021)\citenamefont {Lahiri},
  \citenamefont {Gimeno-Soler}, \citenamefont {Font},\ and\ \citenamefont
  {Mej\'\i{}as}}]{Lahiri_2020}%
  \BibitemOpen
  \bibfield  {author} {\bibinfo {author} {\bibfnamefont {S.}~\bibnamefont
  {Lahiri}}, \bibinfo {author} {\bibfnamefont {S.}~\bibnamefont
  {Gimeno-Soler}}, \bibinfo {author} {\bibfnamefont {J.~A.}\ \bibnamefont
  {Font}},\ and\ \bibinfo {author} {\bibfnamefont {A.~M.}\ \bibnamefont
  {Mej\'\i{}as}},\ }\href {https://doi.org/10.1103/PhysRevD.103.044034}
  {\bibfield  {journal} {\bibinfo  {journal} {Phys. Rev. D}\ }\textbf {\bibinfo
  {volume} {103}},\ \bibinfo {pages} {044034} (\bibinfo {year}
  {2021})}\BibitemShut {NoStop}%
\bibitem [{\citenamefont {Vincent}\ \emph {et~al.}(2016)\citenamefont
  {Vincent}, \citenamefont {Meliani}, \citenamefont {Grandclement},
  \citenamefont {Gourgoulhon},\ and\ \citenamefont {Straub}}]{Vincent_2016}%
  \BibitemOpen
  \bibfield  {author} {\bibinfo {author} {\bibfnamefont {F.}~\bibnamefont
  {Vincent}}, \bibinfo {author} {\bibfnamefont {Z.}~\bibnamefont {Meliani}},
  \bibinfo {author} {\bibfnamefont {P.}~\bibnamefont {Grandclement}}, \bibinfo
  {author} {\bibfnamefont {E.}~\bibnamefont {Gourgoulhon}},\ and\ \bibinfo
  {author} {\bibfnamefont {O.}~\bibnamefont {Straub}},\ }\href@noop {}
  {\bibfield  {journal} {\bibinfo  {journal} {Class. Quant. Grav.}\ }\textbf
  {\bibinfo {volume} {33}},\ \bibinfo {pages} {105015} (\bibinfo {year}
  {2016})}\BibitemShut {NoStop}%
\bibitem [{\citenamefont {Olivares}\ \emph {et~al.}(2020)\citenamefont
  {Olivares}, \citenamefont {Younsi}, \citenamefont {Fromm}, \citenamefont
  {De~Laurentis}, \citenamefont {Porth}, \citenamefont {Mizuno}, \citenamefont
  {Falcke}, \citenamefont {Kramer},\ and\ \citenamefont {Rezzolla}}]{Olivares}%
  \BibitemOpen
  \bibfield  {author} {\bibinfo {author} {\bibfnamefont {H.}~\bibnamefont
  {Olivares}}, \bibinfo {author} {\bibfnamefont {Z.}~\bibnamefont {Younsi}},
  \bibinfo {author} {\bibfnamefont {C.~M.}\ \bibnamefont {Fromm}}, \bibinfo
  {author} {\bibfnamefont {M.}~\bibnamefont {De~Laurentis}}, \bibinfo {author}
  {\bibfnamefont {O.}~\bibnamefont {Porth}}, \bibinfo {author} {\bibfnamefont
  {Y.}~\bibnamefont {Mizuno}}, \bibinfo {author} {\bibfnamefont
  {H.}~\bibnamefont {Falcke}}, \bibinfo {author} {\bibfnamefont
  {M.}~\bibnamefont {Kramer}},\ and\ \bibinfo {author} {\bibfnamefont
  {L.}~\bibnamefont {Rezzolla}},\ }\href@noop {} {\bibfield  {journal}
  {\bibinfo  {journal} {MNRAS}\ }\textbf {\bibinfo {volume} {497}},\ \bibinfo
  {pages} {521} (\bibinfo {year} {2020})}\BibitemShut {NoStop}%
\bibitem [{\citenamefont {Vincent}\ \emph {et~al.}(2021)\citenamefont
  {Vincent}, \citenamefont {Wielgus}, \citenamefont {Abramowicz}, \citenamefont
  {Gourgoulhon}, \citenamefont {Lasota}, \citenamefont {Paumard},\ and\
  \citenamefont {Perrin}}]{Vincent_2021}%
  \BibitemOpen
  \bibfield  {author} {\bibinfo {author} {\bibfnamefont {F.}~\bibnamefont
  {Vincent}}, \bibinfo {author} {\bibfnamefont {M.}~\bibnamefont {Wielgus}},
  \bibinfo {author} {\bibfnamefont {M.}~\bibnamefont {Abramowicz}}, \bibinfo
  {author} {\bibfnamefont {E.}~\bibnamefont {Gourgoulhon}}, \bibinfo {author}
  {\bibfnamefont {J.-P.}\ \bibnamefont {Lasota}}, \bibinfo {author}
  {\bibfnamefont {T.}~\bibnamefont {Paumard}},\ and\ \bibinfo {author}
  {\bibfnamefont {G.}~\bibnamefont {Perrin}},\ }\href@noop {} {\bibfield
  {journal} {\bibinfo  {journal} {A\&A}\ }\textbf {\bibinfo {volume} {646}},\
  \bibinfo {pages} {A37} (\bibinfo {year} {2021})}\BibitemShut {NoStop}%
\bibitem [{\citenamefont {Akiyama~et al.}(2022{\natexlab{b}})}]{EHT_Sgr}%
  \BibitemOpen
  \bibfield  {author} {\bibinfo {author} {\bibfnamefont {K.}~\bibnamefont
  {Akiyama~et al.}} (\bibinfo {collaboration} {Event Horizon Telescope}),\
  }\href@noop {} {\bibfield  {journal} {\bibinfo  {journal} {Astrophys. J.
  Lett.}\ }\textbf {\bibinfo {volume} {930}},\ \bibinfo {pages} {L17} (\bibinfo
  {year} {2022}{\natexlab{b}})}\BibitemShut {NoStop}%
\bibitem [{\citenamefont {Gimeno-Soler}\ \emph {et~al.}(2019)\citenamefont
  {Gimeno-Soler}, \citenamefont {Font}, \citenamefont {Herdeiro},\ and\
  \citenamefont {Radu}}]{Font_2019}%
  \BibitemOpen
  \bibfield  {author} {\bibinfo {author} {\bibfnamefont {S.}~\bibnamefont
  {Gimeno-Soler}}, \bibinfo {author} {\bibfnamefont {J.~A.}\ \bibnamefont
  {Font}}, \bibinfo {author} {\bibfnamefont {C.}~\bibnamefont {Herdeiro}},\
  and\ \bibinfo {author} {\bibfnamefont {E.}~\bibnamefont {Radu}},\ }\href@noop
  {} {\bibfield  {journal} {\bibinfo  {journal} {Phys. Rev. D}\ }\textbf
  {\bibinfo {volume} {99}},\ \bibinfo {pages} {043002} (\bibinfo {year}
  {2019})}\BibitemShut {NoStop}%
\bibitem [{\citenamefont {Cruz-Osorio}\ \emph {et~al.}(2021)\citenamefont
  {Cruz-Osorio}, \citenamefont {Gimeno-Soler}, \citenamefont {Font},
  \citenamefont {De~Laurentis},\ and\ \citenamefont {Mendoza}}]{Font_2021}%
  \BibitemOpen
  \bibfield  {author} {\bibinfo {author} {\bibfnamefont {A.}~\bibnamefont
  {Cruz-Osorio}}, \bibinfo {author} {\bibfnamefont {S.}~\bibnamefont
  {Gimeno-Soler}}, \bibinfo {author} {\bibfnamefont {J.~A.}\ \bibnamefont
  {Font}}, \bibinfo {author} {\bibfnamefont {M.}~\bibnamefont {De~Laurentis}},\
  and\ \bibinfo {author} {\bibfnamefont {S.}~\bibnamefont {Mendoza}},\
  }\href@noop {} {\bibfield  {journal} {\bibinfo  {journal} {Phys. Rev. D}\
  }\textbf {\bibinfo {volume} {103}},\ \bibinfo {pages} {124009} (\bibinfo
  {year} {2021})}\BibitemShut {NoStop}%
\bibitem [{\citenamefont {Teodoro}\ \emph
  {et~al.}(2021{\natexlab{a}})\citenamefont {Teodoro}, \citenamefont
  {Collodel}, \citenamefont {Doneva}, \citenamefont {Kunz}, \citenamefont
  {Nedkova},\ and\ \citenamefont {Yazadjiev}}]{Teodoro_2021}%
  \BibitemOpen
  \bibfield  {author} {\bibinfo {author} {\bibfnamefont {M.~C.}\ \bibnamefont
  {Teodoro}}, \bibinfo {author} {\bibfnamefont {L.~G.}\ \bibnamefont
  {Collodel}}, \bibinfo {author} {\bibfnamefont {D.}~\bibnamefont {Doneva}},
  \bibinfo {author} {\bibfnamefont {J.}~\bibnamefont {Kunz}}, \bibinfo {author}
  {\bibfnamefont {P.}~\bibnamefont {Nedkova}},\ and\ \bibinfo {author}
  {\bibfnamefont {S.}~\bibnamefont {Yazadjiev}},\ }\href@noop {} {\bibfield
  {journal} {\bibinfo  {journal} {Phys. Rev. D}\ }\textbf {\bibinfo {volume}
  {104}},\ \bibinfo {pages} {124047} (\bibinfo {year}
  {2021}{\natexlab{a}})}\BibitemShut {NoStop}%
\bibitem [{\citenamefont {Gimeno-Soler}\ \emph {et~al.}(2021)\citenamefont
  {Gimeno-Soler}, \citenamefont {Font}, \citenamefont {Herdeiro},\ and\
  \citenamefont {Radu}}]{Gimeno-Soler}%
  \BibitemOpen
  \bibfield  {author} {\bibinfo {author} {\bibfnamefont {S.}~\bibnamefont
  {Gimeno-Soler}}, \bibinfo {author} {\bibfnamefont {J.~A.}\ \bibnamefont
  {Font}}, \bibinfo {author} {\bibfnamefont {C.}~\bibnamefont {Herdeiro}},\
  and\ \bibinfo {author} {\bibfnamefont {E.}~\bibnamefont {Radu}},\ }\href@noop
  {} {\bibfield  {journal} {\bibinfo  {journal} {Phys. Rev. D}\ }\textbf
  {\bibinfo {volume} {104}},\ \bibinfo {pages} {103008} (\bibinfo {year}
  {2021})}\BibitemShut {NoStop}%
\bibitem [{\citenamefont {Stuchl{\'\i}k}\ \emph {et~al.}(2000)\citenamefont
  {Stuchl{\'\i}k}, \citenamefont {Slan{\`y}},\ and\ \citenamefont
  {Hled{\'\i}k}}]{Stuchlik_2000}%
  \BibitemOpen
  \bibfield  {author} {\bibinfo {author} {\bibfnamefont {Z.}~\bibnamefont
  {Stuchl{\'\i}k}}, \bibinfo {author} {\bibfnamefont {P.}~\bibnamefont
  {Slan{\`y}}},\ and\ \bibinfo {author} {\bibfnamefont {S.}~\bibnamefont
  {Hled{\'\i}k}},\ }\href@noop {} {\bibfield  {journal} {\bibinfo  {journal}
  {A\&A}\ }\textbf {\bibinfo {volume} {363}},\ \bibinfo {pages} {425} (\bibinfo
  {year} {2000})}\BibitemShut {NoStop}%
\bibitem [{\citenamefont {Slan{\`y}}\ and\ \citenamefont
  {Stuchl{\'\i}k}(2005)}]{Slany}%
  \BibitemOpen
  \bibfield  {author} {\bibinfo {author} {\bibfnamefont {P.}~\bibnamefont
  {Slan{\`y}}}\ and\ \bibinfo {author} {\bibfnamefont {Z.}~\bibnamefont
  {Stuchl{\'\i}k}},\ }\href@noop {} {\bibfield  {journal} {\bibinfo  {journal}
  {Class. Quant. Grav.}\ }\textbf {\bibinfo {volume} {22}},\ \bibinfo {pages}
  {3623} (\bibinfo {year} {2005})}\BibitemShut {NoStop}%
\bibitem [{\citenamefont {Faraji}\ and\ \citenamefont
  {Trova}(2021{\natexlab{a}})}]{Faraji_2021a}%
  \BibitemOpen
  \bibfield  {author} {\bibinfo {author} {\bibfnamefont {S.}~\bibnamefont
  {Faraji}}\ and\ \bibinfo {author} {\bibfnamefont {A.}~\bibnamefont {Trova}},\
  }\href@noop {} {\bibfield  {journal} {\bibinfo  {journal} {Phys. Rev. D}\
  }\textbf {\bibinfo {volume} {104}},\ \bibinfo {pages} {083006} (\bibinfo
  {year} {2021}{\natexlab{a}})}\BibitemShut {NoStop}%
\bibitem [{\citenamefont {Faraji}\ and\ \citenamefont
  {Trova}(2021{\natexlab{b}})}]{Faraji_2021b}%
  \BibitemOpen
  \bibfield  {author} {\bibinfo {author} {\bibfnamefont {S.}~\bibnamefont
  {Faraji}}\ and\ \bibinfo {author} {\bibfnamefont {A.}~\bibnamefont {Trova}},\
  }\href@noop {} {\bibfield  {journal} {\bibinfo  {journal} {A\&A}\ }\textbf
  {\bibinfo {volume} {654}},\ \bibinfo {pages} {A100} (\bibinfo {year}
  {2021}{\natexlab{b}})}\BibitemShut {NoStop}%
\bibitem [{\citenamefont {Ad{\'a}mek}\ and\ \citenamefont
  {Stuchl{\'\i}k}(2013)}]{Adamek}%
  \BibitemOpen
  \bibfield  {author} {\bibinfo {author} {\bibfnamefont {K.}~\bibnamefont
  {Ad{\'a}mek}}\ and\ \bibinfo {author} {\bibfnamefont {Z.}~\bibnamefont
  {Stuchl{\'\i}k}},\ }\href@noop {} {\bibfield  {journal} {\bibinfo  {journal}
  {Class. Quant. Grav.}\ }\textbf {\bibinfo {volume} {30}},\ \bibinfo {pages}
  {205007} (\bibinfo {year} {2013})}\BibitemShut {NoStop}%
\bibitem [{\citenamefont {Stuchl\'\i{}k}\ \emph {et~al.}(2015)\citenamefont
  {Stuchl\'\i{}k}, \citenamefont {Pugliese}, \citenamefont {Schee},\ and\
  \citenamefont {Ku\v{c}\'akov\'a}}]{Stuchlik_2015}%
  \BibitemOpen
  \bibfield  {author} {\bibinfo {author} {\bibfnamefont {Z.}~\bibnamefont
  {Stuchl\'\i{}k}}, \bibinfo {author} {\bibfnamefont {D.}~\bibnamefont
  {Pugliese}}, \bibinfo {author} {\bibfnamefont {J.}~\bibnamefont {Schee}},\
  and\ \bibinfo {author} {\bibfnamefont {H.}~\bibnamefont {Ku\v{c}\'akov\'a}},\
  }\href@noop {} {\bibfield  {journal} {\bibinfo  {journal} {EPJC}\ }\textbf
  {\bibinfo {volume} {75}},\ \bibinfo {pages} {451} (\bibinfo {year}
  {2015})}\BibitemShut {NoStop}%
\bibitem [{\citenamefont {Jetzer}(1992)}]{Jetzer}%
  \BibitemOpen
  \bibfield  {author} {\bibinfo {author} {\bibfnamefont {P.}~\bibnamefont
  {Jetzer}},\ }\href@noop {} {\bibfield  {journal} {\bibinfo  {journal} {Phys.
  Rept.}\ }\textbf {\bibinfo {volume} {220}},\ \bibinfo {pages} {163} (\bibinfo
  {year} {1992})}\BibitemShut {NoStop}%
\bibitem [{\citenamefont {Schunck}\ and\ \citenamefont
  {Mielke}(2003)}]{Schunck}%
  \BibitemOpen
  \bibfield  {author} {\bibinfo {author} {\bibfnamefont {F.~E.}\ \bibnamefont
  {Schunck}}\ and\ \bibinfo {author} {\bibfnamefont {E.~W.}\ \bibnamefont
  {Mielke}},\ }\href@noop {} {\bibfield  {journal} {\bibinfo  {journal} {Class.
  Quant. Grav.}\ }\textbf {\bibinfo {volume} {20}},\ \bibinfo {pages} {R301}
  (\bibinfo {year} {2003})}\BibitemShut {NoStop}%
\bibitem [{\citenamefont {Liebling}\ and\ \citenamefont
  {Palenzuela}(2017)}]{Liebling}%
  \BibitemOpen
  \bibfield  {author} {\bibinfo {author} {\bibfnamefont {S.~L.}\ \bibnamefont
  {Liebling}}\ and\ \bibinfo {author} {\bibfnamefont {C.}~\bibnamefont
  {Palenzuela}},\ }\href {https://doi.org/10.1007/s41114-017-0007-y} {\bibfield
   {journal} {\bibinfo  {journal} {Living Rev. Rel.}\ }\textbf {\bibinfo
  {volume} {20}},\ \bibinfo {pages} {5} (\bibinfo {year} {2017})}\BibitemShut
  {NoStop}%
\bibitem [{\citenamefont {Kaup}(1968)}]{Kaup}%
  \BibitemOpen
  \bibfield  {author} {\bibinfo {author} {\bibfnamefont {D.~J.}\ \bibnamefont
  {Kaup}},\ }\href@noop {} {\bibfield  {journal} {\bibinfo  {journal} {Phys.
  Rev.}\ }\textbf {\bibinfo {volume} {172}},\ \bibinfo {pages} {1331} (\bibinfo
  {year} {1968})}\BibitemShut {NoStop}%
\bibitem [{\citenamefont {Feinblum}\ and\ \citenamefont
  {McKinley}(1968)}]{Feinblum}%
  \BibitemOpen
  \bibfield  {author} {\bibinfo {author} {\bibfnamefont {D.~A.}\ \bibnamefont
  {Feinblum}}\ and\ \bibinfo {author} {\bibfnamefont {W.~A.}\ \bibnamefont
  {McKinley}},\ }\href@noop {} {\bibfield  {journal} {\bibinfo  {journal}
  {Phys. Rev.}\ }\textbf {\bibinfo {volume} {168}},\ \bibinfo {pages} {1445}
  (\bibinfo {year} {1968})}\BibitemShut {NoStop}%
\bibitem [{\citenamefont {Ruffini}\ and\ \citenamefont
  {Bonazzola}(1969)}]{Ruffini}%
  \BibitemOpen
  \bibfield  {author} {\bibinfo {author} {\bibfnamefont {R.}~\bibnamefont
  {Ruffini}}\ and\ \bibinfo {author} {\bibfnamefont {S.}~\bibnamefont
  {Bonazzola}},\ }\href@noop {} {\bibfield  {journal} {\bibinfo  {journal}
  {Phys. Rev.}\ }\textbf {\bibinfo {volume} {187}},\ \bibinfo {pages} {1767}
  (\bibinfo {year} {1969})}\BibitemShut {NoStop}%
\bibitem [{\citenamefont {Kobayashi}\ \emph {et~al.}(1994)\citenamefont
  {Kobayashi}, \citenamefont {Kasai},\ and\ \citenamefont
  {Futamase}}]{Kobayashi}%
  \BibitemOpen
  \bibfield  {author} {\bibinfo {author} {\bibfnamefont {Y.-s.}\ \bibnamefont
  {Kobayashi}}, \bibinfo {author} {\bibfnamefont {M.}~\bibnamefont {Kasai}},\
  and\ \bibinfo {author} {\bibfnamefont {T.}~\bibnamefont {Futamase}},\
  }\href@noop {} {\bibfield  {journal} {\bibinfo  {journal} {Phys. Rev. D}\
  }\textbf {\bibinfo {volume} {50}},\ \bibinfo {pages} {7721} (\bibinfo {year}
  {1994})}\BibitemShut {NoStop}%
\bibitem [{\citenamefont {Yoshida}\ and\ \citenamefont
  {Eriguchi}(1997)}]{Yoshida}%
  \BibitemOpen
  \bibfield  {author} {\bibinfo {author} {\bibfnamefont {S.}~\bibnamefont
  {Yoshida}}\ and\ \bibinfo {author} {\bibfnamefont {Y.}~\bibnamefont
  {Eriguchi}},\ }\href@noop {} {\bibfield  {journal} {\bibinfo  {journal}
  {Phys. Rev. D}\ }\textbf {\bibinfo {volume} {56}},\ \bibinfo {pages} {762}
  (\bibinfo {year} {1997})}\BibitemShut {NoStop}%
\bibitem [{\citenamefont {Schunck}\ and\ \citenamefont
  {Mielke}(1998)}]{Schunck_1998}%
  \BibitemOpen
  \bibfield  {author} {\bibinfo {author} {\bibfnamefont {F.~E.}\ \bibnamefont
  {Schunck}}\ and\ \bibinfo {author} {\bibfnamefont {E.~W.}\ \bibnamefont
  {Mielke}},\ }\href {https://doi.org/10.1016/S0375-9601(98)00778-6} {\bibfield
   {journal} {\bibinfo  {journal} {Phys. Lett. A}\ }\textbf {\bibinfo {volume}
  {249}},\ \bibinfo {pages} {389} (\bibinfo {year} {1998})}\BibitemShut
  {NoStop}%
\bibitem [{\citenamefont {Kleihaus}\ \emph {et~al.}(2005)\citenamefont
  {Kleihaus}, \citenamefont {Kunz},\ and\ \citenamefont {List}}]{Kleihaus}%
  \BibitemOpen
  \bibfield  {author} {\bibinfo {author} {\bibfnamefont {B.}~\bibnamefont
  {Kleihaus}}, \bibinfo {author} {\bibfnamefont {J.}~\bibnamefont {Kunz}},\
  and\ \bibinfo {author} {\bibfnamefont {M.}~\bibnamefont {List}},\ }\href@noop
  {} {\bibfield  {journal} {\bibinfo  {journal} {Phys. Rev. D}\ }\textbf
  {\bibinfo {volume} {72}},\ \bibinfo {pages} {064002} (\bibinfo {year}
  {2005})}\BibitemShut {NoStop}%
\bibitem [{\citenamefont {Kleihaus}\ \emph {et~al.}(2008)\citenamefont
  {Kleihaus}, \citenamefont {Kunz}, \citenamefont {List},\ and\ \citenamefont
  {Schaffer}}]{Kleihaus_2008}%
  \BibitemOpen
  \bibfield  {author} {\bibinfo {author} {\bibfnamefont {B.}~\bibnamefont
  {Kleihaus}}, \bibinfo {author} {\bibfnamefont {J.}~\bibnamefont {Kunz}},
  \bibinfo {author} {\bibfnamefont {M.}~\bibnamefont {List}},\ and\ \bibinfo
  {author} {\bibfnamefont {I.}~\bibnamefont {Schaffer}},\ }\href@noop {}
  {\bibfield  {journal} {\bibinfo  {journal} {Phys. Rev. D}\ }\textbf {\bibinfo
  {volume} {77}},\ \bibinfo {pages} {064025} (\bibinfo {year}
  {2008})}\BibitemShut {NoStop}%
\bibitem [{\citenamefont {Collodel}\ \emph {et~al.}(2017)\citenamefont
  {Collodel}, \citenamefont {Kleihaus},\ and\ \citenamefont
  {Kunz}}]{Collodel_2017}%
  \BibitemOpen
  \bibfield  {author} {\bibinfo {author} {\bibfnamefont {L.~G.}\ \bibnamefont
  {Collodel}}, \bibinfo {author} {\bibfnamefont {B.}~\bibnamefont {Kleihaus}},\
  and\ \bibinfo {author} {\bibfnamefont {J.}~\bibnamefont {Kunz}},\ }\href@noop
  {} {\bibfield  {journal} {\bibinfo  {journal} {Phys. Rev. D}\ }\textbf
  {\bibinfo {volume} {96}},\ \bibinfo {pages} {084066} (\bibinfo {year}
  {2017})}\BibitemShut {NoStop}%
\bibitem [{\citenamefont {Collodel}\ \emph {et~al.}(2019)\citenamefont
  {Collodel}, \citenamefont {Kleihaus},\ and\ \citenamefont
  {Kunz}}]{Collodel_2019}%
  \BibitemOpen
  \bibfield  {author} {\bibinfo {author} {\bibfnamefont {L.~G.}\ \bibnamefont
  {Collodel}}, \bibinfo {author} {\bibfnamefont {B.}~\bibnamefont {Kleihaus}},\
  and\ \bibinfo {author} {\bibfnamefont {J.}~\bibnamefont {Kunz}},\ }\href@noop
  {} {\bibfield  {journal} {\bibinfo  {journal} {Phys. Rev. D}\ }\textbf
  {\bibinfo {volume} {99}},\ \bibinfo {pages} {104076} (\bibinfo {year}
  {2019})}\BibitemShut {NoStop}%
\bibitem [{\citenamefont {Lee}\ and\ \citenamefont {Pang}(1989)}]{Lee}%
  \BibitemOpen
  \bibfield  {author} {\bibinfo {author} {\bibfnamefont {T.}~\bibnamefont
  {Lee}}\ and\ \bibinfo {author} {\bibfnamefont {Y.}~\bibnamefont {Pang}},\
  }\href@noop {} {\bibfield  {journal} {\bibinfo  {journal} {Nucl. Phys. B}\
  }\textbf {\bibinfo {volume} {315}},\ \bibinfo {pages} {477} (\bibinfo {year}
  {1989})}\BibitemShut {NoStop}%
\bibitem [{\citenamefont {Kusmartsev}\ \emph {et~al.}(1991)\citenamefont
  {Kusmartsev}, \citenamefont {Mielke},\ and\ \citenamefont
  {Schunck}}]{Kusmartsev}%
  \BibitemOpen
  \bibfield  {author} {\bibinfo {author} {\bibfnamefont {F.~V.}\ \bibnamefont
  {Kusmartsev}}, \bibinfo {author} {\bibfnamefont {E.~W.}\ \bibnamefont
  {Mielke}},\ and\ \bibinfo {author} {\bibfnamefont {F.~E.}\ \bibnamefont
  {Schunck}},\ }\href@noop {} {\bibfield  {journal} {\bibinfo  {journal} {Phys.
  Rev. D}\ }\textbf {\bibinfo {volume} {43}},\ \bibinfo {pages} {3895}
  (\bibinfo {year} {1991})}\BibitemShut {NoStop}%
\bibitem [{\citenamefont {Sanchis-Gual}\ \emph {et~al.}(2019)\citenamefont
  {Sanchis-Gual}, \citenamefont {Di~Giovanni}, \citenamefont {Zilh{\~a}o},
  \citenamefont {Herdeiro}, \citenamefont {Cerd{\'a}-Dur{\'a}n}, \citenamefont
  {Font},\ and\ \citenamefont {Radu}}]{Sanchis}%
  \BibitemOpen
  \bibfield  {author} {\bibinfo {author} {\bibfnamefont {N.}~\bibnamefont
  {Sanchis-Gual}}, \bibinfo {author} {\bibfnamefont {F.}~\bibnamefont
  {Di~Giovanni}}, \bibinfo {author} {\bibfnamefont {M.}~\bibnamefont
  {Zilh{\~a}o}}, \bibinfo {author} {\bibfnamefont {C.}~\bibnamefont
  {Herdeiro}}, \bibinfo {author} {\bibfnamefont {P.}~\bibnamefont
  {Cerd{\'a}-Dur{\'a}n}}, \bibinfo {author} {\bibfnamefont {J.}~\bibnamefont
  {Font}},\ and\ \bibinfo {author} {\bibfnamefont {E.}~\bibnamefont {Radu}},\
  }\href@noop {} {\bibfield  {journal} {\bibinfo  {journal} {Phys. Rev. Lett.}\
  }\textbf {\bibinfo {volume} {123}},\ \bibinfo {pages} {221101} (\bibinfo
  {year} {2019})}\BibitemShut {NoStop}%
\bibitem [{\citenamefont {Siemonsen}\ and\ \citenamefont
  {East}(2021)}]{Siemonsen:2020hcg}%
  \BibitemOpen
  \bibfield  {author} {\bibinfo {author} {\bibfnamefont {N.}~\bibnamefont
  {Siemonsen}}\ and\ \bibinfo {author} {\bibfnamefont {W.~E.}\ \bibnamefont
  {East}},\ }\href {https://doi.org/10.1103/PhysRevD.103.044022} {\bibfield
  {journal} {\bibinfo  {journal} {Phys. Rev. D}\ }\textbf {\bibinfo {volume}
  {103}},\ \bibinfo {pages} {044022} (\bibinfo {year} {2021})},\ \Eprint
  {https://arxiv.org/abs/2011.08247} {arXiv:2011.08247 [gr-qc]} \BibitemShut
  {NoStop}%
\bibitem [{\citenamefont {Diemer}\ \emph {et~al.}(2013)\citenamefont {Diemer},
  \citenamefont {Eilers}, \citenamefont {Hartmann}, \citenamefont {Schaffer},\
  and\ \citenamefont {Toma}}]{Eilers}%
  \BibitemOpen
  \bibfield  {author} {\bibinfo {author} {\bibfnamefont {V.}~\bibnamefont
  {Diemer}}, \bibinfo {author} {\bibfnamefont {K.}~\bibnamefont {Eilers}},
  \bibinfo {author} {\bibfnamefont {B.}~\bibnamefont {Hartmann}}, \bibinfo
  {author} {\bibfnamefont {I.}~\bibnamefont {Schaffer}},\ and\ \bibinfo
  {author} {\bibfnamefont {C.}~\bibnamefont {Toma}},\ }\href@noop {} {\bibfield
   {journal} {\bibinfo  {journal} {Phys. Rev. D}\ }\textbf {\bibinfo {volume}
  {88}},\ \bibinfo {pages} {044025} (\bibinfo {year} {2013})}\BibitemShut
  {NoStop}%
\bibitem [{\citenamefont {Brihaye}\ \emph {et~al.}(2014)\citenamefont
  {Brihaye}, \citenamefont {Diemer},\ and\ \citenamefont {Hartmann}}]{Brihaye}%
  \BibitemOpen
  \bibfield  {author} {\bibinfo {author} {\bibfnamefont {Y.}~\bibnamefont
  {Brihaye}}, \bibinfo {author} {\bibfnamefont {V.}~\bibnamefont {Diemer}},\
  and\ \bibinfo {author} {\bibfnamefont {B.}~\bibnamefont {Hartmann}},\
  }\href@noop {} {\bibfield  {journal} {\bibinfo  {journal} {Phys. Rev. D}\
  }\textbf {\bibinfo {volume} {89}},\ \bibinfo {pages} {084048} (\bibinfo
  {year} {2014})}\BibitemShut {NoStop}%
\bibitem [{\citenamefont {Grandclement}\ \emph {et~al.}(2014)\citenamefont
  {Grandclement}, \citenamefont {Som{\'e}},\ and\ \citenamefont
  {Gourgoulhon}}]{Grandclement}%
  \BibitemOpen
  \bibfield  {author} {\bibinfo {author} {\bibfnamefont {P.}~\bibnamefont
  {Grandclement}}, \bibinfo {author} {\bibfnamefont {C.}~\bibnamefont
  {Som{\'e}}},\ and\ \bibinfo {author} {\bibfnamefont {E.}~\bibnamefont
  {Gourgoulhon}},\ }\href@noop {} {\bibfield  {journal} {\bibinfo  {journal}
  {Phys. Rev. D}\ }\textbf {\bibinfo {volume} {90}},\ \bibinfo {pages} {024068}
  (\bibinfo {year} {2014})}\BibitemShut {NoStop}%
\bibitem [{\citenamefont {Meliani}\ \emph {et~al.}(2015)\citenamefont
  {Meliani}, \citenamefont {Vincent}, \citenamefont {Grandclément},
  \citenamefont {Gourgoulhon}, \citenamefont {Monceau-Baroux},\ and\
  \citenamefont {Straub}}]{Meliani}%
  \BibitemOpen
  \bibfield  {author} {\bibinfo {author} {\bibfnamefont {Z.}~\bibnamefont
  {Meliani}}, \bibinfo {author} {\bibfnamefont {F.~H.}\ \bibnamefont
  {Vincent}}, \bibinfo {author} {\bibfnamefont {P.}~\bibnamefont
  {Grandclément}}, \bibinfo {author} {\bibfnamefont {E.}~\bibnamefont
  {Gourgoulhon}}, \bibinfo {author} {\bibfnamefont {R.}~\bibnamefont
  {Monceau-Baroux}},\ and\ \bibinfo {author} {\bibfnamefont {O.}~\bibnamefont
  {Straub}},\ }\href {https://doi.org/10.1088/0264-9381/32/23/235022}
  {\bibfield  {journal} {\bibinfo  {journal} {Class. Quant. Grav.}\ }\textbf
  {\bibinfo {volume} {32}},\ \bibinfo {pages} {235022} (\bibinfo {year}
  {2015})}\BibitemShut {NoStop}%
\bibitem [{\citenamefont {Grould}\ \emph {et~al.}(2017)\citenamefont {Grould},
  \citenamefont {Meliani}, \citenamefont {Vincent}, \citenamefont
  {Grandcl{\'e}ment},\ and\ \citenamefont {Gourgoulhon}}]{Grould}%
  \BibitemOpen
  \bibfield  {author} {\bibinfo {author} {\bibfnamefont {M.}~\bibnamefont
  {Grould}}, \bibinfo {author} {\bibfnamefont {Z.}~\bibnamefont {Meliani}},
  \bibinfo {author} {\bibfnamefont {F.}~\bibnamefont {Vincent}}, \bibinfo
  {author} {\bibfnamefont {P.}~\bibnamefont {Grandcl{\'e}ment}},\ and\ \bibinfo
  {author} {\bibfnamefont {E.}~\bibnamefont {Gourgoulhon}},\ }\href@noop {}
  {\bibfield  {journal} {\bibinfo  {journal} {Class. Quant. Grav.}\ }\textbf
  {\bibinfo {volume} {34}},\ \bibinfo {pages} {215007} (\bibinfo {year}
  {2017})}\BibitemShut {NoStop}%
\bibitem [{\citenamefont {Collodel}\ \emph {et~al.}(2018)\citenamefont
  {Collodel}, \citenamefont {Kleihaus},\ and\ \citenamefont
  {Kunz}}]{Collodel_2017a}%
  \BibitemOpen
  \bibfield  {author} {\bibinfo {author} {\bibfnamefont {L.~G.}\ \bibnamefont
  {Collodel}}, \bibinfo {author} {\bibfnamefont {B.}~\bibnamefont {Kleihaus}},\
  and\ \bibinfo {author} {\bibfnamefont {J.}~\bibnamefont {Kunz}},\ }\href
  {https://doi.org/10.1103/PhysRevLett.120.201103} {\bibfield  {journal}
  {\bibinfo  {journal} {Phys. Rev. Lett.}\ }\textbf {\bibinfo {volume} {120}},\
  \bibinfo {pages} {201103} (\bibinfo {year} {2018})}\BibitemShut {NoStop}%
\bibitem [{\citenamefont {Meliani}\ \emph {et~al.}(2017)\citenamefont
  {Meliani}, \citenamefont {Casse}, \citenamefont {Grandcl{\'e}ment},
  \citenamefont {Gourgoulhon},\ and\ \citenamefont {Dauvergne}}]{Meliani_2017}%
  \BibitemOpen
  \bibfield  {author} {\bibinfo {author} {\bibfnamefont {Z.}~\bibnamefont
  {Meliani}}, \bibinfo {author} {\bibfnamefont {F.}~\bibnamefont {Casse}},
  \bibinfo {author} {\bibfnamefont {P.}~\bibnamefont {Grandcl{\'e}ment}},
  \bibinfo {author} {\bibfnamefont {E.}~\bibnamefont {Gourgoulhon}},\ and\
  \bibinfo {author} {\bibfnamefont {F.}~\bibnamefont {Dauvergne}},\ }\href@noop
  {} {\bibfield  {journal} {\bibinfo  {journal} {Class. Quant. Grav.}\ }\textbf
  {\bibinfo {volume} {34}},\ \bibinfo {pages} {225003} (\bibinfo {year}
  {2017})}\BibitemShut {NoStop}%
\bibitem [{\citenamefont {Teodoro}\ \emph
  {et~al.}(2021{\natexlab{b}})\citenamefont {Teodoro}, \citenamefont
  {Collodel},\ and\ \citenamefont {Kunz}}]{Teodoro_2021a}%
  \BibitemOpen
  \bibfield  {author} {\bibinfo {author} {\bibfnamefont {M.~C.}\ \bibnamefont
  {Teodoro}}, \bibinfo {author} {\bibfnamefont {L.~G.}\ \bibnamefont
  {Collodel}},\ and\ \bibinfo {author} {\bibfnamefont {J.}~\bibnamefont
  {Kunz}},\ }\href@noop {} {\bibfield  {journal} {\bibinfo  {journal} {Phys.
  Rev. D}\ }\textbf {\bibinfo {volume} {103}},\ \bibinfo {pages} {104064}
  (\bibinfo {year} {2021}{\natexlab{b}})}\BibitemShut {NoStop}%
\bibitem [{\citenamefont {Teodoro}\ \emph
  {et~al.}(2021{\natexlab{c}})\citenamefont {Teodoro}, \citenamefont
  {Collodel},\ and\ \citenamefont {Kunz}}]{Teodoro_2021b}%
  \BibitemOpen
  \bibfield  {author} {\bibinfo {author} {\bibfnamefont {M.~C.}\ \bibnamefont
  {Teodoro}}, \bibinfo {author} {\bibfnamefont {L.~G.}\ \bibnamefont
  {Collodel}},\ and\ \bibinfo {author} {\bibfnamefont {J.}~\bibnamefont
  {Kunz}},\ }\href {https://doi.org/10.1088/1475-7516/2021/03/063} {\bibfield
  {journal} {\bibinfo  {journal} {JCAP}\ }\textbf {\bibinfo {volume}
  {2021}}\bibinfo  {number} { (03)},\ \bibinfo {pages} {063}}\BibitemShut
  {NoStop}%
\bibitem [{\citenamefont {Font}\ and\ \citenamefont
  {Daigne}(2002)}]{Font_2002}%
  \BibitemOpen
\bibfield  {number} {  }\bibfield  {author} {\bibinfo {author} {\bibfnamefont
  {J.~A.}\ \bibnamefont {Font}}\ and\ \bibinfo {author} {\bibfnamefont
  {F.}~\bibnamefont {Daigne}},\ }\href
  {https://doi.org/10.1046/j.1365-8711.2002.05515.x} {\bibfield  {journal}
  {\bibinfo  {journal} {MNRAS}\ }\textbf {\bibinfo {volume} {334}},\ \bibinfo
  {pages} {383–400} (\bibinfo {year} {2002})}\BibitemShut {NoStop}%
\bibitem [{\citenamefont {Abramowicz}\ \emph {et~al.}(1983)\citenamefont
  {Abramowicz}, \citenamefont {Calvani},\ and\ \citenamefont
  {Nobili}}]{Abramowicz_1983}%
  \BibitemOpen
  \bibfield  {author} {\bibinfo {author} {\bibfnamefont {M.~A.}\ \bibnamefont
  {Abramowicz}}, \bibinfo {author} {\bibfnamefont {M.}~\bibnamefont
  {Calvani}},\ and\ \bibinfo {author} {\bibfnamefont {L.}~\bibnamefont
  {Nobili}},\ }\href@noop {} {\bibfield  {journal} {\bibinfo  {journal}
  {Nature}\ }\textbf {\bibinfo {volume} {302}},\ \bibinfo {pages} {597}
  (\bibinfo {year} {1983})}\BibitemShut {NoStop}%
\bibitem [{\citenamefont {Schauder}\ \emph {et~al.}(1992)\citenamefont
  {Schauder}, \citenamefont {Wei{\ss}},\ and\ \citenamefont
  {Sch{\"o}nauer}}]{Schauder:1992}%
  \BibitemOpen
  \bibfield  {author} {\bibinfo {author} {\bibfnamefont {M.}~\bibnamefont
  {Schauder}}, \bibinfo {author} {\bibfnamefont {R.}~\bibnamefont {Wei{\ss}}},\
  and\ \bibinfo {author} {\bibfnamefont {W.}~\bibnamefont {Sch{\"o}nauer}},\
  }\href@noop {} {\bibfield  {journal} {\bibinfo  {journal} {Universit{\"a}t
  Karlsruhe, Interner Bericht Nr.}\ }\textbf {\bibinfo {volume} {46/92}}
  (\bibinfo {year} {1992})}\BibitemShut {NoStop}%
\bibitem [{\citenamefont {Rezzolla}\ and\ \citenamefont
  {Zanotti}(2013)}]{Rezzolla}%
  \BibitemOpen
  \bibfield  {author} {\bibinfo {author} {\bibfnamefont {L.}~\bibnamefont
  {Rezzolla}}\ and\ \bibinfo {author} {\bibfnamefont {O.}~\bibnamefont
  {Zanotti}},\ }\href@noop {} {\emph {\bibinfo {title} {{Relativistic
  Hydrodynamics}}}}\ (\bibinfo  {publisher} {Oxford University Press},\
  \bibinfo {year} {2013})\BibitemShut {NoStop}%
\end{thebibliography}%

\end{document}